\documentclass[amsmath,amssymb,11pt,nofootinbib]{article}
\pdfoutput=1
\usepackage{jheppub}

\usepackage{hyperref}
\usepackage{amsmath}
\usepackage{graphicx}
\usepackage{xspace}
\usepackage{color}
\usepackage{overpic}
\usepackage{url}
\usepackage[utf8]{inputenc}
\usepackage{epstopdf}
\usepackage{tikz}
\usepackage{caption}
\usepackage{makecell}
\usepackage{bm}
\usepackage{revsymb}
\usepackage{slashed}
\usepackage{float}
\DeclareUnicodeCharacter{2212}{-}
\usepackage{pifont}
\usepackage{blindtext}
\usepackage{tabularx}
\usepackage{subfig}

\newcommand{\bea}{\begin{eqnarray}}
\newcommand{\eea}{\end{eqnarray}}

\newcommand{\eq}[1]{Eq.~\eqref{#1}}

\definecolor{point1}{rgb}{1,0,0}
\definecolor{point2}{rgb}{0.5,0,0.5}
\definecolor{point3}{rgb}{1,0.75,0}
\definecolor{point4}{rgb}{0,0.66,0}

\begin{document}
\preprint{CERN-TH-2020-167, PSI-20-17, ZU-TH 38/20}

\title{Scalar Leptoquarks in Leptonic Processes}

\author[a,b,c]{Andreas Crivellin}
\affiliation[a]{CERN Theory Division, CH--1211 Geneva 23, Switzerland}
\affiliation[b]{Physik-Institut, Universit\"at Z\"urich,
	Winterthurerstrasse 190, CH-8057 Z\"urich, Switzerland}
\affiliation[c]{Paul Scherrer Institut, CH--5232 Villigen PSI, Switzerland}

\author[d]{Christoph Greub}

\author[b,c]{Dario M\"uller}

\author[d]{Francesco Saturnino}
\affiliation[d]{Albert Einstein Center for Fundamental Physics, Institute
	for Theoretical Physics,\\ University of Bern, CH-3012 Bern,
	Switzerland}
	
\emailAdd{andreas.crivellin@cern.ch}
\emailAdd{greub@itp.unibe.ch}
\emailAdd{dario.mueller@psi.ch}
\emailAdd{saturnino@itp.unibe.ch}

\abstract{Leptoquarks are hypothetical new particles, which couple quarks directly to leptons. They experienced a renaissance in recent years as they are prime candidates to explain the so-called \textit{flavor anomalies}, i.e. the deviations between the Standard Model predictions and measurements in $b\to s\ell^{+}\ell^{-}$ and $b\to c\tau\nu$ processes and in the anomalous magnetic moment of the muon. At the one-loop level these particles unavoidably generate effects in the purely leptonic processes like $Z\to\ell^{+}\ell^{-}$, $Z\to\nu\bar\nu$, $W\to\ell\nu$ and $h\to\ell^{+}\ell^{-}$ and can even generate non-zero rates for lepton flavor violating processes such as $\ell\to \ell^{\prime}\gamma$, $Z\to\ell^+\ell^{\prime -}$, $h\to\ell^+\ell^{\prime -}$ and $\ell\to 3\ell^\prime$. In this article we calculate these processes for all five representations of scalar Leptoquarks. We include their most general interaction terms with the Standard Model Higgs boson, which leads to Leptoquark mixing after the former acquires a vacuum expectation value. In our phenomenological analysis we investigate the effects in modified lepton couplings to electroweak gauge bosons, we study the correlations of the anomalous magnetic moment of the muon with $h\to\mu^{+}\mu^{-}$ and $Z\to\mu^{+}\mu^{-}$ as well as the interplay between different lepton flavor violating decays.}


\maketitle

\newpage

\section{Introduction}

Leptoquarks (LQs) are particles with an interaction vertex connecting leptons with quarks. These particles are predicted by Grand Unified Theories~\cite{Pati:1974yy,Georgi:1974sy,Georgi:1974yf,Fritzsch:1974nn} and were systematically classified for the first time in Ref.~\cite{Buchmuller:1986zs} into ten possible representations under the Standard Model (SM) gauge group (five representations of scalar particles and five representations of vector particles). Their tree-level effects in low energy precision and flavor observables were studied comprehensively in Ref.~\cite{Davidson:1993qk}. After the disappearance of the HERA excess~\cite{Adloff:1997fg,Breitweg:1997ff}, which could have been interpreted as a LQ, the interest in LQs decreased until in recent years they experienced a renaissance due to the emergence of the \textit{flavor anomalies}.
\smallskip 

These flavor anomalies are hints for lepton flavor universality (LFU) violating NP in $R(D^{(*)})$~\cite{Lees:2012xj,Lees:2013uzd,Aaij:2015yra,Aaij:2017deq,Aaij:2017uff,Abdesselam:2019dgh}, $b\to s\ell^{+}\ell^{-}$~\cite{CMS:2014xfa,Aaij:2015oid,Abdesselam:2016llu,Aaij:2017vbb,Aaij:2019wad,Aaij:2020nrf} and in the anomalous magnetic moment (AMM) of the muon ($a_\mu$)~\cite{Bennett:2006fi}, with a significance of~$>3\,\sigma$~\cite{Amhis:2016xyh,Murgui:2019czp,Shi:2019gxi,Blanke:2019qrx,Kumbhakar:2019avh}, $>5\sigma$~\cite{Capdevila:2017bsm, Altmannshofer:2017yso,Alguero:2019ptt,Alok:2019ufo,Ciuchini:2019usw,Aebischer:2019mlg, Arbey:2019duh,Kumar:2019nfv} and \mbox{$>3\,\sigma$}~\cite{Aoyama:2020ynm}, respectively\footnote{Also the (apparent) deficit in first row CKM unitarity can be interpreted as a sign of LFU violation~\cite{Coutinho:2019aiy,Crivellin:2020lzu,Crivellin:2020ebi,Kirk:2020wdk}.}. In this context, it has been shown that LQs can explain $b\to s\ell^+\ell^-$ data~\cite{Alonso:2015sja, Calibbi:2015kma, Hiller:2016kry, Bhattacharya:2016mcc, Buttazzo:2017ixm, Barbieri:2015yvd, Barbieri:2016las, Calibbi:2017qbu, Crivellin:2017dsk, Bordone:2018nbg, Kumar:2018kmr, Crivellin:2018yvo, Crivellin:2019szf, Cornella:2019hct, Bordone:2019uzc, Bernigaud:2019bfy,Aebischer:2018acj,Fuentes-Martin:2019ign,Fajfer:2015ycq,  Blanke:2018sro,deMedeirosVarzielas:2019lgb,Varzielas:2015iva,Crivellin:2019dwb,Saad:2020ihm,Saad:2020ucl,Gherardi:2020qhc,1821789}, $R(D^{(*)})$~\cite{Alonso:2015sja, Calibbi:2015kma, Fajfer:2015ycq, Bhattacharya:2016mcc, Buttazzo:2017ixm, Barbieri:2015yvd, Barbieri:2016las, Calibbi:2017qbu, Bordone:2017bld, Bordone:2018nbg, Kumar:2018kmr, Biswas:2018snp, Crivellin:2018yvo, Blanke:2018sro, Heeck:2018ntp,deMedeirosVarzielas:2019lgb, Cornella:2019hct, Bordone:2019uzc,Sahoo:2015wya, Chen:2016dip, Dey:2017ede, Becirevic:2017jtw, Chauhan:2017ndd, Becirevic:2018afm, Popov:2019tyc,Fajfer:2012jt, Deshpande:2012rr, Freytsis:2015qca, Bauer:2015knc, Li:2016vvp, Zhu:2016xdg, Popov:2016fzr, Deshpand:2016cpw, Becirevic:2016oho, Cai:2017wry, Altmannshofer:2017poe, Kamali:2018fhr, Azatov:2018knx, Wei:2018vmk, Angelescu:2018tyl, Kim:2018oih, Crivellin:2019qnh, Yan:2019hpm,Crivellin:2017zlb, Marzocca:2018wcf, Bigaran:2019bqv,Crivellin:2019dwb,Saad:2020ihm,Dev:2020qet,Saad:2020ucl,Altmannshofer:2020axr,Gherardi:2020qhc} and/or $a_\mu$~\cite{Bauer:2015knc,Djouadi:1989md, Chakraverty:2001yg,Cheung:2001ip,Popov:2016fzr,Chen:2016dip,Biggio:2016wyy,Davidson:1993qk,Couture:1995he,Mahanta:2001yc,Queiroz:2014pra,ColuccioLeskow:2016dox,Chen:2017hir,Das:2016vkr,Crivellin:2017zlb,Cai:2017wry,Crivellin:2018qmi,Kowalska:2018ulj,Mandal:2019gff,Dorsner:2019itg,Crivellin:2019dwb,DelleRose:2020qak,Saad:2020ihm,Bigaran:2020jil,Dorsner:2020aaz,Gherardi:2020qhc,1821789,Babu:2020hun}, which makes them prime candidates for extending the SM with new particles.
\smallskip

Therefore, the search for LQ effects in observables other than the flavor anomalies is very well motivated. Complementary to direct LHC searches~\cite{Kramer:1997hh,Kramer:2004df,Faroughy:2016osc, Greljo:2017vvb, Dorsner:2017ufx, Cerri:2018ypt, Bandyopadhyay:2018syt, Hiller:2018wbv, Faber:2018afz, Schmaltz:2018nls,Chandak:2019iwj,Allanach:2019zfr, Buonocore:2020erb,Borschensky:2020hot}, oblique electroweak (EW) parameters and Higgs couplings to gauge bosons can be used to test LQs indirectly~\cite{Keith:1997fv,Dorsner:2016wpm,Bhaskar:2020kdr,Zhang:2019jwp,Gherardi:2020det}, as studied recently in detail in Ref.~\cite{Crivellin:2020ukd}. In this article we focus on the purely leptonic processes $\ell\to \ell^{\prime}\gamma$, $a_\ell$, $Z\to\ell^+\ell^{(\prime)-}$, $Z\to\nu\bar\nu$, $W\to\ell\nu$, $h\to\ell^+\ell^{(\prime)-}$, $\ell\to 3\ell^\prime$ and $\ell\to\ell^\prime\nu\bar\nu$. The correlations between $h\to\tau\mu$ and $\tau\to\mu\gamma$ were studied in Refs.~\cite{Cheung:2015yga,Baek:2015mea}, between $Z\to\mu^{+}\mu^{-}$ and $a_\mu$ in Ref.~\cite{ColuccioLeskow:2016dox} and between $Z$ and $W$ decays in Ref.~\cite{Arnan:2019olv}. While in the references above no LQ mixing, induced via couplings to the SM Higgs, was considered, this has been done for $a_\mu$ in Ref.~\cite{Dorsner:2019itg} and for the case of the singlet-triplet model in Refs.~\cite{Gherardi:2020det,Gherardi:2020qhc}. However, a complete calculation of leptonic processes with scalar LQs, including all possible interaction terms with the SM Higgs, is still missing. This is the purpose of this article.
\smallskip

In the next section we define our conventions before we discuss the self-energies, masses and the renormalization in Sec.~\ref{SE_masses_renormalization}. We then present the analytic results of LQ-induced effects in leptonic amplitudes in Sec.~\ref{observables}. In Sec.~\ref{pheno} we perform our phenomenological analysis, followed by the conclusions. The Appendix contains further helpful results, in particular the generic expressions with exact diagonalization of the LQ mixing matrices.
\medskip

\begin{table}
	\centering
	\renewcommand{\arraystretch}{1.8}
	\begin{tabular}{c|c|c}
		& $\mathcal{G}_{\text{SM}}$ & $\mathcal{L}_{ql\Phi}$\\
		\hline
		$\Phi_1$ & $\bigg(3,1,-\dfrac{2}{3}\bigg)$ & $\left(\lambda_{fj}^{1R}\,\bar{u}^c_f\ell_{j}+\lambda_{fj}^{1L}\,\bar{Q}_{f}^{\,c}i\tau_{2}L_{j}\right)\Phi_{1}^{\dagger}+\text{h.c.}$\\
		$\tilde{\Phi}_1$ & $\bigg(3,1,-\dfrac{8}{3}\bigg)$ & $\tilde{\lambda}_{fj}^{1}\,\bar{d}^{c}_{f}\ell_{j}\tilde{\Phi}_{1}^{\dagger}+\text{h.c.}$\\
		$\Phi_{2}$ & $\bigg(3,2,\dfrac{7}{3}\bigg)$ & $\lambda_{fj}^{2RL}\,\bar{u}_{f}\Phi_{2}^{T}i\tau_{2}L_{j}+\lambda_{fj}^{2LR}\,\bar{Q}_f\ell_j\Phi_{2}+\text{h.c.}$\\
		$\tilde{\Phi}_2$ & $\bigg(3,2,\dfrac{1}{3}\bigg)$ & $\tilde{\lambda}_{fj}^{2}\,\bar{d}_{f}\tilde{\Phi}_{2}^{T}i\tau_{2}L_{j}+\text{h.c.}$ \\
		$\Phi_{3}$ & $\bigg(3,3,-\dfrac{2}{3}\bigg)$ & $\lambda_{fj}^{3}\,\bar{Q}^{\,c}_{f}i\tau_{2}\left(\tau\cdot\Phi_{3}\right)^{\dagger}L_{j}+\text{h.c.}$
	\end{tabular}
	\caption{The five different possible scalar representations of LQs under the SM gauge group and their couplings to quarks and leptons. Note that in our conventions all LQs are $SU(3)_{c}$ triplets. The superscript $T$ refers to transposition in $SU(2)_L$ space, $c$ to charge conjugation and $\tau$ to the Pauli matrices. We did not include LQ couplings to two quarks, which are possible for some representations and which would lead to proton decays. Note that such couplings can always be avoided by assigning quark or lepton number to the SM fermions and to the LQs.}
	\label{eq:SLQ_interaction}
\end{table}

\section{Setup and Conventions}

As outlined in the introduction, LQs are prime candidates to explain the accumulated anomalies in semi-leptonic $B$ meson decays. Since vector LQs, as any massive vector particle, are not renormalizable without a Higgs mechanism, and since we are interested in loop processes, we will study only scalar LQs in the following.
\smallskip

The five different representations of scalar LQs transform under the SM gauge group 
\begin{align}
\mathcal{G}_{\text{SM}}=SU(3)_{c}\times SU(2)_L\times U(1)_Y
\end{align}
as given in Tab.~\ref{eq:SLQ_interaction}. Note that we have two singlets under $SU(2)_L$ ($\Phi_1$ and $\tilde \Phi_1$), two doublets ($\Phi_2$ and $\tilde \Phi_2$) and one triplet $\Phi_3$. The fermion fields $Q^{(c)}$ and $L$ are (charge-conjugated) quark and lepton $SU(2)_{L}$ doublets, while $u^{(c)}$, $d^{(c)}$ and $\ell$ are the corresponding $SU(2)_{L}$ singlets of up-quarks, down-quarks and charged leptons, respectively. The indices $f$ and $j$ refer to flavor and $\tau$ are the Pauli matrices, for which we use the convention
\begin{align}
\tau_{1}=\begin{pmatrix}0&1\\1&0\end{pmatrix}\,,&&
\tau_{2}=\begin{pmatrix}0&-i\\i&0\end{pmatrix}\,,&&
\tau_{3}=\begin{pmatrix}1&0\\0&-1\end{pmatrix}\,.
\label{eq:pauli_matrics}
\end{align}
We defined the hypercharge $Y$ such that the electromagnetic charge is given by
\begin{align}
Q=\frac{1}{2}Y+T_{3}\,,
\end{align}
with $T_{3}$ representing the third component of the weak isospin ($\pm 1/2$ for $SU(2)_L$ doublets and $1,0,-1$ for the $SU(2)_L$ triplet). According to this relation, LQs can be decomposed into the electromagnetic charge eigenstates as
\begin{subequations}
\label{eq:charge_eigenstates}
\begin{align}
	\Phi_{1}&\equiv\Phi_{1}^{-1/3}\,,\\
	\tilde{\Phi}_{1}&\equiv \tilde{\Phi}_{1}^{-4/3}\,,\\
	\Phi_{2}&\equiv\begin{pmatrix} \Phi_{2}^{5/3}\\\Phi_{2}^{2/3}\end{pmatrix}\,,\\ 
	\tilde{\Phi}_{2}&\equiv \begin{pmatrix}\tilde{\Phi}_{2}^{2/3}\\\tilde{\Phi}_{2}^{-1/3}\end{pmatrix}\,,\\
	\tau\cdot\Phi_{3}&\equiv
	\begin{pmatrix}\Phi_{3}^{-1/3}& \sqrt{2}\Phi_{3}^{2/3}\\ \sqrt{2}\Phi_{3}^{-4/3}&-\Phi_{3}^{-1/3}\end{pmatrix}\,,
\label{eq:Phi_3}
\end{align}
\end{subequations}
where the superscripts refer to the electric charge. 
\smallskip

The LQs couple according to their representation under the SM gauge group to gauge bosons, introduced for the first time in Ref.~\cite{Blumlein:1992ej}, where we use the following definition for the covariant derivative
\begin{align}
D_{\mu}\Phi=\big(\partial_{\mu}-ig_{1}\frac{Y}{2}B_{\mu}-ig_{2}T_{k}W_{\mu}^{k}-ig_{s}\frac{\lambda_{a}}{2}G_{\mu}^{a}\big)\Phi\,.
\end{align}
Here, $B_{\mu}$ is the $U(1)_{Y}$ gauge boson, $W_{\mu}$ the one of $SU(2)_{L}$ and $G_{\mu}$ of $SU(3)_{c}$ with the couplings $g_1$, $g_2$ and $g_s$, respectively. The index $k$ runs from 1 to 3, $a$ from 1 to 8. $T_k$ are the generators of $SU(2)$ and $\lambda_{a}$ are the well-known Gell-Mann matrices. For $SU(2)_{L}$ singlets we have $T_k=0$, for doublets we have $T_k=\tau_k/2$ with the Pauli matrices from Eq.~\eqref{eq:pauli_matrics} while the $SU(2)_{L}$ triplet $\Phi_3$ is in the adjoint representation of $SU(2)$. We use
\begin{align}
T_{1}=
\begin{pmatrix}
0&0&0\\0&0&-i\\0&i&0
\end{pmatrix}\,,&&
T_{2}=
\begin{pmatrix}
0&0&i\\0&0&0\\-i&0&0
\end{pmatrix}\,,&&
T_{3}=
\begin{pmatrix}
0&-i&0\\i&0&0\\0&0&0
\end{pmatrix}\,,
\end{align}
where $\Phi_3$ is defined according to \eq{eq:Phi_3}.

\smallskip

\subsection{Leptoquark-Higgs Interactions and Electroweak Symmetry Breaking}

\begin{figure}
	\centering
	\begin{overpic}[scale=.47,,tics=10]
		{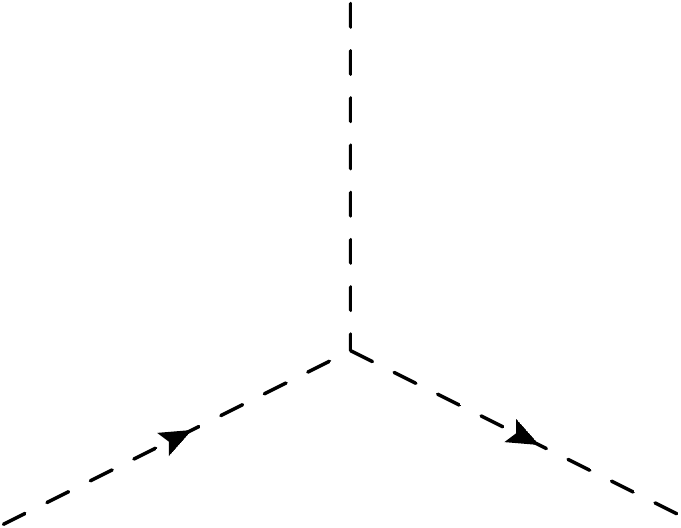}
		\put(3,18){$\Phi_{1}^{-1/3}$}
		\put(80,18){$\tilde{\Phi}_{2}^{-1/3}$}
		\put(43,9){$A_{\tilde 21}$}
		\put(57,60){$h$}
	\end{overpic}
\hspace{0.8cm}
	\begin{overpic}[scale=.47,,tics=10]
		{./Diagrams/LQ_LQ_higgs.pdf}
		\put(3,18){$\Phi_{3}^{-1/3}$}
		\put(80,18){$\tilde{\Phi}_{2}^{-1/3}$}
		\put(43,9){$A_{\tilde{2}3}$}
		\put(57,60){$h$}
	\end{overpic}
\hspace{0.8cm}
	\begin{overpic}[scale=.47,,tics=10]
		{./Diagrams/LQ_LQ_higgs.pdf}
		\put(3,18){$\Phi_{3}^{-2/3}$}
		\put(80,18){$\tilde{\Phi}_{2}^{-2/3}$}
		\put(43,9){$A_{\tilde{2}3}$}
		\put(57,60){$h$}
	\end{overpic}
\vspace{0.5cm}
\newline
	\begin{overpic}[scale=.47,,tics=10]
		{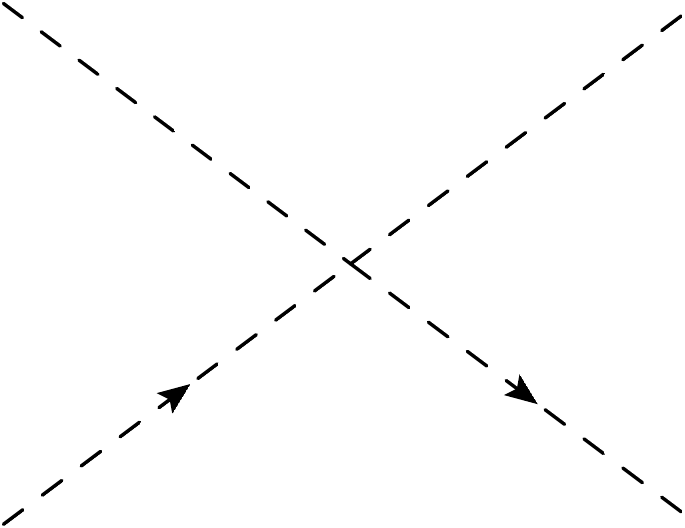}
		\put(3,22){$\Phi_{3}^{-1/3}$}
		\put(80,22){$\Phi_{1}^{-1/3}$}
		\put(47,21){$Y_{13}$}
		\put(68,65){$h$}
		\put(22,65){$h$}
	\end{overpic}
\hspace{1cm}
	\begin{overpic}[scale=.47,,tics=10]
		{./Diagrams/LQ_LQ_higgs_higgs.pdf}
		\put(3,22){$\tilde{\Phi}_{2}^{-2/3}$}
		\put(80,22){$\Phi_{2}^{-2/3}$}
		\put(47,21){$Y_{2\tilde2}$}
		\put(68,65){$h$}
		\put(22,65){$h$}
	\end{overpic}
	\caption{Feynman diagrams depicting the LQ-Higgs interactions induced by the terms in the first two lines of Eq.~\eqref{eq:LQ_mixing}. If the physical Higgs $h$ is replaced by its vev, mixing among the LQ representations is generated.}
	\label{fig:diagramm_HLQLQ_int}
\end{figure}
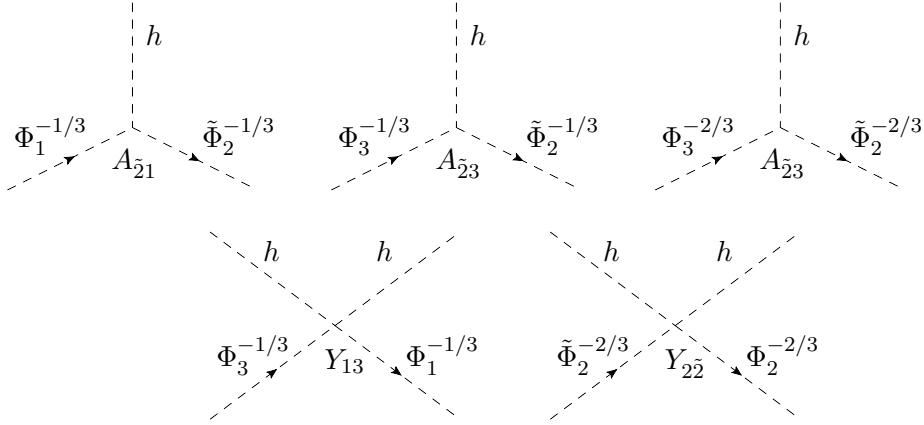

In addition to their couplings to fermions and the gauge interactions, LQs can couple to the SM-like Higgs doublet $H$ (with hypercharge +1) via the Lagrangian~\cite{Hirsch:1996qy}
\begin{align}
\begin{aligned}
	\mathcal{L}_{H\Phi}&=\left(-A_{\tilde 2 1}\big( \tilde{\Phi}_2^{\dagger} H\big)\Phi_1+A_{\tilde{2}3}\big(\tilde{\Phi}_2^{\dagger}\big(\tau\cdot\Phi_{3}\big)H\big)+Y_{2\tilde 2}\big(\Phi_{2}^{\dagger}H\big)\big(Hi\tau_{2}\tilde{\Phi}_{2}\big)\right.\\
	&\left.+Y_{3 \tilde 1}\big(H i\tau_{2} \left(\tau\cdot\Phi_{3}\right)^\dagger H\big)\tilde{\Phi}_1+Y_{13} \big(H^{\dagger}\left(\tau\cdot\Phi_{3} \right)H \big)\Phi_{1}^{\dagger}+\text{h.c.}\right)\\
	&-Y_{22}\big(Hi\tau_{2}\Phi_{2}\big)\big(Hi\tau_{2}\Phi_{2}\big)^{\dagger}-Y_{\tilde{2}\tilde{2}}\big(Hi\tau_{2}\tilde{\Phi}_{2}\big)\big(Hi\tau_{2}\tilde{\Phi}_{2}\big)^{\dagger}\\
	&-iY_{33}\varepsilon_{IJK}H^{\dagger}\tau_{I}H\Phi_{3,K}^{\dagger}\Phi_{3,J}\\
	&-\sum_{k=1}^{3}\big(m_{k}^2+Y_{k} H^{\dagger}H\big)\Phi_{k}^{\dagger}\Phi_{k}-\sum_{k=1}^{2}\big(\tilde{m}_{k}^2+Y_{\tilde{k}} H^{\dagger}H\big)\tilde{\Phi}_{k}^{\dagger}\tilde{\Phi}_{k}\,.
\end{aligned}
\label{eq:LQ_mixing}
\end{align}
Here $m_{k}^2$ and $\tilde m_{k}^2$ represent the $SU(2)_L$ invariant mass terms of the LQs before EW symmetry breaking and $\varepsilon_{IJK}$ is the three-dimensional Levi-Civita tensor with $\varepsilon_{123}=1$. For simplicity, we omitted the color indices, which are always contracted among the LQs. Note that $A_{\tilde 21}$ and $A_{\tilde 2 3}$ have mass dimension one, while the $Y$ couplings are dimensionless\footnote{We did not include terms with three or four LQ fields since they do not contribute at the one-loop level to the observables computed in this article.}. The LQ-Higgs interactions depicted in Fig.~\ref{fig:diagramm_HLQLQ_int} lead to mixing among the LQ representations after EW symmetry breaking.
\smallskip

Once the Higgs acquires its vacuum expectation value (vev) $v\approx 174$ GeV, this generates the mass matrices
\begin{equation}
\mathcal{L}_{\mathcal{M}}^{\rm LQ}=-\sum_{Q}\Phi^\dagger_Q \mathcal{M}^{Q}\Phi_Q
\end{equation}
in the weak basis, with $Q=\{-1/3,2/3,-4/3,5/3\}$ and
\begin{subequations}
\label{eq:LQ_mixing_matrices}
\begin{align}
\mathcal{M}^{-1/3}&=
\begin{pmatrix}
m_{1}^2+v^{2}Y_{1}& vA_{\tilde 21}^*&v^2Y_{13}\\
vA_{\tilde 21}&\tilde{m}_{2}^2+v^{2}Y_{\tilde{2}}&vA_{\tilde{2}3}\\
v^2Y_{13}^*&vA_{\tilde{2}3}^*&m_{3}^2+v^{2}Y_{3}
\end{pmatrix}\,,\\
\mathcal{M}^{2/3}&=
\begin{pmatrix}
m_{2}^2+v^{2}Y_{2}&v^2Y_{2\tilde2}&0\\
v^2Y_{2\tilde2}^*&\tilde{m}_{2}^2+v^{2}\big(Y_{\tilde{2}\tilde{2}}+Y_{\tilde{2}}\big)& -\sqrt{2}vA_{\tilde{2}3}\\
0& -\sqrt{2}vA_{\tilde{2}3}^*&m_{3}^2+v^{2}\big(Y_{3}+Y_{33}\big)
\end{pmatrix}\,,
\\
\mathcal{M}^{-4/3}&=
\begin{pmatrix}
\tilde{m}_{1}^2+v^{2}Y_{\tilde{1}}&\sqrt{2}v^2Y_{3 \tilde 1}^*\\
\sqrt{2}v^2Y_{3 \tilde 1}&m_{3}^2+v^{2}\big(Y_{3}-Y_{33}\big)
\end{pmatrix}\,,
\\
\mathcal{M}^{5/3}&=m_{2}^2+v^{2}\big(Y_{22}+Y_{2}\big)\,,
\end{align}
\end{subequations}
where the eigenstates of the electric charge 
\begin{subequations}
\label{eq:LQ_basis}
\begin{align}
\Phi_{-1/3}&\equiv
\begin{pmatrix}
	\Phi_{1}^{-1/3}\\ \tilde{\Phi}_2^{-1/3}\\ \Phi_{3}^{-1/3}
\end{pmatrix}\,,\\
\Phi_{2/3}&\equiv
\begin{pmatrix}
	\Phi_{2}^{2/3}\\ \tilde{\Phi}_{2}^{2/3}\\ \Phi_3^{2/3}
\end{pmatrix}\,,\\
\Phi_{-4/3}&\equiv
\begin{pmatrix}
	\tilde{\Phi}_{1}^{-4/3}\\ \Phi_{3}^{-4/3}
\end{pmatrix}\,,\\
\Phi_{5/3}&\equiv\Phi_{2}^{5/3}\,,
\end{align}
\end{subequations}
are assembled from the LQ field components of Eq.~(\ref{eq:charge_eigenstates}).
\smallskip

To work in the physical basis with mass eigenstates, in which the amplitudes are calculated, we need to diagonalize the mass matrices in \eq{eq:LQ_mixing_matrices}. This can be achieved by a unitary transformation
\begin{align}
\hat{\mathcal{M}}^{Q}=W^{Q}\mathcal{M}^{Q}W^{Q\dagger}\,,
\end{align}
such that $\hat{\mathcal{M}}^{Q}$ is diagonal. This means that the interaction eigenstates in \eqref{eq:LQ_basis} are written as
\begin{align}
W^{Q}\Phi_{Q}\equiv\hat{\Phi}^{Q}\,,
\end{align}
where $\hat{\Phi}^{Q}$ are the mass eigenstates. The analytic expressions for the diagonalization matrices $W^{-1/3}$ and $W^{2/3}$ are very lengthy or must be computed numerically. Therefore, we diagonalize the mass matrices perturbatively up to $\mathcal{O}(v^2/m_{\rm{LQ}}^2)$, where $m$ are the $SU(2)_L$ invariant mass terms of the LQs. The analytic expressions for the perturbative $W^{Q}$ read
\begin{subequations}
\label{Wexpanded}
\begin{align}
W^{-1/3}&\approx\begin{pmatrix}
1\!-\!\frac{v^2|A_{\tilde 21}|^2}{2(m_{1}^2-\tilde{m}_2^2)^2}& \frac{v A_{\tilde 21}^{*}}{m_{1}^2-\tilde{m}_{2}^2} & \frac{v^2(Y_{13}(m_{1}^2-\tilde{m}_2^2)+A_{\tilde 21}^{*}A_{\tilde{2}3})}{(m_{1}^2-m_{3}^2)(m_{1}^2-\tilde{m}_{2}^2)}\\
\frac{-v A_{\tilde 21}}{m_{1}^2-\tilde{m}_{2}^2} & 1\!-\!\frac{v^2}{2}\Big(\!\frac{|A_{\tilde 21}|^2}{(m_{1}^2-\tilde{m}_{2}^2)^2}\!+\!\frac{|A_{\tilde{2}3}|^2}{(m_{3}^2-\tilde{m}_{2}^2)^2}\!\Big)&
\frac{-v A_{\tilde{2}3}}{m_{3}^2-\tilde{m}_{2}^2}\\
\frac{-v^2(Y_{13}^{*}(m_{3}^2-\tilde{m}_2^2)+A_{\tilde 21}A_{\tilde{2}3}^{*})}{(m_{1}^2-m_{3}^2)(m_{3}^2-\tilde{m}_{2}^2)}&\frac{v A_{\tilde{2}3}^{*}}{m_3^2-\tilde{m}_{2}^2}& 1\!-\!\frac{v^2 |A_{\tilde{2}3}|^2}{2(m_{3}^2-\tilde{m}_{2}^2)^2}
\end{pmatrix}\,,\\
W^{2/3}&\approx\begin{pmatrix}
1& \frac{v^2 Y_{\tilde 22}}{m_{2}^2-\tilde{m}_{2}^2} & 0\\
\frac{-v^2 Y_{\tilde 22}^{*}}{m_{2}^2-\tilde{m}_{2}^2} & 1\!-\!\frac{v^{2}|A_{\tilde{2}3}|^2}{(m_{3}^2-\tilde{m}_{2}^2)^2} & \frac{-\sqrt{2}vA_{\tilde{2}3}}{\tilde{m}_{2}^2-m_{3}^2}\\
0 & \frac{\sqrt{2}vA_{\tilde{2}3}^{*}}{\tilde{m}_{2}^2-m_{3}^2} & 1-\frac{v^2|A_{\tilde{2}3}|^2}{(m_{3}^2-\tilde{m}_{2}^2)^2 }
\end{pmatrix}\,,\\
W^{-4/3}&\approx\begin{pmatrix}
1 & \frac{\sqrt{2}v^2 Y_{3 \tilde 1}^{*}}{\tilde{m}_{1}^2-m_{3}^2}\\
\frac{-\sqrt{2}v^2 Y_{3 \tilde 1}}{\tilde{m}_{1}^2-m_{3}^2}& 1
\end{pmatrix}\,.
\end{align}
\end{subequations}
Then the physical LQ masses are
\begin{subequations}
\label{eq:masses_diagonal}
\begin{align}
\left( M_a^{ - 1/3} \right)^2 &\approx \left(\! m_1^2\! +\! v^2\bigg( \!{Y_1} \!- \!\frac{|{A_{\tilde 21}}{|^2}}{{\tilde m_2^2 - m_1^2}} \bigg),\;\tilde m_2^2\!+\! {v^2}\bigg(\! {{Y_{\tilde 2}} \!+\! \frac{{|{A_{\tilde 21}}{|^2}}}{{\tilde m_2^2 - m_1^2}} \!+\! \frac{{|{A_{\tilde{2}3}}{|^2}}}{{\tilde m_2^2 - m_3^2}}} \bigg),\;\right.\nonumber \\  & 
\quad\left. m_3^2  + {v^2}\bigg(\! {Y_3} \!-\! \frac{{|{A_{\tilde{2}3}}{|^2}}}{{\tilde m_2^2 - m_3^2}} \bigg) \!\right)_a, \\
{\left( {M_a^{  2/3}} \right)^2} &\approx \left( \!m_2^2 \!+\! {v^2}{Y_2},\;\tilde m_2^2 \!+\! {v^2}\bigg( {{Y_{\tilde 2\tilde 2}} \!+\! {Y_{\tilde 2}}\! +\! \frac{{2|{A_{\tilde{2}3}}{|^2}}}{{\tilde m_2^2 - m_3^2}}} \bigg),\right. \nonumber\\
&\quad\left.\;m_3^2 \!+\! {v^2}\bigg( \!{{Y_3}\!+Y_{33} \!-\! \frac{{2|{A_{\tilde{2}3}}{|^2}}}{{\tilde m_2^2 - m_3^2}}} \bigg)\! \right)_a ,\\
{\left( {M_a^{ - 4/3}} \right)^2} &\approx \Big( {\tilde m_1^2 + {v^2}{Y_{\tilde 1}},\;m_3^2 + {v^2}\big({Y_3}-Y_{33}\big)} \Big)_a ,\\
{\left( {M^{5/3}} \right)^2} &\approx m_2^2 + {v^2}\big( {{Y_{22}} + {Y_2}} \big) \,,
\end{align}
\end{subequations}
keeping terms up to order $v^2$. The index $a$ runs from 1~to~3 for $Q=-1/3$ and $Q=2/3$ and from 1~to~2 for $Q=-4/3$, respectively.
\smallskip

\subsection{Leptoquark-Fermion Couplings}

EW symmetry breaking also leads to non-diagonal quark mass matrices in the weak basis, originating from the SM Yukawa couplings. Note that we can work in the basis with a diagonal lepton Yukawa coupling in the approximation of massless neutrinos. We therefore apply the following unitary rotation matrices on the left-handed quark fields
\begin{align}
u_{L}\to U^{u_L}\,u_{L}\,, \qquad d_{L}\to U^{d_L}\,d_{L}\,,
\end{align}
while the right-handed rotations can be absorbed by a redefinition of the LQ-quark-lepton couplings and are therefore unphysical. We now choose to work in the so-called down basis such that 
\begin{align}
U_{ji}^{u_L *}=V_{ij}\,, \qquad U_{ij}^{d_{L}}=\delta_{ij}\,,
\end{align}
with $V_{ij}$ being the CKM matrix. This means that CKM elements only appear in couplings involving up-type quarks.
\smallskip

We now decompose the LQ-fermion interactions in Tab.~\ref{eq:SLQ_interaction} into their $SU(2)_L$ components and write them in terms of mass eigenstates
\begin{align}
\begin{split}
\mathcal{L}_{ql\Phi}&=\big[\bar{u}^{c}_i\big(\Gamma_{u_{i}^{c}\ell_{j}}^{R,a}P_{R}+\Gamma_{u_{i}^{c}\ell_{j}}^{L,a}P_{L}\big)\ell_{j}+\Gamma_{d_{i}^{c}\nu_{j}}^{L,a}\bar{d}^{c}_{i}P_{L}\nu_{j}+\Gamma_{d_{i}\nu_{j}}^{L,a*}\bar{\nu}_{j}P_{R}d_{i}\big]\hat{\Phi}_{a}^{-1/3\,\dagger}\\
&\quad+\big[\bar{d}_{i}\big(\Gamma_{d_{i}\ell_{j}}^{R,a}P_{R}+\Gamma_{d_{i}\ell_{j}}^{L,a}P_{L}\big)\ell_{j}+\Gamma_{u_{i}^{c}\nu_{j}}^{L,a*}\bar{\nu}_{j}P_{R}u_{i}^{c}+\Gamma_{u_{i}\nu_{j}}^{L,a}\bar{u}_{i}P_{L}\nu_{j}\big]\hat{\Phi}_{a}^{2/3}\\
&\quad+\big[\bar{d}_{i}^{c}\big(\Gamma_{d_{i}^{c}\ell_{j}}^{R,a}P_{R}+\Gamma_{d_{i}^{c}\ell_{j}}^{L,a}P_{L}\big)\ell_{j}\big]\hat{\Phi}_{a}^{-4/3\,\dagger}+\big[\bar{u}_{i}\big(\Gamma_{u_{i}\ell_{j}}^{R}P_{R}+\Gamma_{u_{i}\ell_{j}}^{L}P_{L}\big)\ell_{j}\big]\hat{\Phi}^{5/3}\\
&\quad+\text{h.c.}\,,
\end{split}
\label{eq:mixing_interaction_Lagrangian}
\end{align}
with
\begin{align}
\begin{aligned}
\Gamma_{u_{i}^{c}\ell_{j}}^{R,a}&=\lambda^{1R}_{ij}W_{a1}^{-1/3}\,, &
\Gamma_{u_{i}^{c}\ell_{j}}^{L,a}&=V_{ik}^{*}\Big(\lambda^{1L}_{kj}W_{a1}^{-1/3}-\lambda_{kj}^{3}W_{a3}^{-1/3}\Big)\,,\\
\Gamma_{d_{i}\nu_{j}}^{L,a*}&=-\tilde{\lambda}^{2*}_{ij}W_{a2}^{-1/3}\,, &
\Gamma_{d_{i}^{c}\nu_{j}}^{L,a}&=-\Big(\lambda_{ij}^{1L}W_{a1}^{-1/3}+\lambda_{ij}^{3}W_{a3}^{-1/3}\Big)\,,\\
\Gamma_{d_{i}\ell_{j}}^{R,a}&= \lambda_{ij}^{2LR}W_{a1}^{2/3*}\,, &
\Gamma_{d_{i}\ell_{j}}^{L,a}&=\tilde{\lambda}^{2}_{ij}W_{a2}^{2/3*}\,,\\
\Gamma_{u_{i}^{c}\nu_{j}}^{L,a*}&=\sqrt{2}V_{ik}\lambda_{kj}^{3*}W_{a3}^{2/3*}\,,&
\Gamma_{u_{i}\nu_{j}}^{L,a}&=-\lambda_{ij}^{2RL}W_{a1}^{2/3*}\,,\\
\Gamma_{d_{i}^{c}\ell_{j}}^{R,a}&=\tilde{\lambda}^{1}_{ij}W_{a1}^{-4/3}\,,&
\Gamma_{d_{i}^{c}\ell_{j}}^{L,a}&=-\sqrt{2}\lambda_{ij}^{3}W_{a2}^{-4/3}\,,\\
\Gamma_{u_{i}\ell_{j}}^{R}&=V_{ik}\lambda^{2LR}_{kj}\,,&
\Gamma_{u_{i}\ell_{j}}^{L}&=\lambda_{ij}^{2RL}\,.
\end{aligned}
\label{eq:Gammas_LQ_mixing}
\end{align}
Note that the index $a$ runs from 1 to 3 for $Q=-1/3$ and $Q=2/3$, while for $Q=-4/3$ only from 1 to 2. Due to our choice of basis, the CKM matrix appears in all couplings involving left-handed up-type quarks. Similarly, also the PMNS matrix would enter in all couplings involving neutrinos in case they were taken to be massive. However, all processes that we are interested in can be calculated for massless neutrinos such that the PMNS matrix drops out. Nonetheless, we will return to the PMNS matrix in the next section when we discuss possible contributions to Majorana mass terms and the renormalization of the $W\ell\nu$ vertex.
\smallskip

\subsection{Leptoquark-Higgs Couplings}

Let us finally consider the couplings of the SM Higgs to LQs. The interaction terms are also affected by the LQ rotations induced by EW symmetry breaking. Again, we express \eq{eq:LQ_mixing} in terms of mass eigenstates as
\begin{align}
\begin{aligned}
\mathcal{L}_{H\Phi}&=-\tilde{\Gamma}^{1/3}_{ab}h\hat{\Phi}_{a}^{-1/3\,\dagger}\hat{\Phi}_{b}^{-1/3}-\tilde{\Gamma}^{2/3}_{ab}h\hat{\Phi}_{a}^{2/3\,\dagger}\hat{\Phi}_{b}^{2/3}-\tilde{\Gamma}^{4/3}_{cd}h\hat{\Phi}_{c}^{-4/3\dagger}\hat{\Phi}_{d}^{-4/3}\\
&\quad-\Gamma^{5/3}h\hat{\Phi}^{5/3\,\dagger}\hat{\Phi}^{5/3}-\tilde{\Lambda}^{1/3}_{ab}h^2\hat{\Phi}_{a}^{-1/3\,\dagger}\hat{\Phi}_{b}^{-1/3}-\tilde{\Lambda}^{2/3}_{ab}h^2\hat{\Phi}_{a}^{2/3\,\dagger}\hat{\Phi}_{b}^{2/3}\\
&\quad-\tilde{\Lambda}^{4/3}_{cd}h^2\hat{\Phi}_{c}^{-4/3\,\dagger}\hat{\Phi}_{d}^{-4/3}-\Lambda^{5/3}h^2\hat{\Phi}^{5/3\,\dagger}\hat{\Phi}^{5/3}\,,
\end{aligned}
\end{align}
with $h$ as the physical Higgs field, $a,b=\{1,2,3\}$ and $c,d=\{1,2\}$. The couplings are defined as
\begin{align}
\begin{aligned}
\tilde{\Gamma}^{1/3}&=W^{-1/3}\Gamma^{1/3}W^{-1/3\,\dagger}\,,
&\quad\quad&\tilde{\Lambda}^{1/3}=W^{-1/3}\Lambda^{1/3}W^{-1/3\,\dagger}\,,\\
\tilde{\Gamma}^{2/3}&=W^{2/3}\Gamma^{2/3}W^{2/3\,\dagger}\,,
&&\tilde{\Lambda}^{2/3}=W^{2/3}\Lambda^{2/3}W^{2/3\,\dagger}\,,\\
\tilde{\Gamma}^{4/3}&=W^{-4/3}\Gamma^{4/3}W^{-4/3\,\dagger}\,,
&&\tilde{\Lambda}^{4/3}=W^{-4/3}\Lambda^{4/3}W^{-4/3\,\dagger}\,,\\
\Gamma^{5/3}&=\sqrt{2}v\big(Y_{22}+Y_{2}\big)\,,
&&\Lambda^{5/3}=\frac{1}{2}\big(Y_{22}+Y_{2}\big)\,,
\end{aligned}
\end{align}
where the $\Gamma^{Q}$ and $\Lambda^{Q}$ matrices read
\begin{subequations}
\begin{align}
\Gamma^{1/3}&=\frac{1}{\sqrt{2}}\!\begin{pmatrix}2vY_{1}&A_{\tilde{2}1}^{*}&2vY_{13}\\ A_{\tilde{2}1}&2vY_{\tilde{2}} & A_{\tilde{2}3} \\ 2vY_{13}^{*}& A_{\tilde{2}3}^{*}& 2vY_{3}\end{pmatrix}
&&
\Lambda^{1/3}=\frac{1}{2}\!\begin{pmatrix}Y_{1} & 0 & Y_{13}\\ 0& Y_{\tilde{2}} & 0\\ Y_{13}^{*} & 0 & Y_{3}\end{pmatrix}\\
\Gamma^{2/3}&=\frac{1}{\sqrt{2}}\!\begin{pmatrix}2vY_{2} & 2vY_{2\tilde 2} & 0\\ 2vY_{2\tilde 2}^{*}& 2v\big(Y_{\tilde{2}}+Y_{\tilde{2}\tilde{2}}\big) &-\sqrt{2}A_{\tilde{2}3}\\ 0 &- \sqrt{2}A_{\tilde{2}3}^{*} & 2v(Y_{3}+Y_{33})\end{pmatrix} &&
\Lambda^{2/3}=\frac{1}{2}\!\begin{pmatrix}Y_{2} & Y_{2\tilde 2} & 0\\ Y_{2\tilde 2}^{*} & Y_{\tilde{2}}+Y_{\tilde{2}\tilde{2}} & 0\\ 0& 0& Y_{3}+Y_{33}\end{pmatrix}\\
\Gamma^{4/3}&=\frac{1}{\sqrt{2}}\!\begin{pmatrix}2vY_{\tilde{1}} & 2vY_{3\tilde{1}}^{*} \\ 2vY_{3\tilde{1}} & 2v(Y_{3}-Y_{33})\end{pmatrix}&&
\Lambda^{4/3}=\frac{1}{2}\!\begin{pmatrix}Y_{\tilde{1}} & Y_{3\tilde{1}}^{*} \\ Y_{3\tilde{1}} & Y_{3}-Y_{33}\end{pmatrix} \,.
\end{align}
\end{subequations}
The expanded expressions for $\tilde{\Gamma}^{Q}$ and $\tilde{\Lambda}^{Q}$ are given in the Appendix \ref{app:matrices}.
\medskip

\begin{boldmath}
	\section{Self-Energies, Masses and Renormalization}
	\label{SE_masses_renormalization}
\end{boldmath}

Self-energies of SM fermions after $SU(2)_{L}$ breaking are directly related to their masses and enter the calculations of effective fermion-fermion-gauge-boson and fermion-fermion-Higgs couplings. In this section, we will first calculate the self-energies, then discuss the issue of renormalization and how the self-energies are included in the calculation of modified gauge-boson and Higgs couplings. 
\smallskip

First, let us define the mass and kinetic terms of the charged lepton and neutrino Lagrangian in momentum space
\begin{equation}
\mathcal{L}^{\ell\nu} = \delta _{fi}\Big( \bar \ell_f\left(  \slashed{p}- m_{f}^{\ell} \right)\ell _i + \bar \nu _f\slashed{p}\nu _i - \frac{m_f^\nu }{2}\bar \nu_f^c\nu _i \Big)\,.
\label{Lnuell}
\end{equation}
We allowed for the possibility of Majorana mass terms for the neutrinos, which can be generated via LQs. We then moved to the physical basis in which all mass matrices are diagonal, such that the CKM matrix $V$ (the PMNS matrix $\hat{V}$) appears in the $Wud$ ($W\ell\nu$) vertex. Considering only the leptonic part, we have explicitly
\begin{equation}
\mathcal{L}_{W}^{\ell\nu}=\frac{{{g_2}}}{{\sqrt 2 }}\hat{V}_{fi}^{}{{\bar \ell }_f}{\gamma ^\mu }{\nu _i}W_\mu ^ -\,.
\label{eq:PMNS_Wlv}
\end{equation}
\smallskip

We define the self-energies of charged leptons as follows
\begin{equation}
\begin{gathered}
\begin{overpic}[scale=.45,,tics=10]
{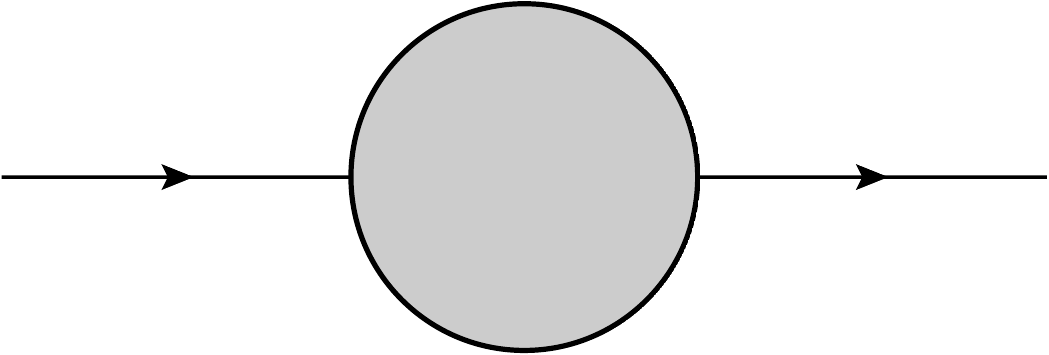}
\put(10,22){$\ell_{i}$}
\put(82,22){$\ell_{f}$}
\put(11,8){$p$}
\put(83,8){$p$}
\end{overpic}
\end{gathered}
\hspace{0.3cm}
=-i\Sigma_{fi}^{\ell}(p^2)\,,
\end{equation}
and decompose $\Sigma_{fi}^{\ell}(p^2)$ as
\begin{equation}
\Sigma_{fi}^{\ell}(p^2)=\slashed{p}\Big(\Sigma_{fi}^{\ell LL}(p^2)P_{L}+\Sigma_{fi}^{\ell RR}(p^2)P_{R}\Big)+\Sigma_{fi}^{\ell RL}(p^2)P_{L}+\Sigma_{fi}^{\ell LR}(p^2)P_{R}\,,
\end{equation}
and similarly for neutrinos, where only the $LL$ self-energy exists, but a possible contribution to the neutrino mass term arises.
\smallskip

We now expand $\Sigma_{fi}^{\ell AB}(p^2)$ with $A,B=\{L,R\}$ in terms of $p^2/m_{\rm{LQ}}^2$, where $m$ represents the LQ mass. Only the leading terms in this expansion (i.e. the ones independent of $p^2$) are UV divergent and non-decoupling. Furthermore, they are the only relevant ones in the calculation of $Z\ell\ell$, $Z\nu\nu$, $W\ell\nu$ and $h\ell\ell$ vertices to be discussed later. The terms linear in $p^2/m^2$ are only necessary to calculate $\ell\to\ell^\prime\gamma$. However, as they are finite and do not affect the renormalization of any parameter, they can be included in the calculation of $\ell\to\ell^\prime\gamma$ in a straightforward way and we do not give the explicit results here. The ones for $\Sigma_{fi}^{\ell,\nu AB}\equiv \Sigma_{fi}^{\ell,\nu AB}(0)$ are given in the Appendix~\ref{app:self_energies}.
\smallskip

\begin{figure}
	\centering
	\begin{overpic}[scale=.8,,tics=10]
		{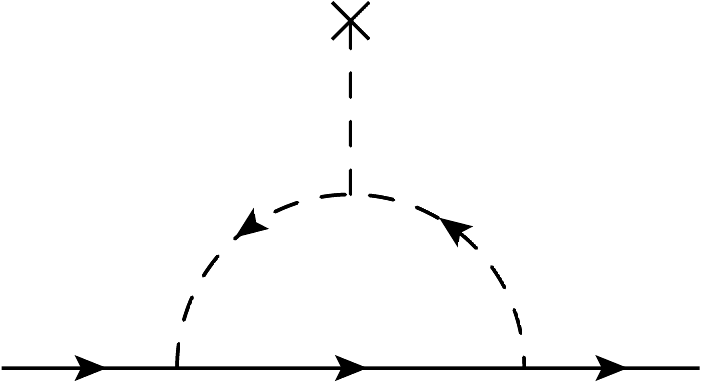}
		\put(5,5){$\nu$}
		\put(92,5){$\nu^{c}$}
		\put(53,42){$v$}
		\put(46,6){$d_{k}$}
		\put(22,21){$\tilde{\Phi}_{2}$}
		\put(68,21){$\Phi_{3,1}$}
	\end{overpic}
	\hspace{1cm}
	\begin{overpic}[scale=.8,,tics=10]
		{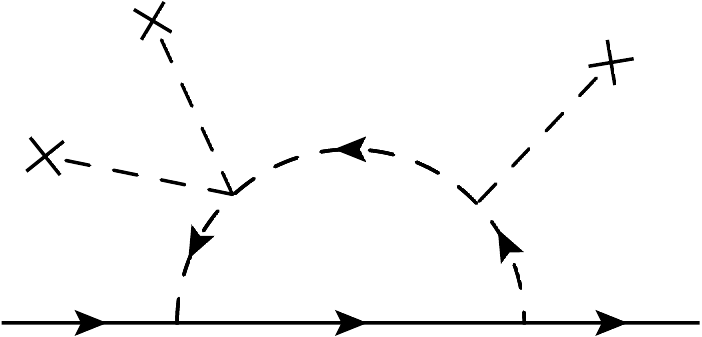}
		\put(5,5){$\nu$}
		\put(92,5){$\nu^{c}$}
		\put(48,6){$u_l$}
		\put(16,11){$\Phi_{2}$}
		\put(46.5,30){$\tilde{\Phi}_{2}$}
		\put(75,11){$\Phi_{3}$}
		\put(13,26){$v$}
		\put(27,38){$v$}
		\put(76,34){$v$}
	\end{overpic}
	\caption{One-loop self-energy diagrams generating Majorana-like neutrino masses. On the left-hand side, we have a down-type quark in the loop. In the case of up-type quarks, the leading contribution only occurs at $\mathcal{O}(v^3)$.}
	\label{fig:diagramm_neutrino_mass}
\end{figure}
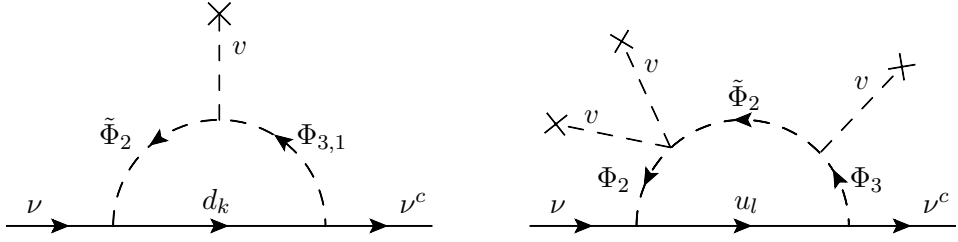

\subsection{Neutrino Masses}

The contribution to the Majorana mass term of the neutrinos can be calculated by considering the $\bar\nu^c_f\nu_i$ two-point function. We have generically
\begin{align}
m^{\nu{\rm LQ}}_{ij}=&\frac{-m_{q_k}N_{c}\big(\Gamma_{q_{k}\nu_{i}}^{L}\Gamma_{q_{k}\nu_{j}}^{L}+\Gamma_{q_{k}\nu_{j}}^{L*}\Gamma_{q_{k}\nu_{i}}^{L*}\big)}{16\pi^2}\mathcal{I}_{0}\Big(\frac{\mu^2}{M^2},\frac{m_{q_k}^2}{M^2}\Big)\,,
\end{align}
where we neglected the external momenta. An implicit sum over all internal quarks $u,d,u^c$ and $d^c$ as well as over their flavors and the corresponding LQs is understood. The loop function $\mathcal{I}_0$ is given in the Appendix~\ref{app:self_energies}.
\smallskip

After summation one can expand this expression in terms of $v/m_{\rm{LQ}}$. In this way, one recovers the two diagrams shown in Fig.~\ref{fig:diagramm_neutrino_mass} and finds
\begin{align}
\begin{aligned}
m^{\nu{\rm LQ}}_{ij}&\approx\frac{m_{d_k}N_{c}v}{16\pi^2\tilde{m}_{2}^2}\bigg(\big(\lambda_{ki}^{1L}\tilde{\lambda}_{kj}^{2}A_{\tilde 21}+\lambda_{kj}^{1L*}\tilde{\lambda}_{ki}^{2*}A_{\tilde 21}^{*}\big)\mathcal{H}_{1}\Big(\frac{m_{1}^2}{\tilde{m}_{2}^2}\Big)\\
&\quad+\big(\lambda_{ki}^{3}\tilde{\lambda}_{kj}^{2}A_{\tilde{2}3}+\lambda_{kj}^{3*}\tilde{\lambda}_{ki}^{2*}A_{\tilde{2}3}^{*}\big)\mathcal{H}_{1}\Big(\frac{m_{3}^2}{\tilde{m}_{2}^2}\Big)\bigg)+\mathcal{O}(m_d \, v^3/m^4)\\
&\quad+\frac{m_{u_l}N_{c}v^3}{8\pi^2}\frac{\lambda_{lj}^{2RL}V_{lk}^{*}\lambda_{ki}^{3}A_{\tilde{2}3}Y_{2\tilde 2}+\lambda_{li}^{2RL*}V_{lk}\lambda_{kj}^{3*}A_{\tilde{2}3}^{*}Y_{2\tilde 2}^{*}} {m_{2}^2(\tilde{m}_{2}^2-m_{3}^2)}\bigg(\!\mathcal{H}_{1}\Big(\frac{\tilde{m}_{2}^2}{m_{2}^2}\Big)\!-\mathcal{H}_{1}\Big(\frac{m_{3}^2}{m_{2}^2}\Big)\!\bigg)\\
&\quad+\mathcal{O}(m_{u_l}v^4/m^5)\,,
\end{aligned}
\end{align}
where the first two lines agree with Ref.~\cite{Dorsner:2017wwn}, originating from down-type quark contributions. The third line, generated by couplings to up-type quarks, was not given previously in the literature. Note that for the latter, the leading contribution only appears at $\mathcal{O}(v^3)$, see Fig.~\ref{fig:diagramm_neutrino_mass}, while for down-type quarks already a $v^1$ term exists and higher orders in $v$ do not generate new, independent coupling structures. The loop function $\mathcal{H}_1$ is given in the Appendix \ref{app:loop_functions}.
\smallskip

\subsection{Renormalization}

With these expressions at hand, we can include the loop effects into the Lagrangian of \eq{Lnuell} to obtain
\begin{align}
\mathcal{L}^{\ell\nu} &= {{\bar \ell }_f}\left( \slashed{p}{\Big( {{\delta _{fi}} - \Sigma _{fi}^{\ell LL}{P_L} - \Sigma _{fi}^{\ell RR}{P_R}} \Big) - m_f^{\ell\left( 0 \right)}{\delta _{fi}} - \Sigma _{fi}^{\ell LR}{P_R} - \Sigma _{fi}^{\ell RL}{P_L}} \right){\ell _i}\nonumber\\
&\quad+ {{\bar \nu }_f}\slashed{p}\Big( {{\delta _{fi}} - \Sigma _{fi}^{\nu LL}} \Big){\nu _i} - \frac{{m_f^{\nu(0)}  + m_{fi}^{\nu {\rm{LQ}}}}}{2}\bar \nu _f^c{\nu _i}\,.
\end{align}
The superscript $(0)$ indicates the bare (unrenormalized) quantities. Now we have to make the kinetic terms canonical again and render the mass matrices diagonal in order to arrive at the physical basis. We start with the kinetic terms, which are made diagonal and correctly normalized once the shifts
\begin{subequations}
\begin{align}
{\ell _{fL}} &\to \left( {{\delta _{fi}} + \frac{1}{2}\Sigma _{fi}^{\ell LL}} \right){\ell _{iL}}\,,\\
{\ell _{fR}} &\to \left( {{\delta _{fi}} + \frac{1}{2}\Sigma _{fi}^{\ell RR}} \right){\ell _{iR}}\,,\\
{\nu _{f}} &\to \left( {{\delta _{fi}} + \frac{1}{2}\Sigma _{fi}^{\nu LL}} \right){\nu _{i}}\,,
\end{align}
\end{subequations}
have been applied. These shifts enter in all observables with external lepton fields, i.e. they also lead to effects in gauge-boson couplings to leptons. Therefore, we include them in this way in our calculations later on.
\smallskip

In addition, these field redefinitions affect the mass terms for charged leptons, which then read~\cite{Crivellin:2010gw}
\begin{equation}
\mathcal{L}^{\ell}_{m} =  - {{\bar \ell }_f}\left( {m_i^{\ell\left( 0 \right)}{\delta _{fi}} + \frac{1}{2}\Sigma _{fi}^{\ell LL}m_i^{\ell\left( 0 \right)} + \frac{1}{2}m_f^{\ell\left( 0 \right)}\Sigma _{fi}^{\ell RR}}  + \Sigma _{fi}^{\ell LR}{P_R} + \Sigma _{fi}^{\ell RL}{P_L} \right){\ell _i}\,,
\end{equation}
and for neutrinos we have
\begin{equation}
\mathcal{L}^{\nu}_m =  - \frac{1}{2}\bar{\nu}_{f}^{c}\left( {m_f^{\nu (0)}{\delta _{fi}} + \frac{1}{2}m_f^{\nu (0)}\Sigma _{fi}^{\nu LL} + \frac{1}{2}\Sigma _{fi}^{\nu LL*}m_i^{\nu (0)} + m_{fi}^{\nu {\rm{LQ}}}} \right){\nu _i}\,.
\end{equation}
These matrices can now be diagonalized as
\begin{equation}
\begin{aligned}
U_{f'f}^{\ell L*}\left( {\Big( {{\delta _{f'j}} + \frac{1}{2}\Sigma _{f'j}^{\ell LL}} \Big)m_j^{\left( 0 \right)}\Big( {{\delta _{ji'}} + \frac{1}{2}\Sigma _{ji'}^{\ell RR}} \Big) + \Sigma _{ji'}^{\ell LR}} \right)U_{i'i}^{\ell R} &= m_i^\ell \delta _{fi}\,,\\
U_{f'f}^\nu \left( {m_{f'}^{\nu (0)}{\delta _{f'i'}} + \frac{1}{2}m_{f'}^{\nu (0)}\Sigma _{f'i'}^{\nu LL} + \frac{1}{2}\Sigma _{f'i'}^{\nu LL*}m_{i'}^{\nu (0)} + m_{f'i'}^{\nu {\rm{LQ}}}} \right)U_{i'i}^\nu  &= m_i^\nu {\delta _{fi}}\,,
\end{aligned}
\end{equation}
with $m_i^\ell $ and $m_i^\nu$ being the physical masses. The unitary matrix $U^{\ell L}$ is given by
\begin{align}
U^{\ell L} &= \left( {\begin{array}{*{20}{c}}
	1&{\frac{{\Sigma _{12}^{\ell LR}}}{{m_2^\ell }} + \frac{1}{2}\Sigma _{12}^{\ell LL}}&{\frac{{\Sigma _{13}^{\ell LR}}}{{m_3^\ell }} + \frac{1}{2}\Sigma _{13}^{\ell LL}}\\
	{ - \frac{{\Sigma _{12}^{\ell LR*}}}{{m_2^\ell }} - \frac{1}{2}\Sigma _{12}^{\ell LL*}}&1&{\frac{{\Sigma _{23}^{\ell LR}}}{{m_3^\ell }} + \frac{1}{2}\Sigma _{23}^{\ell LL}}\\
	{ - \frac{{\Sigma _{13}^{\ell LR*}}}{{m_3^\ell }} - \frac{1}{2}\Sigma _{13}^{\ell LL*}}&{ - \frac{{\Sigma _{23}^{\ell LR*}}}{{m_3^\ell }} - \frac{1}{2}\Sigma _{23}^{\ell LL*}}&1
	\end{array}} \right)\,.
\label{ULrot}
\end{align}
We used the lepton mass hierarchy to simplify $U^{\ell L}$ and the fact that the self-energies are just corrections to a diagonal matrix to get an explicit expression. $U^{\ell R}$ is simply obtained by exchanging $L$ and $R$.
\smallskip

These unitary rotations (or at leading order the unit matrix plus anti-hermitian corrections) do not have a physical effect in the sense that they cannot be measured in observables. In fact, they correspond to unphysical rotations, in case of $U^{\ell R}$, or they can be absorbed by a renormalization of the PMNS matrix, in case of $U^{\ell L}$ and $U^{\nu}$. This can also be seen by applying these rotations to gauge bosons vertices, where they drop out for the $Z$ interaction terms and only enter the $W\ell\nu$ vertex in the combination
\begin{equation}
\hat{V}_{fi}^{} = U_{f'f}^{\ell L*}\hat{V}_{f'i'}^{\left( 0 \right)}U_{i'i}^\nu\,,
\end{equation}
where $\hat{V}$ on the left-hand side of the equation is identified with the PMNS matrix, see Eq.~\ref{eq:PMNS_Wlv}.
\smallskip

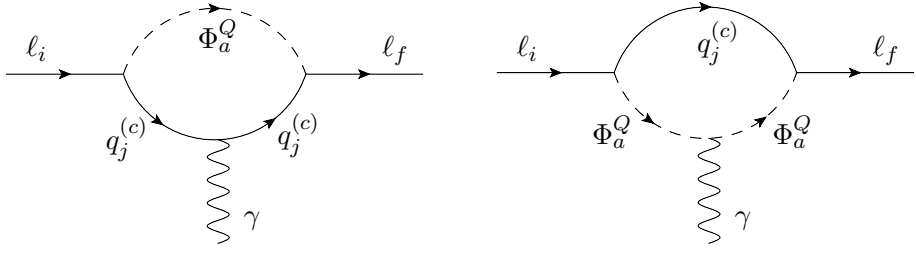
\begin{figure}[t]
	\centering
	\begin{overpic}[scale=.55,,tics=10]
		{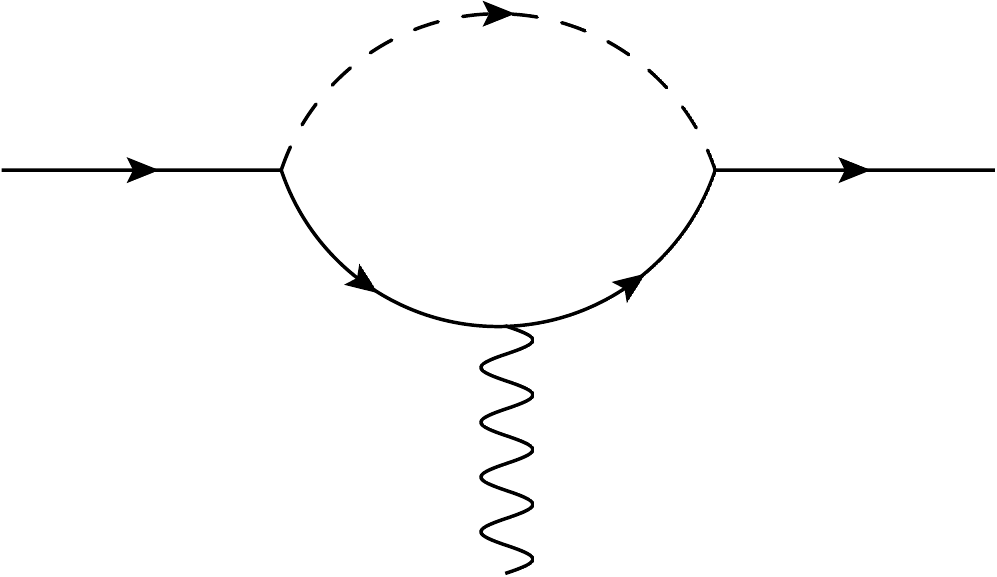}
		\put(5,45){$\ell_{i}$}
		\put(90,45){$\ell_{f}$}
		\put(57,5){$\gamma$}
		\put(46,47){$\Phi_{a}^{Q}$}
		\put(24,23){$q_{j}^{(c)}$}
		\put(65,24){$q_{j}^{(c)}$}
	\end{overpic}
	\hspace{0.7cm}
	\begin{overpic}[scale=.55,,tics=10]
		{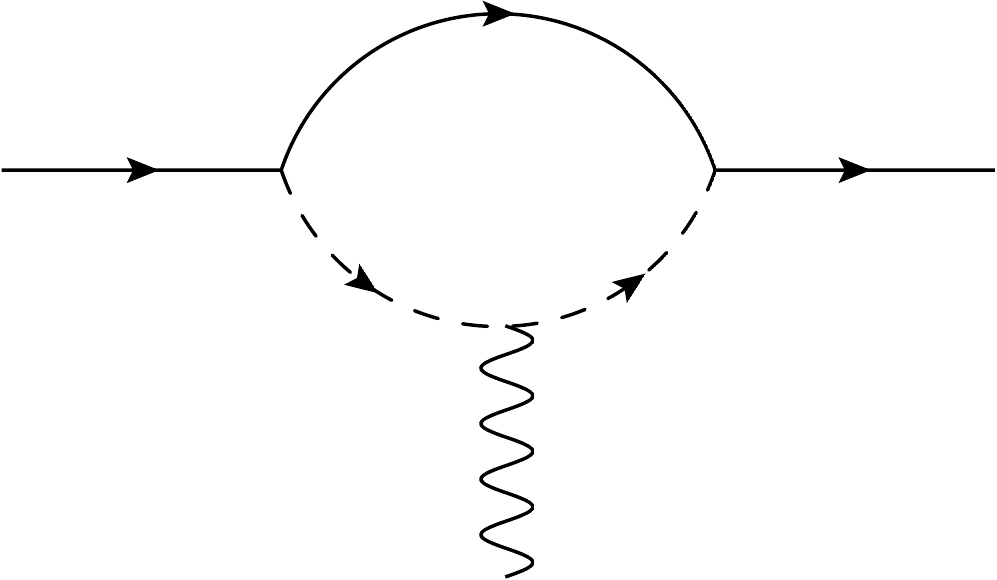}
		\put(5,45){$\ell_{i}$}
		\put(90,45){$\ell_{f}$}
		\put(57,5){$\gamma$}
		\put(48,46){$q_{j}^{(c)}$}
		\put(23,24){$\Phi_{a}^{Q}$}
		\put(66,24){$\Phi_{a}^{Q}$}
	\end{overpic}
	\caption{The vertex diagrams which contribute to $\ell_{i}\to\ell_{f}\gamma$. Depending on the electric charge of the LQ, we have (charge-conjugated) up- or down-type quarks in the loop.}
	\label{fig:AMM_LQ}
\end{figure}

Finally, let us consider the $h\ell\ell$ vertex. Here we have
\begin{equation}
 - h{{\bar \ell }_f}U_{f'f}^{\ell L*}\left( {\Big( {{\delta _{f'j}} + \frac{1}{2}\Sigma _{f'j}^{\ell LL}} \Big)Y_j^{\ell\left( 0 \right)}\Big( {{\delta _{ji'}} + \frac{1}{2}\Sigma _{ji'}^{\ell RR}} \Big) + \Lambda _{ji'}^{\ell LR}} \right)U_{i'i}^{\ell R}{P_R}{\ell _i}\,,
 \end{equation}
where $Y_j^{\ell\left( 0 \right)}=m^{\ell(0)}_j/v$ and $\Lambda _{ji'}^{\ell LR}$ represents the genuine vertex correction. Therefore, the effective Yukawa coupling measured in $h\to\ell^{+}\ell^{\prime -}$ decays can be expressed in terms of the physical lepton mass and $\Sigma _{fi}^{\ell LR}$ as follows
\begin{equation}
Y_{fi}^{\ell\,{\rm{eff}}} = \frac{{m_i^\ell {\delta _{fi}} - \Sigma _{fi}^{\ell LR}}}{v} + \Lambda _{fi}^{\ell LR}\,.
\end{equation}
\medskip

\section{Calculation of the One-Loop Effects}
\label{observables}

In this section, we compute the amplitudes governing the various purely leptonic observables. For this we take into account the Higgs-induced mixing among the different LQ representations. We will consider amplitudes involving the following fields:
\begin{enumerate}
\begin{minipage}{\linewidth}
	\item $\ell\ell\gamma$
	\vspace{-1mm}
	\item $Z\ell\ell$ and $Z\nu\nu$
	\vspace{-1mm}
	\item $W\ell\nu$
		\vspace{-1mm}
	\item $h\ell\ell$
		\vspace{-1mm}
	\item $4\ell$
		\vspace{-1mm}
	\item $2\ell 2\nu$
\end{minipage}
\end{enumerate}
For our purpose, the gauge bosons and the Higgs can be both on- and off-shell while the leptons are all on-shell. We set all lepton masses to zero, except for $\ell_{i}\to \ell_{f}\gamma$, where we expand up to the first non-vanishing order. In addition, we expanded the loop integrals in $m_{\rm EW}/m_{\rm{LQ}}$, where $m_{\rm EW}\approx v$ can denote $m_W,\,m_Z,\,m_H$ or $m_t$. Furthermore, we expanded the mass eigenvalues of the LQs and the mixing matrices in $v/m_{\rm{LQ}}$, while the results obtained with exact diagonalization of the LQ mass matrices are given in the Appendix. Note that we do not include Higgs or gauge-boson self-energies in our calculations. Such effects are flavor universal, drop out at leading order if branching ratios are considered and are already included in the oblique parameters~\cite{Peskin:1991sw,Altarelli:1990zd} as studied in Ref.~\cite{Crivellin:2020ukd}.
\medskip

\begin{boldmath}
\subsection{$\ell\ell\gamma$}
\end{boldmath}

In case of an on-shell photon, we define the effective Hamiltonian as
\begin{align}
\mathcal{H}^{\ell\ell\gamma}_{\rm{eff}}=C^{L}_{\ell_{f}\ell_{i}}O^{L}_{\ell_{f}\ell_{i}}+C^{R}_{\ell_{f}\ell_{i}}O^{R}_{\ell_{f}\ell_{i}}\,,
\label{eq:Heff_llgamma}
\end{align}
with
\begin{align}
O_{\ell_{f}\ell_{i}}^{L(R)}=\frac{e}{16\pi^2}\big[\bar{\ell}_{f}\sigma^{\mu\nu}P_{L(R)}\ell_{i}\big]F_{\mu\nu}\,.
\end{align}
Note that we have $C^{R}_{\ell_{f}\ell_{i}} =C^{L*}_{\ell_{i}\ell_{f}}$ due to the hermiticity of the Hamiltonian.
\smallskip

The coefficients are induced by the diagrams in Fig.~\ref{fig:AMM_LQ} and for a single LQ representation only are given by
\begin{subequations}
	\allowdisplaybreaks
	\label{eq:CL_AMM}
	\begin{align}
	C_{\ell_{f}\ell_{i}}^{L,\Phi_1}&\approx\frac{N_c(m_{\ell_f}\lambda_{jf}^{1L*}\lambda_{ji}^{1L}+m_{\ell_i}\lambda_{jf}^{1R*}\lambda_{ji}^{1R})}{24m_{1}^2}\nonumber\\
	&\quad-\frac{N_{c}m_{t}}{12 m_{1}^2}\lambda_{3f}^{1R*}V_{3k}^{*}\lambda_{ki}^{1L}\bigg(\mathcal{E}_{1}\Big(\frac{m_{t}^2}{m_{1}^2}\Big)-\frac{v^2Y_{1}}{m_{1}^2}\mathcal{E}_{2}\Big(\frac{m_{t}^2}{m_{1}^2}\Big)\bigg)\,,\\
	C_{\ell_{f}\ell_{i}}^{L,\tilde{\Phi}_{1}}&\approx\frac{-m_{\ell_i}\tilde{\lambda}_{jf}^{1*}\tilde{\lambda}_{ji}^{1*}}{12\tilde{m}_{1}^2}\,,\\
	C_{\ell_{f}\ell_{i}}^{L,\Phi_2}&\approx \frac{-N_{c}(m_{\ell_f}\lambda_{jf}^{2LR*}\lambda_{ji}^{2LR}+m_{\ell_i}\lambda_{jf}^{2RL*}\lambda_{ji}^{2RL})}{8m_{2}^2}\nonumber\\
	&\quad+\frac{N_{c}m_{t}}{12 m_{2}^2}V_{3k}^{*}\lambda_{kf}^{2LR*}\lambda_{3i}^{2RL}\bigg(\mathcal{E}_{3}\Big(\frac{m_{t}^2}{m_{2}^2}\Big)-\frac{v^2\big(Y_{22}+Y_{2}\big)}{m_{2}^2}\mathcal{E}_{4}\Big(\frac{m_{t}^2}{m_{2}^2}\Big)\bigg)\,,\\
	C_{\ell_{f}\ell_{i}}^{L,\tilde{\Phi}_2}&\approx 0\,,\\
	C_{\ell_{f}\ell_{i}}^{L,\Phi_3}&\approx\frac{-N_{c}m_{\ell_f}\lambda_{jf}^{3*}\lambda_{ji}^{3}}{8m_{3}^2}\,,
	\end{align}
\end{subequations}
where the quark index $j$ runs from 1 to 3. We expanded the results up to the first non-vanishing order in external momenta and masses. Note that the Wilson coefficients are composed by two parts: a contribution which is proportional to $m_{\ell_{f,i}}$ and a contribution proportional to the quark mass, originating from a chirality flip on the internal quark line. The latter term appears only if a LQ couples simultaneously to left- and right-handed up- or down-type quarks. E.g. for the AMM of the muon this effect dominates in cases where we couple to third generation quarks, i.e. generates a relative enhancement by a factor $m_{t}/m_{\mu}\sim 1600$ or $m_{b}/m_{\mu}\sim 40$, respectively. Therefore, these terms are the most important ones from the phenomenological point of view. And for our results with $m_{t}$ and $m_b$ we also include the $\mathcal{O}(v^2/m_{\rm{LQ}}^2)$ terms, originating from the Higgs-LQ interaction, while we only present the leading order effects for the $m_{\ell_{f,i}}$ terms.
\smallskip

Turning to the contributions with multiple LQ representations, i.e. the terms involving LQ mixing, we also focus on the terms proportional to $m_{b,t}$ and we find
\begin{align}
	\begin{aligned}
		C_{\ell_{f}\ell_{i}}^{L}&\approx \frac{-N_{c}m_{t}v^2}{12 m_{1}^2}\bigg[\lambda_{3f}^{1R*}V_{3k}^{*}\lambda_{ki}^{1L}\frac{|A_{\tilde 21}|^2}{\tilde{m}_{2}^4}\mathcal{E}_{5}\Big(\frac{m_{t}^2}{\tilde{m}_{2}^2},\frac{m_{1}^2}{\tilde{m}_{2}^2}\Big)\\
		&\quad+\lambda_{3f}^{1R*}V_{3k}^{*}\lambda_{ki}^{3}\bigg(\frac{Y_{13}}{m_{3}^2}\mathcal{E}_{6}\Big(\frac{m_{t}^2}{m_{3}^2},\frac{m_{1}^2}{m_{3}^2}\Big)+ \frac{A_{\tilde{2}3}A_{\tilde 21}^{*}}{\tilde{m}_{2}^4} \mathcal{E}_{7}\Big(\frac{m_{t}^2}{\tilde{m}_{2}^2},\frac{m_{1}^2}{\tilde{m}_{2}^2},\frac{m_{3}^2}{\tilde{m}_{2}^2}\Big)\bigg)\bigg]\\
		&\quad+\frac{N_{c}m_{b}v^2}{12}\left[\frac{\lambda_{3f}^{2LR*}\tilde{\lambda}_{3i}^{2}Y_{2\tilde{2}}^{*}}{\tilde{m}_{2}^4}\mathcal{E}_{8}\Big(\frac{m_{b}^2}{\tilde{m}_{2}^2},\frac{m_{2}^2}{\tilde{m}_{2}^2}\Big) +\frac{\tilde{\lambda}_{3f}^{1*}\lambda_{3i}^{3}Y_{3\tilde{1}}^{*}}{m_{3}^4}\mathcal{E}_{9}\Big(\frac{m_{b}^2}{\tilde{m}_{1}^2},\frac{\tilde{m}_{1}^2}{m_{3}^2}\Big)\right]\,.
	\end{aligned}
	\label{eq:CL_AMM_mixing}
\end{align}
The involved loop functions are given explicitly in the Appendix \ref{app:loop_functions} and the general analytical results in Appendix~\ref{app:ellellgamma}. Note that we assumed the quarks of the first two generations to be massless and that we integrated out the bottom and top quark together with the LQs. This means that \eq{eq:CL_AMM} and \eq{eq:CL_AMM_mixing} should be understood to be at the low scale, such that the mixing of the four-fermion operators into the magnetic one is already included, reproducing the logarithms.
\smallskip

Considering $\ell_i\to\ell_{f}\gamma^*$ transitions with a momentum configuration $q^2=(p_i-p_f)^2$, we define the amplitude
\begin{equation}
\mathcal{A} (\ell _i \to \ell _f \gamma ^* ) = -e{q^2} \,\bar{u}(p_f,m_f) \, \slashed{\varepsilon}^*(q) \left(\delta_{fi}+ {\widehat{\Xi} _{fi}^L}P_L + {\widehat{\Xi} _{fi}^R}P_R \right) u(p_i,m_i)\,.
\label{eq:offshell_photon_amp}
\end{equation}
We first give the separate contributions of each LQ representation
\begin{subequations}
\label{eq:llgamma_LQ}
\begin{align}
\Xi_{fi}^{L,\Phi_1}&\approx\frac{-N_{c}\lambda_{kf}^{1L*}V_{jk}V_{jl}^{*}\lambda_{li}^{1L}}{288\pi^2 m_{1}^2}\bigg(\!\mathcal{F}_{1}\Big(\frac{m_{u_j}^2}{m_{1}^2}\Big)-\frac{v^2Y_{1}}{m_{1}^2}\mathcal{F}_{2}\Big(\frac{m_{u_j}^2}{m_{1}^2}\!\Big)+\frac{v^2|A_{\tilde{2}1}|^2}{\tilde{m}_{2}^4}\mathcal{F}_{3}\Big(\frac{m_{u_j}^2}{m_{1}^2},\frac{m_{1}^2}{\tilde{m}_{2}^2}\Big)\!\bigg) \,,\\
\Xi_{fi}^{R,\Phi_1}&\approx\frac{-N_{c}\lambda_{jf}^{1R*}\lambda_{ji}^{1R}}{288\pi^2 m_{1}^2}\bigg(\mathcal{F}_{1}\Big(\frac{m_{u_j}^2}{m_{1}^2}\Big)-\frac{v^2Y_{1}}{m_{1}^2}\mathcal{F}_{2}\Big(\frac{m_{u_j}^2}{m_{1}^2}\Big)+\frac{v^2|A_{\tilde{2}1}|^2}{\tilde{m}_{2}^4}\mathcal{F}_{3}\Big(\frac{m_{u_j}^2}{m_{1}^2},\frac{m_{1}^2}{\tilde{m}_{2}^2}\Big)\bigg) \,, \\
\Xi_{fi}^{R,\tilde{\Phi}_1}&\approx\frac{N_{c}\tilde{\lambda}_{jf}^{1*}\tilde{\lambda}_{ji}^{1}}{144\pi^2\tilde{m}_{1}^2}\bigg(\mathcal{F}_{4}\Big(\frac{m_{d_j}^2}{\tilde{m}_{1}^2}\Big)-\frac{v^2Y_{\tilde{1}}}{\tilde{m}_{1}^2}\mathcal{F}_{8}\Big(\frac{m_{d_j}^2}{\tilde{m}_{1}^2}\Big)\bigg) \,,\\
\Xi_{fi}^{L,\Phi_2}&\approx\frac{N_{c}\lambda_{jf}^{2RL*}\lambda_{ji}^{2RL}}{288\pi^2 m_{2}^2}\bigg(\mathcal{F}_{5}\Big(\frac{m_{u_j}^2}{m_{2}^2}\Big)-\frac{v^2(Y_{2}+Y_{22})}{m_{2}^2}\mathcal{F}_{6}\Big(\frac{m_{u_j}^2}{m_{2}^2}\Big)\bigg) \,, \\
\Xi_{fi}^{R,\Phi_2}&\approx\frac{N_{c}}{288\pi^2 m_{2}^2}\bigg(V_{jk}^{*}\lambda_{kf}^{2LR*}V_{jl}\lambda_{li}^{2LR}\bigg(\mathcal{F}_{5}\Big(\frac{m_{u_j}^2}{m_{2}^2}\Big)-\frac{v^2(Y_{2}+Y_{22})}{m_{2}^2}\mathcal{F}_{6}\Big(\frac{m_{u_j}^2}{m_{2}^2}\Big)\bigg)\nonumber \,, \\
&\quad-2\lambda_{jf}^{2LR*}\lambda_{ji}^{2LR}\bigg(\mathcal{F}_{7}\Big(\frac{m_{d_j}^2}{m_{2}^2}\Big)-\frac{2v^2Y_{2}}{m_{2}^2}\mathcal{F}_{8}\Big(\frac{m_{d_j}^2}{m_{2}^2}\Big)\bigg)\bigg) \,, \\
\Xi_{fi}^{L,\tilde{\Phi}_2}&\approx\frac{-N_{c}\tilde{\lambda}_{jf}^{2*}\tilde{\lambda}_{ji}^{2}}{144\pi^2\tilde{m}_{2}^2}\bigg(\!\mathcal{F}_{7}\Big(\frac{m_{d_j}^2}{\tilde{m}_{2}^2}\Big)\!-\frac{v^2(Y_{\tilde{2}}+Y_{\tilde{2}\tilde{2}})}{\tilde{m}_{2}^2}\mathcal{F}_{4}\Big(\frac{m_{d_j}^2}{\tilde{m}_{2}^2}\Big)\!+\frac{v^2|A_{\tilde{2}3}|^2}{\tilde{m}_{2}^4}\mathcal{F}_{9}\Big(\frac{m_{d_j}^2}{\tilde{m}_{2}^2},\frac{m_{3}^2}{\tilde{m}_{2}^2}\Big)\!\bigg) \,, \\
\Xi_{fi}^{L,\Phi_3}&\approx\frac{-N_{c}}{288\pi^2 m_{3}^2}\bigg(\!\lambda_{kf}^{3*}V_{jk}V_{jl}^{*}\lambda_{li}^{3}\bigg(\!\mathcal{F}_{1}\Big(\frac{m_{u_j}^2}{m_{3}^2}\Big)\!-\frac{v^2Y_{3}}{m_{3}^2}\mathcal{F}_{2}\Big(\frac{m_{u_j}^2}{m_{3}^2}\Big)\!+\frac{v^2|A_{\tilde{2}3}|^2}{\tilde{m}_{2}^4}\mathcal{F}_{3}\Big(\frac{m_{u_j}^2}{m_{3}^2},\frac{m_{3}^2}{\tilde{m}_{2}^2}\Big)\!\bigg)\nonumber \,,\\
&\quad-4\lambda_{jf}^{3*}\lambda_{ji}^{3}\bigg(\mathcal{F}_{4}\Big(\frac{m_{d_j}^2}{m_{3}^2}\Big)-\frac{v^2(Y_{3}-Y_{33})}{m_{3}^2}\mathcal{F}_{8}\Big(\frac{m_{d_j}^2}{m_{3}^2}\Big)\bigg)\bigg) \,.
\end{align}
\end{subequations}
If we include LQ Higgs interactions, we find a new structure originating from $\Phi_{1}$-$\Phi_{3}$ mixing
\begin{align}
\begin{aligned}
\tilde{\Xi}_{fi}^{L}&\approx\frac{N_{c}v^2\lambda_{kf}^{3*}V_{jk}V_{jl}^{*}\lambda_{li}^{1L}}{288\pi^2} \bigg(\frac{Y_{13}^{*}}{m_{3}^4}\mathcal{F}_{10}\Big(\frac{m_{u_j}^2}{m_{3}^2},\frac{m_{1}^2}{m_{3}^2}\Big)+\frac{A_{\tilde{2}1}A_{\tilde{2}3}^{*}}{\tilde{m}_{2}^6}\mathcal{F}_{11}\Big(\frac{m_{u_j}^2}{\tilde{m}_{2}^2},\frac{m_{1}^2}{\tilde{m}_{2}^2},\frac{m_{3}^2}{\tilde{m}_{2}^2}\Big)\bigg)\\
&\quad+\frac{N_{c}v^2\lambda_{kf}^{1L*}V_{jk}V_{jl}^{*}\lambda_{li}^{3}}{288\pi^2}\bigg(\frac{Y_{13}}{m_{3}^4}\mathcal{F}_{10}\Big(\frac{m_{u_j}^2}{m_{3}^2},\frac{m_{1}^2}{m_{3}^2}\Big)+\frac{A_{\tilde{2}3}A_{\tilde{2}1}^{*}}{\tilde{m}_{2}^6}\mathcal{F}_{11}\Big(\frac{m_{u_j}^2}{\tilde{m}_{2}^2},\frac{m_{1}^2}{\tilde{m}_{2}^2},\frac{m_{3}^2}{\tilde{m}_{2}^2}\Big)\bigg)\,,
\end{aligned}
\label{eq:llgamma_LQ_mixing}
\end{align}
at $\mathcal{O}(v^2/m_{\rm{LQ}}^2)$. The quark index $j$ runs from 1 to 3 and the loop functions are given in the Appendix~\ref{app:loop_functions}. Note that we again assumed that the quarks can be integrated out at the same scale as the LQs. This means that the expressions should be understood to be at the low scale and include the mixing of two-quark-two-lepton operators into four-fermion ones. Therefore, in case the quark is lighter than the corresponding leptonic process, one has to insert the scale of that process (rather than the quark mass)  into the logarithms of the loop functions in Appendix~\ref{app:loop_functions}.
\medskip

\begin{boldmath}
\subsection{$Z\ell\ell$ and $Z\nu\nu$}
\label{sec:Zll}
\end{boldmath}

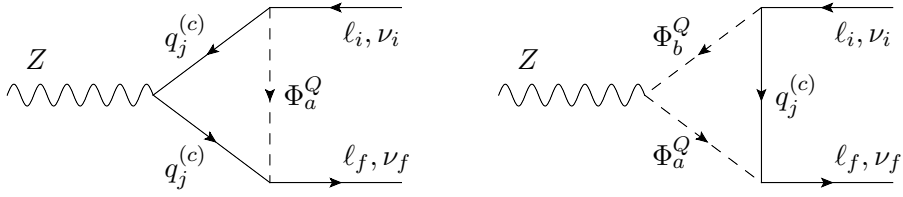
\begin{figure}[t]
	\centering
	\begin{overpic}[scale=.55,,tics=10]
		{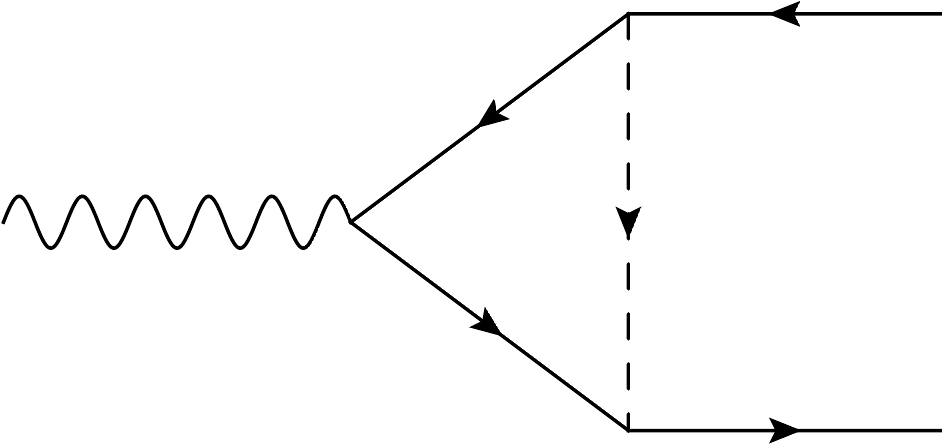}
		\put(5,30){$Z$}
		\put(85,37){$\ell_{i},\nu_{i}$}
		\put(85,6){$\ell_{f},\nu_{f}$}
		\put(40,36){$q_{j}^{(c)}$}
		\put(40,3){$q_{j}^{(c)}$}
		\put(70,21){$\Phi_{a}^{Q}$}
	\end{overpic}
	\hspace{1cm}
	\begin{overpic}[scale=.55,,tics=10]
		{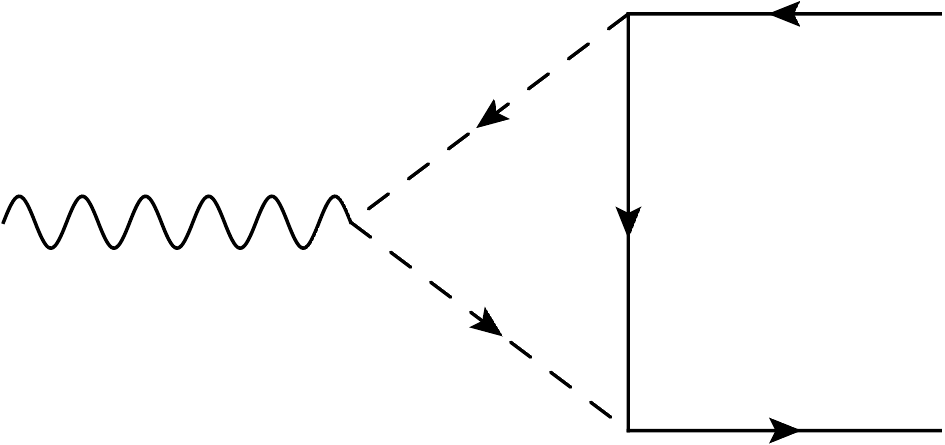}
		\put(5,30){$Z$}
		\put(85,37){$\ell_{i},\nu_{i}$}
		\put(85,6){$\ell_{f},\nu_{f}$}
		\put(39,36){$\Phi_{b}^{Q}$}
		\put(39,6){$\Phi_{a}^{Q}$}
		\put(70,21){$q_{j}^{(c)}$} 
	\end{overpic}
	\caption{Vertex diagrams which contribute to $Z\to\ell_{f}^{-}\ell_{i}^{+}$ and $Z\to\nu_{f}\bar{\nu}_{i}$. Note that in case of mixing among LQs, the Z coupling, unlike the photon, can connect different representations with each other.}
	\label{fig:Zll_Zvvv}
\end{figure}

We now compute the LQ effects on the $Z\to\ell_{f}^{-}\ell_{i}^{+}$ and $Z\to\nu_{f}\bar{\nu}_{i}$ amplitudes, depicted in Fig.~\ref{fig:Zll_Zvvv}
\begin{subequations}
\label{eq:def_Zll_and_Zvv}
\begin{align}
\mathcal{A} (Z\to \ell_{f}^{-} \ell_{i}^{+})&=\frac{g_2}{c_{w}}\bar{u}(p_f,m_{\ell_f})\slashed{\varepsilon}(q)\left(\Lambda^{L}_{\ell_f\ell_i}\!\big(q^2\big)P_L+\Lambda_{\ell_f\ell_i}^{R}\!\big(q^2\big)P_R\right)v(p_i,m_{\ell_i}) \,,\label{eq:Zll_amplitude}\\
\mathcal{A} (Z\to \nu_{f}\bar{\nu}_{i})&= \frac{g_2}{c_w}\Theta_{\nu_{f}\nu_{i}}\!\big(q^2\big) \bar{u}(p_f) \slashed{\varepsilon}(q) P_L v(p_i) \,,
\end{align}
\end{subequations}
with $\varepsilon^\mu(q)$ as the polarization vector of the $Z$ boson and $q^2=(p_{f}+p_{i})^2$. In addition, there is an magnetic form factor for $Z\to \ell^+\ell^-$~\cite{Bernreuther:1996dr}. However, we do not give the form factor of this amplitude explicitly, since it does not interfere with the SM for $m_\ell=0$. We perform this calculation for vanishing lepton masses and decompose the form factors as
\begin{subequations}
\label{eq:effectice_Z_couplings}
\begin{align}
\Lambda_{\ell_f\ell_i}^{L(R)}\!\big(q^2\big)&=\Lambda_{\text{SM}}^{L(R)}(q^2)\delta_{fi}+\sum_{\Phi}\Delta_{L(R),fi}^{\Phi}\big(q^2\big)+\tilde{\Delta}_{L(R),fi}\,,\\
\Theta_{\nu_{f}\nu_{i}}\big(q^2\big)&=\Theta_{\text{SM}}(q^2)\delta_{fi} + \sum_{\Phi}\Theta_{fi}^{\Phi}\big(q^2\big)+\tilde{\Theta}_{fi} \,.
\end{align}
\end{subequations}
The $\Delta_{L(R),fi}^{\Phi}\big(q^2\big)$ and $\Theta_{fi}^{\Phi}\big(q^2\big)$ contain the part with no LQ mixing, grouped into $\Phi=\{\Phi_1,\tilde{\Phi}_1,\Phi_2,\tilde{\Phi}_2,\Phi_3\}$, while the $\tilde{\Delta}_{L(R),fi}$ and $\tilde{\Theta}_{fi}$ contain the part induced by LQ mixing. In our conventions, the tree-level SM couplings read
\begin{align}
\Lambda_{\text{SM}}^{L}=s_w^2-\frac{1}{2}\,, &&\Lambda_{\text{SM}}^{R}=s_w^2\,,&&\Theta_{\text{SM}}=\frac{1}{2}\,,
\end{align}
with $s_w$ ($c_w$) being the sine (cosine) of the Weinberg angle. Beyond tree-level, also the SM couplings receive momentum dependent corrections, which are included in the predictions for EW observables that we study later in the phenomenological analysis.
\smallskip

In our calculation we only include contributions of $\mathcal{O}(m_{\text{EW}}^2/m_{\rm{LQ}}^2)$, i.e. effects from the top quark, the $Z$ mass as well as the ones induced by LQ mixing, while setting all other masses to zero\footnote{Similar results for the diquark contribution to $Z \to \ell^+ \ell^-$ have been obtained in Ref.~\cite{Djouadi:1989me}.}. In case where the $Z$ boson has a squared momentum $q^2$ we find
\begin{subequations}
\allowdisplaybreaks
\begin{align}
\Delta_{L,fi}^{\Phi_1}(q^2) &\approx\frac{-N_{c}\lambda_{kf}^{1L*}V_{3k}V_{3j}^{*}\lambda_{ji}^{1L}}{32\pi^2}\bigg[\mathcal{H}_{0}\Big(\frac{m_{t}^2}{m_{1}^2}\Big)\!-\frac{q^2}{18m_{1}^2}\Big(\!11-10s_{w}^2\!+2(3-4s_{w}^2)\log\!\Big(\frac{m_{t}^2}{m_{1}^2}\Big)\!\Big)\nonumber\\
&\hspace{4.2cm}-\frac{q^2}{180 m_{1}^2}\frac{q^2}{m_{t}^2}\Big(16s_{w}^2-9\Big)\bigg]\nonumber\\
&\!\!-\sum_{j=1}^2\frac{N_{c}\lambda_{kf}^{1L*}V_{jk}V_{jl}^{*}\lambda_{li}^{1L}}{864\pi^2}\frac{q^2}{m_{1}^2}\bigg[3-3i\pi(4s_{w}^2-3)-5s_{w}^2+3(4s_{w}^2-3)\log\!\Big(\frac{q^2}{m_{1}^2}\Big)\bigg] \,,\\
\begin{split}
\Delta_{R,fi}^{\Phi_1}(q^2)&\approx\frac{N_{c}\lambda_{3f}^{1R*}\lambda_{3i}^{1R}}{32\pi^2 }\bigg[\mathcal{H}_{0}\Big(\frac{m_{t}^2}{m_{1}^2}\Big)-\frac{q^2}{18m_{1}^2}\Big(3+10s_{w}^2 +8s_{w}^2\log\!\Big(\frac{m_{t}^2}{m_{1}^2}\Big)\Big)
\\
&\hspace{2.8cm}+\frac{q^2}{180 m_{1}^2}\frac{q^2}{m_{t}^2}\Big(16s_{w}^2-3\Big)\bigg]
\end{split}
\\
&+\sum_{j=1}^2\frac{N_{c}\lambda_{jf}^{1R*}\lambda_{ji}^{1R}}{864\pi^2}\frac{q^2}{m_{1}^2}\bigg[5s_{w}^2+12i\pi s_{w}^2-12s_{w}^2\log\!\Big(\frac{q^2}{m_{1}^2}\Big)\bigg] \,,\nonumber \\
\Delta_{R,fi}^{\tilde{\Phi}_1}(q^2)&\approx\sum_{j=1}^3\frac{N_{c}\tilde{\lambda}_{jf}^{1*}\tilde{\lambda}_{ji}^{1}}{432\pi^2}\frac{q^2}{\tilde{m}_{1}^2}\bigg[s_{w}^2-3i\pi s_{w}^2 +3s_{w}^2\log\!\Big(\frac{q^2}{\tilde{m}_{1}^2}\Big)\bigg] \,, \\
\Delta_{L,fi}^{\Phi_2}(q^2)&\approx\frac{-N_{c}\lambda_{3f}^{2RL*}\lambda_{3i}^{2RL}}{32\pi^2}\bigg[\mathcal{H}_{0}\Big(\frac{m_{t}^2}{m_{2}^2}\Big)-\frac{q^2}{9m_{2}^2}\Big(1+7s_{w}^2+4s_{w}^2\log\!\Big(\frac{m_{t}^2}{m_{2}^2}\Big)\Big)
\phantom{\,.}\nonumber\\
&\hspace{3.6cm}+\frac{q^2}{180m_{2}^2}\frac{q^2}{m_{t}^2}\Big(16s_{w}^2-3\Big)\bigg]\\
&-\sum_{j=1}^2\frac{N_{c}\lambda_{jf}^{2RL*}\lambda_{ji}^{2RL}}{1728\pi^2}\frac{q^2}{m_{2}^2}\bigg[3-2s_{w}^2+24i\pi s_{w}^2-24s_{w}^2\log\!\Big(\frac{q^2}{m_{2}^2}\Big)\bigg]\nonumber \,, \\
\Delta_{R,fi}^{\Phi_2}(q^2)&\approx\frac{N_{c}V_{3k}^{*}\lambda_{kf}^{2LR*}V_{3j}\lambda_{ji}^{2LR}}{32\pi^2}\bigg[\mathcal{H}_{0}\Big(\frac{m_{t}^2}{m_{2}^2}\Big)-\frac{q^2}{9m_{2}^2}\Big(6-7s_{w}^2+(3-4s_{w}^2)\log\!\Big(\frac{m_{t}^2}{m_{2}^2}\Big)\Big)
\phantom{\,.}\nonumber \\
&\hspace{4.4cm}-\frac{q^2}{180 m_{2}^2}\frac{q^2}{m_{t}^2}\Big(16s_{w}^2-9\Big)\bigg]\nonumber\\
&+\sum_{j=1}^2\frac{N_{c}V_{jk}^{*}\lambda_{kf}^{2LR*}V_{jl}\lambda_{li}^{2LR}}{1728\pi^2}\frac{q^2}{m_{2}^2}\bigg[3+2s_{w}^2\!-6i\pi(4s_{w}^2\!-3) +6(4s_{w}^2\!-3)\log\!\Big(\frac{q^2}{m_{2}^2}\Big)\!\bigg]
\phantom{\,.}\nonumber\\
&-\sum_{j=1}^3\frac{N_{c}\lambda_{jf}^{2LR*}\lambda_{ji}^{2LR}}{1728\pi^2}\frac{q^2}{m_{2}^2}\bigg[3-6i\pi(2s_{w}^2-3)-8s_{w}^2+6(2s_{w}^2-3)\log\!\Big(\frac{q^2}{m_{2}^2}\Big)\bigg] \,,\displaybreak[0]\\
\Delta_{L,fi}^{\tilde{\Phi}_2}(q^2)&\approx\sum_{j=1}^3\frac{-N_{c}\tilde{\lambda}_{jf}^{2*}\tilde{\lambda}_{ji}^{2}}{1728\pi^2}\frac{q^2}{\tilde{m}_{2}^2}\bigg[3-8s_{w}^2-12i\pi s_{w}^2 +12s_{w}^2\log\!\Big(\frac{q^2}{\tilde{m}_{2}^2}\Big)\bigg] \,, \\
\Delta_{L,fi}^{\Phi_3}(q^2)&\approx
\frac{-N_{c}V_{3k}\lambda_{kf}^{3*}V_{3j}^{*}\lambda_{ji}^{3}}{32\pi^2}\bigg[\mathcal{H}_{0}\Big(\frac{m_{t}^2}{m_{3}^2}\Big)\!-\frac{q^2}{18m_{3}^2}\Big(11-10s_{w}^2\!+2(3-4s_{w}^2)\log\!\Big(\frac{m_{t}^2}{m_{3}^2}\Big)\!\Big)
\phantom{\,.}\nonumber\\
&\hspace{3.9cm}-\frac{q^2}{180 m_{3}^2}\frac{q^2}{m_{t}^2}\Big(16s_{w}^2-9\Big)\bigg]\nonumber\\
&-\sum_{j=1}^2\frac{N_{c}V_{jk}\lambda_{kf}^{3*}V_{jl}^{*}\lambda_{li}^{3}}{864\pi^2}\frac{q^2}{m_{3}^2}\bigg[3-3i\pi(4s_{w}^2-3)-5s_{w}^2+3(4s_{w}^2-3)\log\!\Big(\frac{q^2}{m_{3}^2}\Big)\bigg]
\phantom{\,.}\nonumber \\
&-\sum_{j=1}^3\frac{N_{c}\lambda_{jf}^{3*}\lambda_{ji}^{3}}{216\pi^2}\frac{q^2}{m_{3}^2}\bigg[3i\pi(2s_{w}^2-3)-2s_{w}^2-3(2s_w^2-3)\log\!\Big(\frac{q^2}{m_{3}^2}\Big)\bigg] \,,
\end{align}
\end{subequations}
where we expanded the results in $q^2/m^2_{\rm LQ}$ and $m_t^2/m^2_{\rm LQ}$. Finally, the contributions from LQ mixing read
\begin{subequations}
\begin{align}
\tilde{\Delta}_{L,fi}&\approx\sum_{j=1}^{3}\frac{-v^2 N_{c}}{64\pi^2 \tilde{m}_{2}^4} \bigg[\lambda_{jf}^{1L*}\lambda_{ji}^{1L}|A_{\tilde 21}|^2\mathcal{H}_{3}\Big(\frac{m_{1}^2}{\tilde{m}_{2}^2}\Big)+\left(\lambda_{jf}^{3*}\lambda_{ji}^{3}+2\tilde{\lambda}_{ji}^{2*}\tilde{\lambda}_{jf}^{2}\right)|A_{\tilde{2}3}|^2\mathcal{H}_{3}\Big(\frac{m_{3}^2}{\tilde{m}_{2}^2}\Big)\nonumber\\
&\quad+\Big(\lambda_{jf}^{1L}\lambda_{ji}^{3*}A_{\tilde 21}A_{\tilde{2}3}^{*}+\lambda_{jf}^{3}\lambda_{ji}^{1L*}A_{\tilde{2}3}A_{\tilde 21}^{*}\Big)\mathcal{H}_{4}\Big(\frac{m_{1}^2}{\tilde{m}_{2}^2},\frac{m_{3}^2}{\tilde{m}_{2}^2}\Big)\bigg] \,,\\
\tilde{\Delta}_{R,fi}&\approx\sum_{j=1}^{3}\frac{-v^2N_{c}}{64\pi^2 \tilde{m}_{2}^4}\lambda_{jf}^{1R*}\lambda_{ji}^{1R}|A_{\tilde 21}|^2\mathcal{H}_{3}\Big(\frac{m_{1}^2}{\tilde{m}_{2}^2}\Big) \,,
\end{align}
\end{subequations}
where the $\mathcal{H}$-functions are given in Appendix~\ref{app:loop_functions}.
\smallskip

Now we turn to the $Z\to\nu_{f}\bar{\nu}_{i}$ amplitudes, where we show the contributions again separated by each representation
\begin{subequations}
\begin{align}
\Theta_{fi}^{\Phi_1}(q^2)&\approx\sum_{j=1}^3\frac{N_{c}\lambda_{jf}^{1L*}\lambda_{ji}^{1L}}{864\pi^2}\frac{q^2}{m_{1}^2}\bigg[3-s_{w}^2+3i\pi(3-2s_{w}^2)-3(3-2s_{w}^2)\log\!\Big(\frac{q^2}{m_{1}^2}\Big)\bigg]\,, \\
\Theta_{fi}^{\Phi_2}(q^2)&\approx\frac{-N_{c}\lambda_{3f}^{2RL*}\lambda_{3i}^{2RL}}{32\pi^2}\bigg[\mathcal{H}_{0}\Big(\frac{m_{t}^2}{m_{2}^2}\Big)-\frac{2q^2}{9m_{2}^2}\Big(1+3s_{w}^2+2s_{w}^2\log\!\Big(\frac{m_{t}^2}{m_{2}^2}\Big)\!\Big)\nonumber\\
&\hspace{3.6cm}+\frac{q^2}{180m_{2}^2}\frac{q^2}{m_{t}^2}\Big(16s_{w}^2-3\Big)\bigg]\\
&\quad+\sum_{j=1}^2\frac{N_{c}\lambda_{jf}^{2RL*}\lambda_{ji}^{2RL}}{1728\pi^2}\frac{q^2}{m_{2}^2}\bigg[3-4s_{w}^2-24i\pi s_{w}^2+24s_{w}^2\log\!\Big(\frac{q^2}{m_2^2}\Big)\bigg] \,,\nonumber\\
\Theta_{fi}^{\tilde{\Phi}_2}(q^2)&\approx\sum_{j=1}^3\frac{N_{c}\tilde{\lambda}_{jf}^{2*}\tilde{\lambda}_{ji}^{2}}{1728\pi^2}\frac{q^2}{\tilde{m}_{2}^2}\bigg[3+2s_{w}^2+12i\pi s_{w}^2 -12s_{w}^2\log\!\Big(\frac{q^2}{\tilde{m}_{2}^2}\Big)\bigg] \,,\displaybreak[0] \\
\Theta_{fi}^{\Phi_3}(q^2)&\approx\frac{-N_{c}V_{3k}\lambda_{ki}^{3*}V_{3j}^{*}\lambda_{jf}^{3}}{16\pi^2}\bigg[\mathcal{H}_{0}\Big(\frac{m_{t}^2}{m_{3}^2}\Big)-\frac{q^2}{18m_{3}^2}\Big(13-12s_{w}^2+(6-8s_{w}^2)\log\!\Big(\frac{m_{t}^2}{m_{3}^2}\Big)\Big)
\nonumber\\
&\hspace{4cm}-\frac{q^2}{180 m_{3}^2}\frac{q^2}{m_{t}^2}\Big(16s_{w}^2-9\Big)\bigg]\nonumber\\
&\quad+\sum_{j=1}^2\frac{N_{c}V_{jk}\lambda_{ki}^{3*}V_{jl}^{*}\lambda_{lf}^{3}}{432\pi^2}\frac{q^2}{m_{3}^2}\bigg[2s_{w}^2-3i\pi(3-4s_{w}^2)-3(4s_{w}^2-3)\log\!\Big(\frac{q^2}{m_3^2}\Big)\bigg]
\phantom{\,.}\nonumber\\
&\quad+\sum_{j=1}^3\frac{N_{c}\lambda_{jf}^{3*}\lambda_{ji}^{3}}{864\pi^2}\frac{q^2}{m_{3}^2}\bigg[3-s_{w}^2+3i\pi(3-2s_{w}^2)-3(3-2s_{w}^2)\log\!\Big(\frac{q^2}{m_{3}^2}\Big)\bigg] \,.
\end{align}
\end{subequations}
Finally, we again have the contributions from LQ mixing
\begin{align}
\begin{aligned}
\tilde{\Theta}_{fi}&\approx\sum_{j=1}^{3}\frac{-v^2 N_{c}}{64\pi^2\tilde{m}_{2}^4}\bigg[\left(\lambda_{jf}^{1L*}\lambda_{ji}^{1L}+\tilde{\lambda}_{ji}^{2*}\tilde{\lambda}_{jf}^{2}\right)|A_{\tilde 21}|^2\mathcal{H}_{3}\Big(\frac{m_{1}^2}{\tilde{m}_{2}^2}\Big)
\\
&\quad+\left(5\lambda_{jf}^{3*}\lambda_{ji}^{3}+\tilde{\lambda}_{ji}^{2*}\tilde{\lambda}_{jf}^{2}\right)|A_{\tilde{2}3}|^2\mathcal{H}_{3}\Big(\frac{m_{3}^2}{\tilde{m}_{2}^2}\Big) \\
&\quad-\Big(\lambda_{jf}^{1L}\lambda_{ji}^{3*}A_{\tilde 21}A_{\tilde{2}3}^{*}+\lambda_{jf}^{3}\lambda_{ji}^{1L*}A_{\tilde{2}3}A_{\tilde 21}^{*}\Big)\mathcal{H}_{4}\Big(\frac{m_{1}^2}{\tilde{m}_{2}^2},\frac{m_{3}^2}{\tilde{m}_{2}^2}\Big)
\bigg] \,.
\end{aligned}
\end{align}
\smallskip

In case of zero momentum transfer, i.e. $q^2=0$, the form factors correspond to effective $Z\ell\ell$ and $Z\nu{\nu}$ couplings. We define them for later purposes in an effective Lagrangian
\begin{subequations}
\label{eq:Zll_eff_Lag}
\begin{align}
\mathcal{L}^{Z\ell\ell}_{\rm{int}}& =\frac{g_2}{c_w}\big[\bar{\ell}_{f} \left(\Lambda_{\ell_f\ell_i}^{L}(0)\gamma^{\mu}P_{L}+\Lambda^{R}_{\ell_f\ell_i}(0)\gamma^{\mu}P_{R}\right)\ell_{i}\big]Z_{\mu}\,,\\
\mathcal{L}^{Z\nu\nu}_{\text{int}}& =\frac{g_2}{c_w}\Theta_{\nu_{f}\nu_{i}}(0)\left[\bar{\nu}_{f}\gamma_{\mu}P_{L}\nu_{i}\right]Z^{\mu}\,,
\end{align}
\end{subequations}
where only the $\tilde{\Delta}$, $\tilde{\Theta}$ and the top contributions remain.
\medskip

\begin{boldmath}
\subsection{$W\ell\nu$}
\label{sec:Wlnu}
\end{boldmath}

\begin{figure}[t]
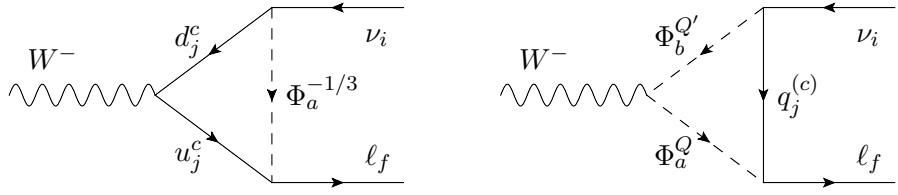

	\centering
	\begin{overpic}[scale=.55,,tics=10]
		{./Diagrams/z_decay.pdf}
		\put(5,30){$W^{-}$}
		\put(90,37){$\nu_{i}$}
		\put(90,6){$\ell_{f}$}
		\put(42,36){$d_{j}^{c}$}
		\put(42,7){$u_{j}^{c}$}
		\put(70,21){$\Phi_{a}^{-1/3}$}
	\end{overpic}
	\hspace{1cm}
	\begin{overpic}[scale=.55,,tics=10]
		{./Diagrams/z_decay2.pdf}
		\put(5,30){$W^{-}$}
		\put(90,37){$\nu_{i}$}
		\put(90,6){$\ell_{f}$}
		\put(39,36){$\Phi_{b}^{Q^{\prime}}$}
		\put(39,6){$\Phi_{a}^{Q}$}
		\put(70,21){$q_{j}^{(c)}$} 
	\end{overpic}
	\caption{Vertex diagrams contributing to $W^{-}\to\ell_{f}^{-}\bar{\nu_{i}}$. In the case of massless down-type quarks the diagram on the left-hand side is only present with charge-conjugated quarks, since the $W$ boson couples to purely to left-handed quarks.}
	\label{fig:diag_wlnu}
\end{figure}

We define the amplitude of this process, also considered for generic new scalars and fermions in Ref.~\cite{Arnan:2019olv}, as follows
\begin{align}
\mathcal{A} (W^{-}\to \ell_{f}^{-} \bar{\nu}_{i})= \frac{g_2}{\sqrt{2}}\Lambda_{\ell_{f}\nu_{i}}^{W}\!\big(q^2\big) \bar{u}(p_{\ell_f},m_{\ell_f}) \slashed{\varepsilon}(q) P_L \, v(p_{\nu_i}) \ ,
\label{eq:ampl_wlnu}
\end{align}
with
\begin{align}
\Lambda^W_{\ell_{f}\nu_{i}}\!\big(q^2\big)&= \Lambda^{W}_{\text{SM}}(q^2)\delta_{fi} + \sum_{\Phi}\Lambda^{\Phi}_{fi}\big(q^2\big)  + \tilde{\Lambda}_{fi}\ .
\label{eq:wlnu_eff}
\end{align}
The diagrams are shown in Fig.~\ref{fig:diag_wlnu}. The form factors $\Lambda_{fi}^{\Phi}\big(q^2\big)$ again contain the parts with no LQ mixing, grouped by representation with $\Phi=\{\Phi_1,\tilde{\Phi}_1,\Phi_2,\tilde{\Phi}_2,\Phi_3\}$, while $\tilde{\Lambda}_{fi}$ contains the part with LQ mixing. In the SM we have at tree-level
\begin{align}
\Lambda^{W}_{\text{SM}}=1\,.
\end{align}

\smallskip
The single LQ contributions read
\begin{subequations}
\allowdisplaybreaks
\begin{align}
\Lambda_{fi}^{\Phi_1}(q^2)&\approx \frac{N_{c}V_{3k}\lambda_{kf}^{1L*}V_{3j}^{*}\lambda_{ji}^{1L}}{64\pi^2}\bigg[\frac{m_{t}^2}{m_{1}^2}\Big(1+2\log\!\Big(\frac{m_{t}^2}{m_{1}^2}\Big)\Big)-\frac{4q^2}{3m_{1}^2}\Big(1+\log\!\Big(\frac{m_{t}^2}{m_{1}^2}\Big)\Big)+\frac{q^2}{2m_{1}^2}\frac{q^2}{m_{t}^2}\bigg]\nonumber\\
&\quad-\sum_{j=1}^2\frac{N_{c}V_{jl}\lambda_{lf}^{1L*}V_{jk}^{*}\lambda_{ki}^{1L}}{144\pi^2}\frac{q^2}{m_{1}^2}\bigg[3\log\!\Big(\frac{q^2}{m_{1}^2}\Big)+3i\pi-1\bigg]\,,\\
\Lambda_{fi}^{\Phi_2}(q^2)&\approx\sum_{j=1}^3\frac{N_{c}\,q^2}{288\pi^2}\frac{\lambda_{jf}^{2RL*}\lambda_{ji}^{2RL}}{m_{2}^2} \,,\\
\Lambda_{fi}^{\tilde{\Phi}_2}(q^2)&\approx\sum_{j=1}^3\frac{N_{c}\,q^2}{288\pi^2}\frac{\tilde{\lambda}_{jf}^{2*}\tilde{\lambda}_{ji}^{2}}{\tilde{m}_{2}^2}\,,\\
\Lambda_{fi}^{\Phi_3}(q^2)&\approx\frac{-N_{c}V_{3k}\lambda_{kf}^{3*}V_{3j}\lambda_{ji}^{3}}{64\pi^2}\bigg[\frac{m_{t}^2}{m_{3}^2}\Big(1+2\log\!\Big(\frac{m_{t}^2}{m_{3}^2}\Big)\!\Big)-\frac{4q^2}{9m_{3}^2}\Big(4+3\log\!\Big(\frac{m_{t}^2}{m_{3}^2}\Big)\!\Big)+\frac{q^2}{2m_{3}^2}\frac{q^2}{m_{t}^2}\bigg]
\nonumber\\
&\quad+\sum_{j=1}^2\frac{N_{c}}{144\pi^2}V_{jl}\lambda_{lf}^{3*}V_{jk}^{*}\lambda_{ki}^{3}\frac{q^2}{m_{3}^2}\Big(3\log\!\Big(\frac{q^2}{m_{3}^2}\Big)+3i\pi+1\Big)\\
&\quad-\sum_{j=1}^3\frac{N_{c}}{144\pi^2}\frac{q^2}{m_{3}^2}\lambda_{jf}^{3*}\lambda_{ji}^{3}\,.\nonumber
\end{align}
\end{subequations}
Additionally, we have the $\mathcal{O}(v^2/m_{\rm{LQ}}^2)$ effects from LQ mixing
\begin{align}
\begin{aligned}
\tilde{\Lambda}_{fi}&\approx 
\sum_{j=1}^{3}\frac{v^2N_{c}}{64\pi^2} \Bigg[\left(\tilde{\lambda}^{2*}_{jf}\tilde{\lambda}_{ji}^{2}-4\lambda_{jf}^{3*}\lambda_{ji}^{3}\right)\frac{|A_{\tilde{2}3}|^2}{\tilde{m}_{2}^4}\mathcal{H}_{3}\Big(\frac{m_{3}^2}{\tilde{m}_{2}^2}\Big)-\tilde{\lambda}^{2*}_{jf}\tilde{\lambda}_{ji}^{2}\frac{|A_{\tilde 21}|^2}{\tilde{m}_{2}^4}\mathcal{H}_{3}\Big(\frac{m_{1}^2}{\tilde{m}_{2}^2}\Big)\\ &\quad+2\frac{\lambda_{ji}^{1L}\lambda_{jf}^{3*}A_{\tilde{2}3}^{*}A_{\tilde 21}+\lambda_{jf}^{1L*}\lambda_{ji}^{3}A_{\tilde{2}3}A_{\tilde 21}^{*}}{\tilde{m}_{2}^4}\mathcal{H}_{4}\Big(\frac{m_{1}^2}{\tilde{m}_{2}^2},\frac{m_{3}^2}{\tilde{m}_{2}^2}\Big)\\
&\quad+2\frac{Y_{13}\lambda_{ji}^{1L*}\lambda_{jf}^{3} -Y_{13}^{*}\lambda_{ji}^{3*}\lambda_{jf}^{1L}}{m_{3}^2}\mathcal{H}_{5}\Big(\frac{m_{1}^2}{m_{3}^2}\Big)
\\
&\quad+2\frac{\lambda_{ji}^{1L}\lambda_{jf}^{3*}A_{\tilde{2}3}^{*}A_{\tilde
21}-\lambda_{jf}^{1L*}\lambda_{ji}^{3}A_{\tilde{2}3}A_{\tilde 21}^{*}}{\tilde{m}_{2}^4}\mathcal{H}_{6}\Big(\frac{m_{1}^2}{\tilde{m}_{2}^2},\frac{m_{3}^2}{\tilde{m}_{2}^2}\Big)
\Bigg] \,,
\end{aligned}
\label{WlnuMixing}
\end{align}
with the loop functions given in the Appendix~\ref{app:loop_functions}.
Note that the terms in the last two lines in \eq{WlnuMixing} are anti-hermitian in flavor space. Therefore, like the anti-hermitian part of the self-energy contributions, see \eq{ULrot}, they are not physical. In fact, we checked that the terms originating from LQ mixing for $Z\ell\ell$, $Z\nu\nu$ and $W\ell\nu$ respect the structure required by the dim-6 operators with manifest $SU(2)_L$ invariance if the anti-hermitian terms are absorbed by the PMNS matrix.
\smallskip

At the level of effective couplings, we have to evaluate the contributions at $q^2=0$, which can be treated in the context of an effective Lagrangian
\begin{align}
\mathcal{L}_{{\rm int}}^{W\ell\nu}=\frac{g}{\sqrt{2}}\Lambda_{\ell_{f}\nu_{i}}^{W}(0) \big[\bar{\ell}_{f}\gamma^{\mu}P_{L}\nu_{i}\big]W_{\mu}^{-}\,.
\end{align}
The effective coupling $\Lambda_{\ell_{f}\nu_{i}}^{W}(0)$ then only receives LQ effects from loop-induced top quarks and from LQ mixing.
\medskip

\begin{boldmath}
	\subsection{$h \ell \ell$}
	\label{sec:hll}
\end{boldmath}

\begin{figure}[t]
	\centering
	\begin{overpic}[scale=.55,,tics=10]
		{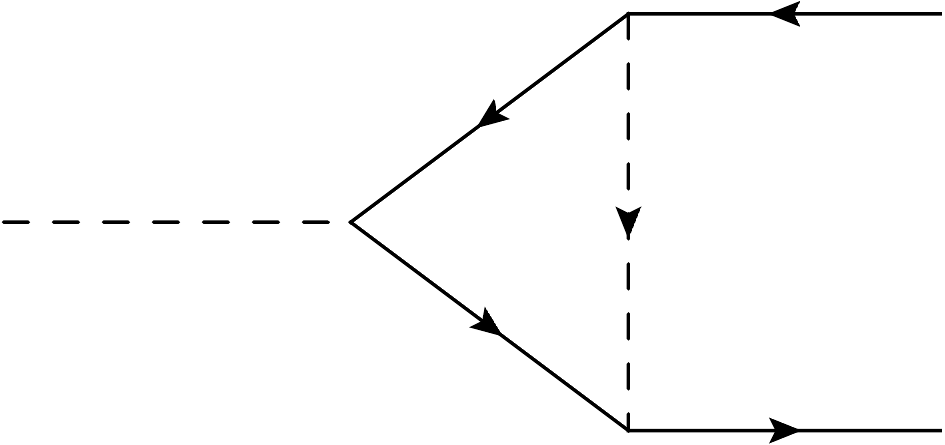}
		\put(5,27){$h$}
		\put(90,37){$\ell_{i}$}
		\put(90,6){$\ell_{f}$}
		\put(40,36){$q_{j}^{(c)}$}
		\put(40,3){$q_{j}^{(c)}$}
		\put(70,21){$\Phi_{a}^{Q}$}
	\end{overpic}
	\hspace{1cm}
	\begin{overpic}[scale=.55,,tics=10]
		{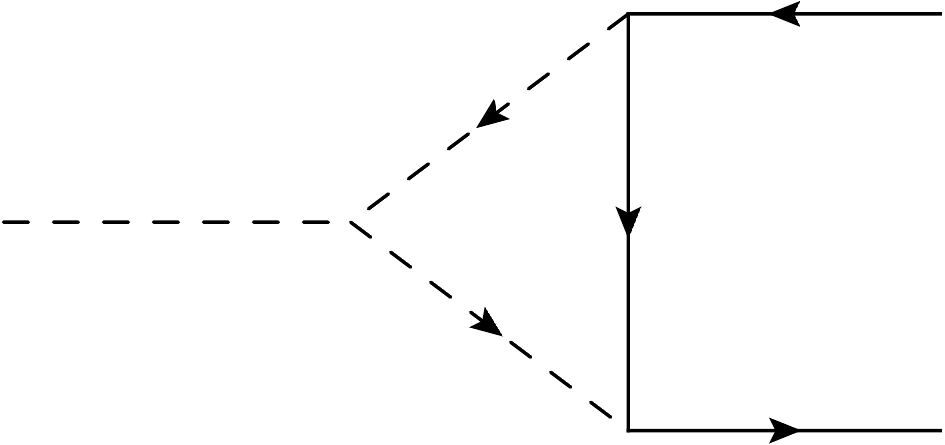}
		\put(5,27){$h$}
		\put(90,37){$\ell_{i}$}
		\put(90,6){$\ell_{f}$}
		\put(39,36){$\Phi_{b}^{Q}$}
		\put(39,6){$\Phi_{a}^{Q}$}
		\put(70,21){$q_{j}^{(c)}$}
	\end{overpic}
	\caption{Vertex diagrams generating $h\to\ell_{f}^{-}\ell_{i}^{+}$ at the 1-loop level.}
	\label{fig:higgs_to_ll}
\end{figure}
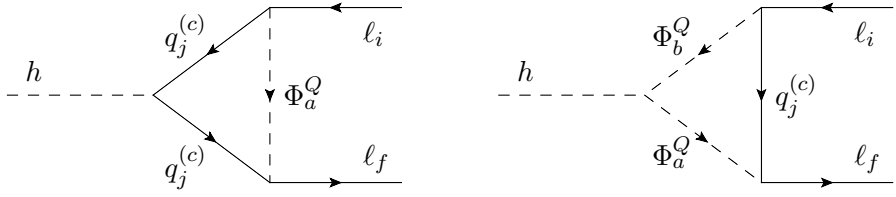

Let us turn next to the Higgs decays $h\to\ell_{f}^{-}\ell_{i}^{+}$. We define the amplitude analogously to the leptonic $W$ and $Z$ decays as
\begin{align}
\mathcal{A}(h\to\ell_{f}^{-}\ell_{i}^{+})=-\frac{m_{fi}}{\sqrt{2}\,v}\bar{u}(p_{f},m_{\ell_f})\Big(\Upsilon_{\ell_{f}\ell_{i}}^{L}(q^2)P_{L}+\Upsilon_{\ell_{f}\ell_{i}}^{R}(q^2)P_{R}\Big)v(p_{i},m_{\ell_i}) \,,
\label{eq:amplitude_hll}
\end{align}
with
\begin{subequations}
\begin{align}
\Upsilon_{\ell_{f}\ell_{i}}^{L}(q^2)&=\delta_{fi}+\sum_{\Phi}\Upsilon_{L,fi}^{\Phi}(q^2)+\tilde{\Upsilon}_{L,fi}\,,\\ \Upsilon_{\ell_{f}\ell_{i}}^{R}(q^2)&=\delta_{fi}+\sum_{\Phi}\Upsilon_{R,fi}^{\Phi}(q^2)+\tilde{\Upsilon}_{R,fi}\,.
\end{align}
\end{subequations}
The sum over $\Phi$ refers to the LQ representations $\Phi=\{\Phi_{1},\tilde{\Phi}_{1},\Phi_{2},\tilde{\Phi}_{2},\Phi_{3}\}$, $\tilde{\Upsilon}_{L(R),fi}$ contain the terms which are only generated by LQ mixing and
\begin{align}
m_{fi}=\max[m_{\ell_{f}},m_{\ell_i}]\,,&&q^2=(p_{f}+p_{i})^2\,.
\end{align}
Note that due to hermicity
\begin{align}
\Upsilon_{\ell_{f}\ell_i}^{R}=\Upsilon_{\ell_{i}\ell_{f}}^{L*}\,.
\end{align}
If $f\neq i$ we can safely neglect the lighter lepton mass. The corresponding Feynman diagrams are shown in Figure \ref{fig:higgs_to_ll}.
\smallskip

We expand again in $v^2/m_{\rm{LQ}}^2$ and set the lepton masses to zero. In the phenomenologically most relevant case of an internal top quark, we additionally use the fact that for Higgs decays $m_{t}^2>m_{h}^2\equiv q^2$, finding
\begin{subequations}
\begin{align}
\Upsilon_{L,fi}^{\Phi_{1}}(q^2)&\approx\frac{N_{c}\lambda_{3f}^{1R*}V_{3k}^{*}\lambda_{ki}^{1L}}{64\pi^2 m_{1}^2}\frac{m_{t}}{m_{fi}} \bigg(m_{t}^2\mathcal{J}_{t}\Big(\frac{q^2}{m_{t}^2},\frac{m_{t}^2}{m_{1}^2}\Big)+8{v^{2}Y_{1}}\bigg) \,,\\
\Upsilon_{L,fi}^{\Phi_{2}}(q^2)&\approx\frac{N_{c}V_{3k}^{*}\lambda_{kf}^{2LR*}\lambda_{3i}^{2RL}}{64\pi^2 m_{2}^2}\frac{m_{t}}{m_{fi}} \bigg(m_{t}^2\mathcal{J}_{t}\Big(\frac{q^2}{m_{t}^2},\frac{m_{t}^2}{m_{2}^2}\Big)+8{v^2\big(Y_{22}+Y_{2}\big)}\bigg)\,.
\end{align}
\end{subequations}
The mixing-induced terms read up to $\mathcal{O}(v^2/m_{\rm{LQ}}^2)$
\begin{align}
\begin{aligned}
\tilde{\Upsilon}_{L,fi}&\approx\sum_{j=1}^3\frac{-v^{2}N_{c}}{8\pi^2}\bigg[\frac{m_{t}}{m_{fi}}\bigg(\lambda_{3f}^{1R*}V_{3k}^{*}\lambda_{ki}^{1L}\frac{|A_{\tilde{2}1}|^2}{\tilde{m}_{2}^4}\mathcal{J}_{1}\Big(\frac{m_{1}^2}{\tilde{m}_{2}^2}\Big)\\
&\quad-\lambda_{3f}^{1R*}V_{3k}^{*}\lambda_{ki}^{3} \Big(\frac{Y_{13}}{m_{1}^2}\mathcal{H}_{1}\Big(\frac{m_{3}^2}{m_{1}^2}\Big)+\frac{A_{\tilde{2}3}A_{\tilde{2}1}^{*}}{\tilde{m}_{2}^4}\mathcal{J}_{2}\Big(\frac{m_{1}^2}{\tilde{m}_{2}^2},\frac{m_{3}^2}{\tilde{m}_{2}^2}\Big)\Big)\bigg)\bigg]\,.
\end{aligned}
\end{align}
The loop functions that we used in this section can be found in the Appendix~\ref{app:loop_functions}. In Appendix~\ref{app:matrices} we additionally present the generic results for light quarks, i.e. for the case where $m_{q_j}^{2}\ll q^2 \equiv m_{h}^2$.
\medskip


\begin{boldmath}
\subsection{$4\ell$}
\end{boldmath}

To describe processes involving four charged leptons, we define the effective Hamiltonian as
\begin{align}
\mathcal{H}_{\text{eff}}^{4\ell}=\mathcal{H}_{\text{eff}}^{\ell\ell\gamma}+\sum_{f,i,a,b}\Big(C_{fiab}^{V\,LL}O_{fiab}^{V\,LL}+C_{fiab}^{V\,LR}O_{fiab}^{V\,LR}+C_{fiab}^{S\,LL}O_{fiab}^{S\,LL}+L\leftrightarrow R\,\Big)\,,
\label{eq:4l_effective_hamiltonian}
\end{align}
with the effective operators
\begin{align}
\begin{aligned}
O_{fiab}^{V\,LL}&=\big[\bar{\ell}_{f}\gamma^{\mu}P_{L}\ell_{i}\big]\big[\bar{\ell}_{a}\gamma_{\mu}P_{L}\ell_{b}\big]\,,\\
O_{fiab}^{V\,LR}&=\big[\bar{\ell}_{f}\gamma^{\mu}P_{L}\ell_{i}\big]\big[\bar{\ell}_{a}\gamma_{\mu}P_{R}\ell_{b}\big]\,,\\
O_{fiab}^{S\,LL}&=\big[\bar{\ell}_{f}P_{L}\ell_{i}\big]\big[\bar{\ell}_{a}P_{L}\ell_{b}\big]\,.
\end{aligned}
\label{eq:4l_effective_operators}
\end{align}
Note that we sum over all flavor indices. Therefore, all other operators can be reduced to the ones in \eqref{eq:4l_effective_hamiltonian}, using Fierz identities. As an advantage, we do not need to distinguish between decays involving the same or different flavors.
\smallskip

\begin{figure}[t]
	\centering
	\begin{overpic}[scale=.47,,tics=10]
		{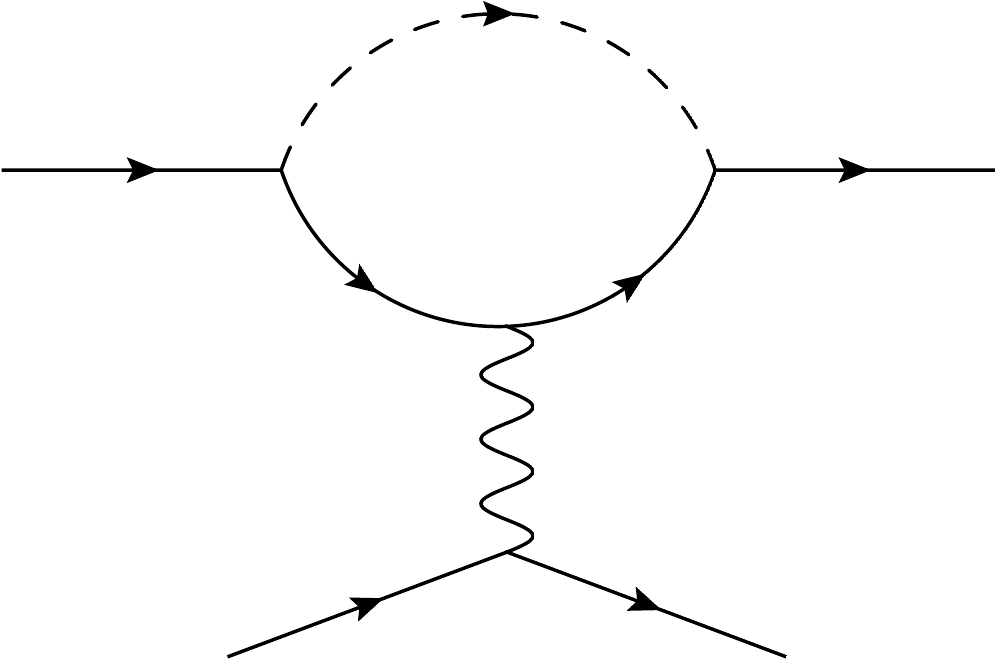}
		\put(5,52){$\ell_{i}$}
		\put(90,53){$\ell_{f}$}
		\put(25,6){$\ell_{b}$}
		\put(70,6){$\ell_{b}$}
		\put(47,37){$q^{(c)}$}
		\put(56,19){$Z,\gamma$}
		\put(45,55){$\Phi_{a}^{Q}$}
	\end{overpic}
\hfill
	\begin{overpic}[scale=.47,,tics=10]
		{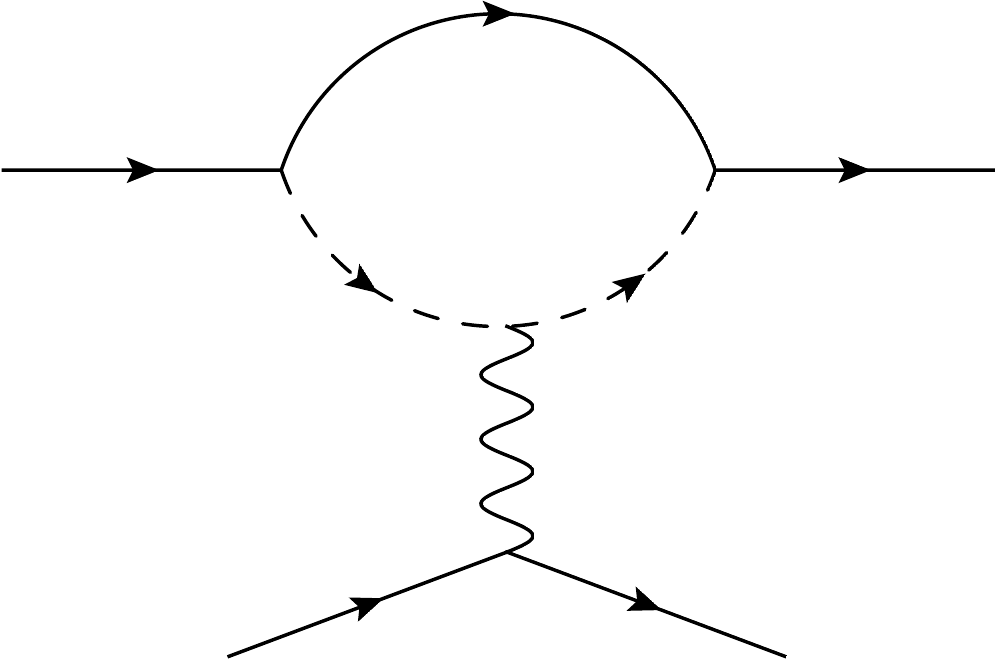}
		\put(5,52){$\ell_{i}$}
		\put(90,53){$\ell_{f}$}
		\put(25,6){$\ell_{b}$}
		\put(70,6){$\ell_{b}$}
		\put(46,37.5){$\Phi_{a}^{Q}$}
		\put(56,19){$Z,\gamma$}
		\put(47,54.5){$q^{(c)}$} 
	\end{overpic}
\hfill
	\begin{overpic}[scale=.50,,tics=10]
		{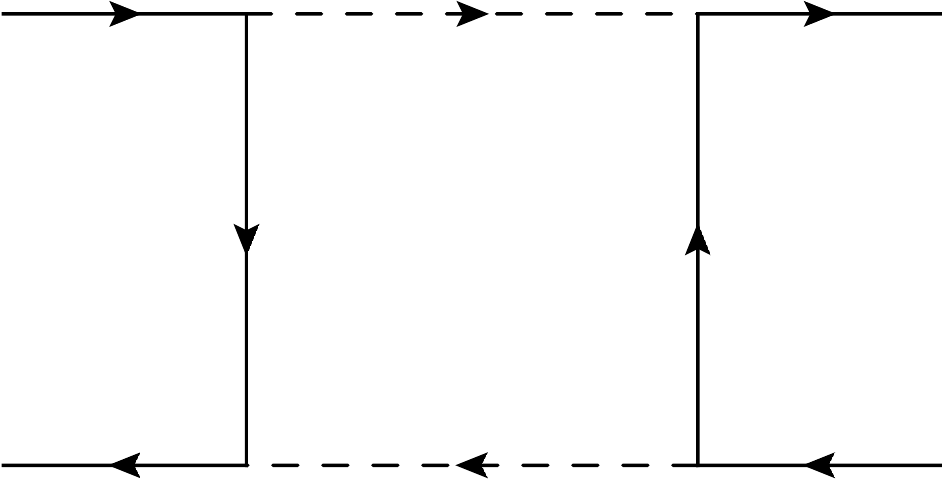}
		\put(5,52){$\ell_{i}$}
		\put(5,5){$\ell_{f}$}
		\put(90,52){$\ell_{a}$}
		\put(90,5){$\ell_{b}$}
		\put(45,39){$\Phi_{a}^{Q}$}
		\put(45,6){$\Phi_{a}^{Q}$}
		\put(79,24){$q^{(c)}$}
		\put(12,24){$q^{(c)}$} 
	\end{overpic}
	\caption{Feynman diagrams giving rise to $\ell_{f}\ell_{i}\ell_{a}\ell_{b}$ amplitudes. Left and center: Penguin diagrams with off-shell $Z$ boson or photon exchange. Right: Box diagram involving two LQs.}
	\label{fig:l_lll_decays}
\end{figure}
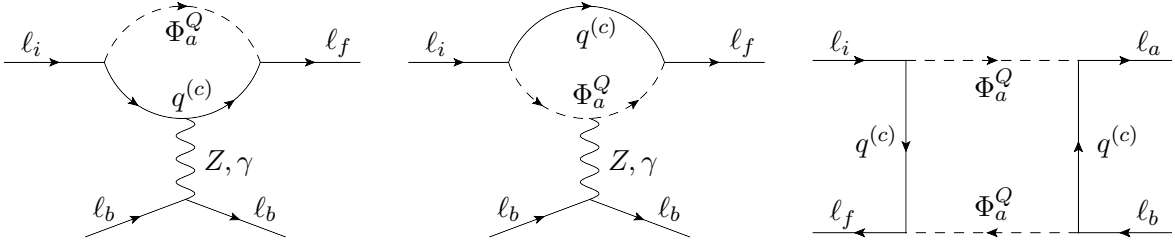

There are two types of diagrams which give a contribution to these operators: penguins and boxes, see Fig~\ref{fig:l_lll_decays}. Starting with the photon penguin, we have
\begin{align}
\begin{aligned}
C_{fiab}^{V\,LL}&=-\pi\alpha\Big(\Xi_{fi}^{L}\Xi_{ab}^{L}+\Xi_{fb}^{L}\Xi_{ai}^{L}\Big)\,,\\
C_{fiab}^{V\,LR}&=-2\pi\alpha\Xi_{fi}^{L}\Xi_{ab}^{R}\,,
\end{aligned}
\end{align}
with
\begin{align}
\Xi_{fi}^{L(R)}=\delta_{fi}+\widehat{\Xi}_{fi}^{L(R)}\,.
\end{align}
The $Z$ boson gives an analogous contribution
\begin{align}
\begin{aligned}
C_{fiab}^{V\,LL}&=\sqrt{2}G_{F}\big(\Lambda_{\ell_{f}\ell_{i}}^{L}(0)\Lambda^{L}_{\ell_{a}\ell_{b}}(0)+\Lambda^{L}_{\ell_{f}\ell_{b}}(0)\Lambda^{L}_{\ell_{a}\ell_{i}}(0)\big)\,,\\
C_{fiab}^{V\,LR}&=\frac{4G_{F}}{\sqrt{2}}\Lambda_{\ell_{f}\ell_{i}}^{L}(0)\Lambda_{\ell_{a}\ell_{b}}^{R}(0)\,.
\end{aligned}
\end{align}
The coefficients $C_{fiab}^{V\,RL}$ and $C_{fiab}^{V\,RR}$ are obtained in a straightforward way by simply exchanging $L\leftrightarrow R$.
\smallskip

The box diagrams generate the following contributions
\begin{subequations}
\begin{align}
C_{fiab}^{V\,LL}&=\frac{N_c}{256\pi^2}\bigg[\frac{\lambda_{ji}^{1L}\lambda_{kb}^{1L}\big(\lambda_{jf}^{1L*}\lambda_{ka}^{1L*}+\lambda_{kf}^{1L*}\lambda_{ja}^{1L*}\big)}{m_{1}^2}+5\frac{\lambda_{ji}^{3}\lambda_{kb}^{3}\big(\lambda_{jf}^{3*}\lambda_{ka}^{3*}+\lambda_{kf}^{3*}\lambda_{ja}^{3*}\big)}{m_{3}^2}\nonumber\\
&\quad+\frac{\lambda_{ji}^{3}\lambda_{kb}^{1L}\big(\lambda_{jf}^{1L*}\lambda_{ka}^{3*}+\lambda_{kf}^{3*}\lambda_{ja}^{1L*}\big)+\lambda_{ji}^{1L}\lambda_{kb}^{3}\big(\lambda_{jf}^{3*}\lambda_{ka}^{1L*}+\lambda_{kf}^{1L*}\lambda_{ja}^{3*}\big)}{m_{3}^2}\mathcal{H}_{1}\Big(\frac{m_{1}^2}{m_{3}^2}\Big)\nonumber\\
&\quad+\frac{\lambda_{ji}^{2RL}\lambda_{kb}^{2RL}\big(\lambda_{jf}^{2RL*}\lambda_{ka}^{2RL*}+\lambda_{kf}^{2RL*}\lambda_{ja}^{2RL*}\big)}{m_{2}^2}+\frac{\tilde{\lambda}_{ji}^{2}\tilde{\lambda}_{kb}^{2}\big(\tilde{\lambda}_{jf}^{2*}\tilde{\lambda}_{ka}^{2*}+\tilde{\lambda}_{kf}^{2*}\tilde{\lambda}_{ja}^{2*}\big)}{\tilde{m}_{2}^2}\bigg] \,, \\
C_{fiab}^{V\,LR}&=\frac{N_c}{128\pi^2}\bigg[\frac{\lambda_{jf}^{1L*}\lambda_{ji}^{1L}\lambda_{ka}^{1R*}\lambda_{kb}^{1R}}{m_{1}^2}+\frac{\lambda_{jf}^{2RL*}\lambda_{ji}^{2RL}\lambda_{ka}^{2LR*}\lambda_{kb}^{2LR}}{m_{2}^2}\bigg] \,,\\
C_{fiab}^{V\,RR}&=\frac{N_c}{256\pi^2}\bigg[\frac{\lambda_{ji}^{1R}\lambda_{kb}^{1R}\big(\lambda_{jf}^{1R*}\lambda_{ka}^{1R*}+\lambda_{kf}^{1R*}\lambda_{ja}^{1R*}\big)}{m_{1}^2}\nonumber\\
&\quad+2\frac{\lambda_{ji}^{2LR}\lambda_{kb}^{2LR}\big(\lambda_{jf}^{2LR*}\lambda_{ka}^{2LR*}+\lambda_{kf}^{2LR*}\lambda_{ja}^{2LR*}\big)}{m_{2}^2}\!+\frac{\tilde{\lambda}_{ji}^{1}\tilde{\lambda}_{kb}^{1}\big(\tilde{\lambda}_{jf}^{1*}\tilde{\lambda}_{ka}^{1*}+\tilde{\lambda}_{kf}^{1*}\tilde{\lambda}_{ja}^{1*}\big)}{\tilde{m}_{1}^2}\bigg] \,,
\end{align}
\label{eq:WC_l->3l}
\end{subequations}
where the loop function $\mathcal{H}_1$ is again given in Appendix~\ref{app:loop_functions}. The indices $j$ and $k$ run from 1 to 3. Note that we only consider the leading effects in $v/m$. In scenarios where the $\lambda$-couplings are smaller than the gauge couplings ($e\approx 0.3$ and $g_{2}\approx 0.6$), the box contributions are typically less important than the gauge boson penguins.
\medskip

\begin{boldmath}
\subsection{$2\ell 2\nu$}
\end{boldmath}

For these fields we use the effective Hamiltonian
\begin{align}
\mathcal{H}_{\text{eff}}^{2\ell 2\nu}=D_{\ell_{f}\ell_{i}}^{L,ab}O_{\ell_{f}\ell_{i}}^{L,ab}+D_{\ell_{f}\ell_{i}}^{R,ab}O_{\ell_{f}\ell_{i}}^{R,ab}
\end{align}
with
\begin{align}
O_{\ell_{f}\ell_{i}}^{L(R),ab}=\big[\bar{\ell}_{f}\gamma_{\mu}P_{L(R)}\ell_{i}\big]\big[\bar{\nu}_{a}\gamma^{\mu}P_{L}\nu_{b}\big]\,.
\end{align}
There are three types of contributions: $Z$ penguins, $W$ penguins and boxes. The $Z$ boson yields
\begin{align}
D_{\ell_{f}\ell_{i}}^{L,ab}=\frac{8G_{F}}{\sqrt{2}}\Lambda_{\ell_{f}\ell_{i}}^{L}(0)\Theta_{\nu_{a}\nu_{b}}(0)\,,&&
D_{\ell_{f}\ell_{i}}^{R,ab}=\frac{8G_{F}}{\sqrt{2}}\Lambda_{\ell_{f}\ell_{i}}^{R}(0)\Theta_{\nu_{a}\nu_{b}}(0)\,,
\end{align}
while we have for the $W$ boson
\begin{align}
D_{\ell_{f}\ell_{i}}^{L,ab}=\frac{4G_{F}}{\sqrt{2}}\Lambda_{\ell_{i}\nu_{a}}^{W*}(0)\Lambda_{\ell_{f}\nu_{b}}^{W}(0)\,.
\end{align}
The box diagrams yield
\begin{subequations}
\begin{align}
D_{\ell_{f}\ell_{i}}^{L,ab}&=\frac{N_c}{64\pi^2}\bigg[\frac{\lambda_{jf}^{1L*}\lambda_{ji}^{1L}\lambda_{ka}^{1L*}\lambda_{kb}^{1L}}{m_{1}^2}+\frac{\lambda_{jf}^{3*}\lambda_{ji}^{3}\lambda_{ka}^{3*}\lambda_{kb}^{3}}{m_{3}^2}\nonumber\\
&\quad-\frac{\lambda_{jf}^{3*}\lambda_{ji}^{1L}\lambda_{ka}^{1L*}\lambda_{kb}^{3}+\lambda_{jf}^{1L*}\lambda_{ji}^{3}\lambda_{ka}^{3*}\lambda_{kb}^{1L}}{m_{3}^2}\mathcal{H}_{1}\Big(\frac{m_{1}^2}{m_{3}^2}\Big)\bigg]\,,\\
D_{\ell_{f}\ell_{i}}^{R,ab}&=\frac{N_c}{64\pi^2}\bigg[\frac{\lambda_{jf}^{1R*}\lambda_{ji}^{1R}\lambda_{ka}^{1L*}\lambda_{kb}^{1L}}{m_{1}^2}+\frac{\lambda_{jf}^{2LR*}\lambda_{ji}^{2LR}\lambda_{ka}^{2RL*}\lambda_{kb}^{2RL}}{m_{2}^2}\bigg]\,.
\end{align}
\end{subequations}
Again the indices $j$ and $k$ run from 1 to 3 and we only considered the leading order LQ effects in $v/m_{\rm{LQ}}$.
\medskip

\begin{figure}
	\renewcommand\tabularxcolumn[1]{m{#1}}
	\setkeys{Gin}{width=\linewidth, 
		height=.6\linewidth,
		keepaspectratio}
	\centering
	\resizebox{0.96\textwidth}{!}{
		\begin{tabularx}{1.05\linewidth}{@{} XX @{}}
			\qquad\qquad
			\subfloat{\includegraphics[width=\linewidth]{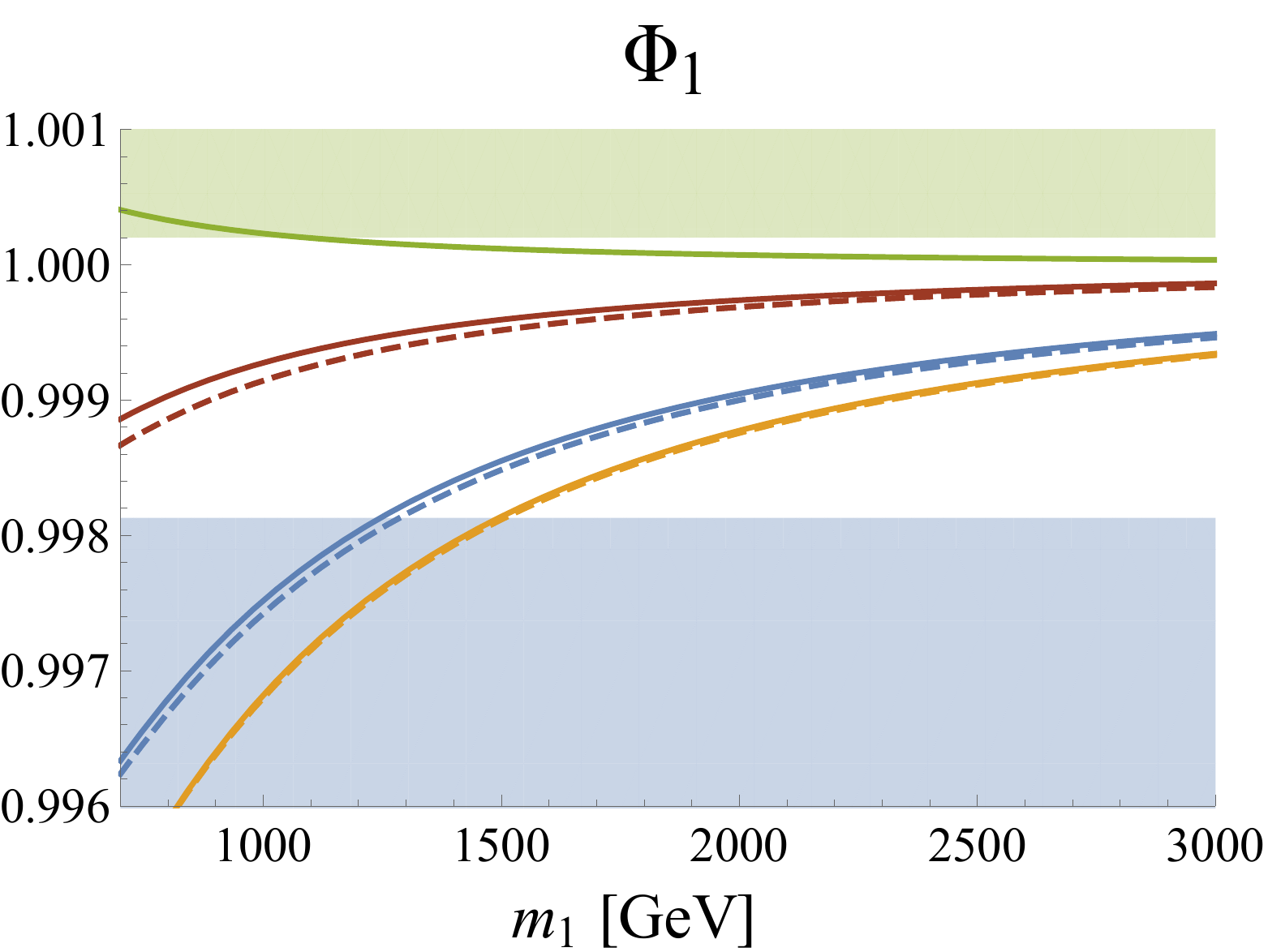}}
			
			\vspace{0.1cm}
			\qquad\qquad
			\subfloat{\includegraphics[width=0.925\linewidth]{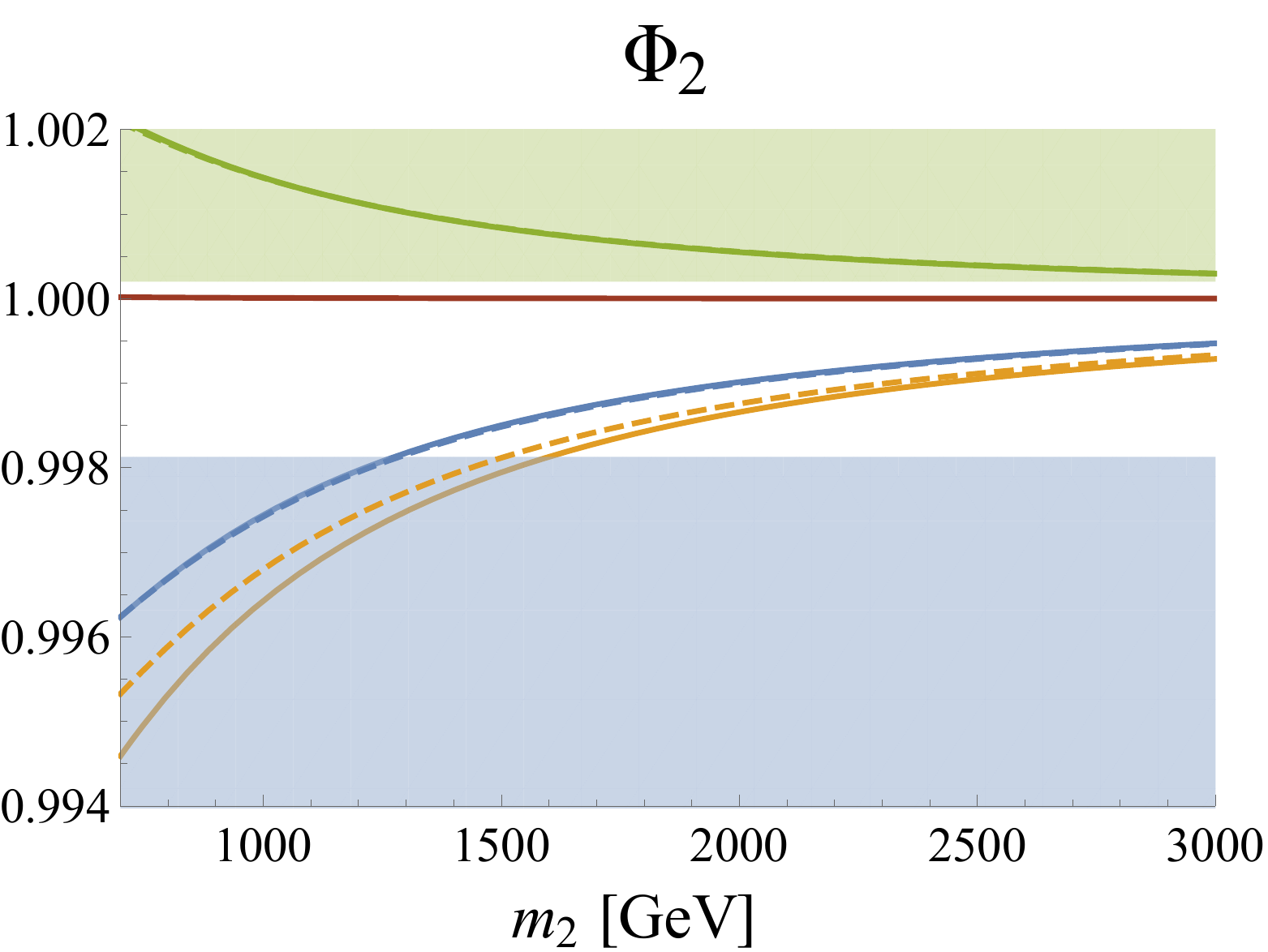}}
			
			\vspace{0.1cm}
			\qquad\qquad
			\subfloat{\includegraphics[width=\linewidth]{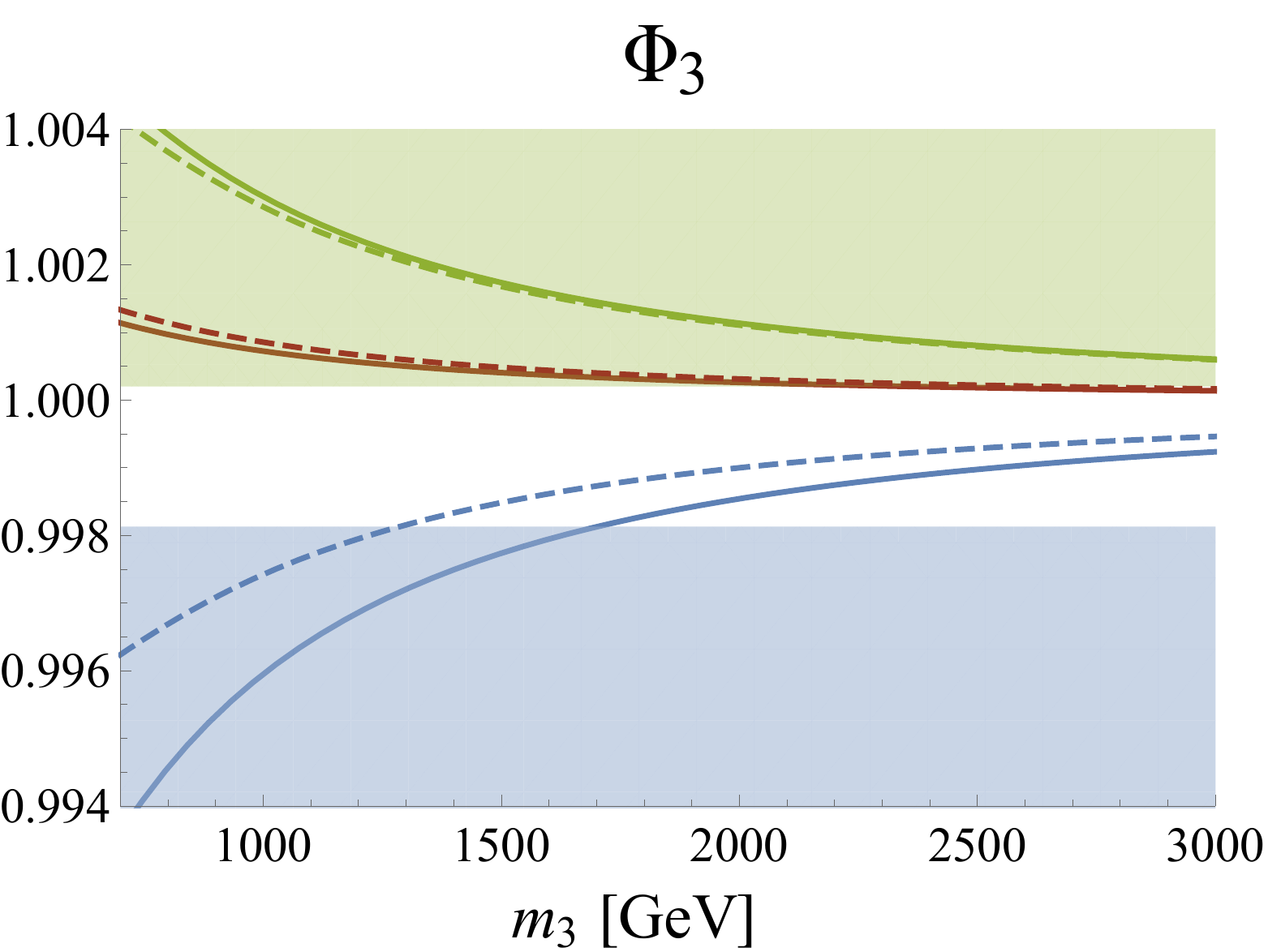}}
			
			&
			
			\qquad
			\subfloat{\includegraphics[width=\linewidth]{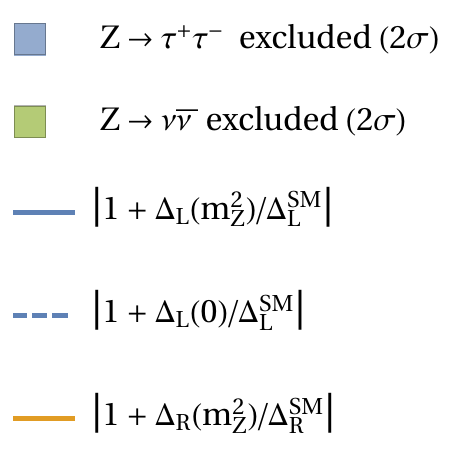}}
			
			\vspace{-4mm}
			\qquad
			\subfloat{\includegraphics[width=\linewidth]{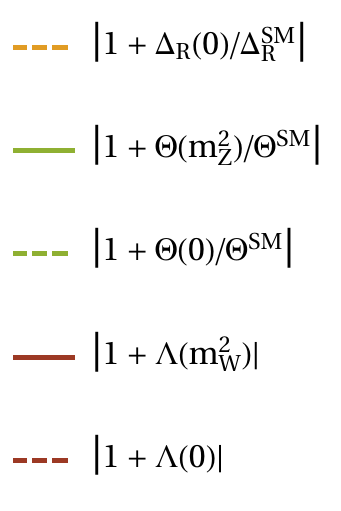}}
			
		\end{tabularx}
	}
	\caption{LQ effects in $Z\ell\ell$ , $Z\nu\nu$ and $W\ell\nu$ couplings for the scalar LQ representations which give rise to $m_{t}^2$ effects ($\Phi_1$, $\Phi_2$ and $\Phi_3$) as a function of the LQ mass. We neglected LQ mixing and considered only the couplings of third generation quarks to a single lepton flavor with unit strength, i.e. $\lambda_{3\ell}=1$. Here, $\Delta_{L,R}$, $\Theta$ and $\Lambda$ stand for the corrections in $Z\ell\ell$, $Z\nu\nu$ and $W\ell\nu$ couplings, respectively (see Sec.~\ref{sec:Zll}). The solid (dashed) lines refer to the couplings entering on-shell decays (effective couplings at $q^2=0$). The green region is excluded by LEP data~\cite{ALEPH:2005ab} from $Z\to \nu\bar{\nu}$ decays. The blue region is excluded by $Z\to\tau^{+}\tau^{-}$ {which is more constraining than $Z\to\mu^{+}\mu^{-}$ (not shown explicitly). Note that we also do not show $Z\to e^{+}e^{-}$ exclusions here for the sake of clarity since couplings to electrons are usually much smaller in setups motivated by the $B$ anomalies, leading to suppressed effects.}}
	\label{fig:eff_couplings}
\end{figure}

\section{Phenomenology}
\label{pheno}
Let us now study the phenomenology of scalar LQs in leptonic processes. Due to the large number of observables and the many free parameters, we will choose some exemplary processes of special interest and use simplifying assumptions for the couplings in order to show the effects and the possible correlations between observables. In particular, we will consider:
\begin{itemize}
	\item EW gauge-boson couplings to leptons: the effects of scalar LQs in (effective) off-shell $Z\ell\ell$, $Z\nu\nu$ and $W\ell\nu$ couplings and the associated gauge-boson decays.
	\item Muonic observables: correlations between the AMM of the muon, $Z\to\ell^{+}\ell^{-}$, effective $W\mu\nu$ couplings and $h\to\mu^{+}\mu^{-}$.
	\item Charged lepton flavor violation: correlations between $\tau\to\mu\gamma$, $Z\to\tau\mu$ and $\tau\to 3\mu$ as well as the analogues in $\mu\to e$ transitions.
\end{itemize} 
\medskip

\begin{figure}
	\centering
	\includegraphics[width=0.8\textwidth]{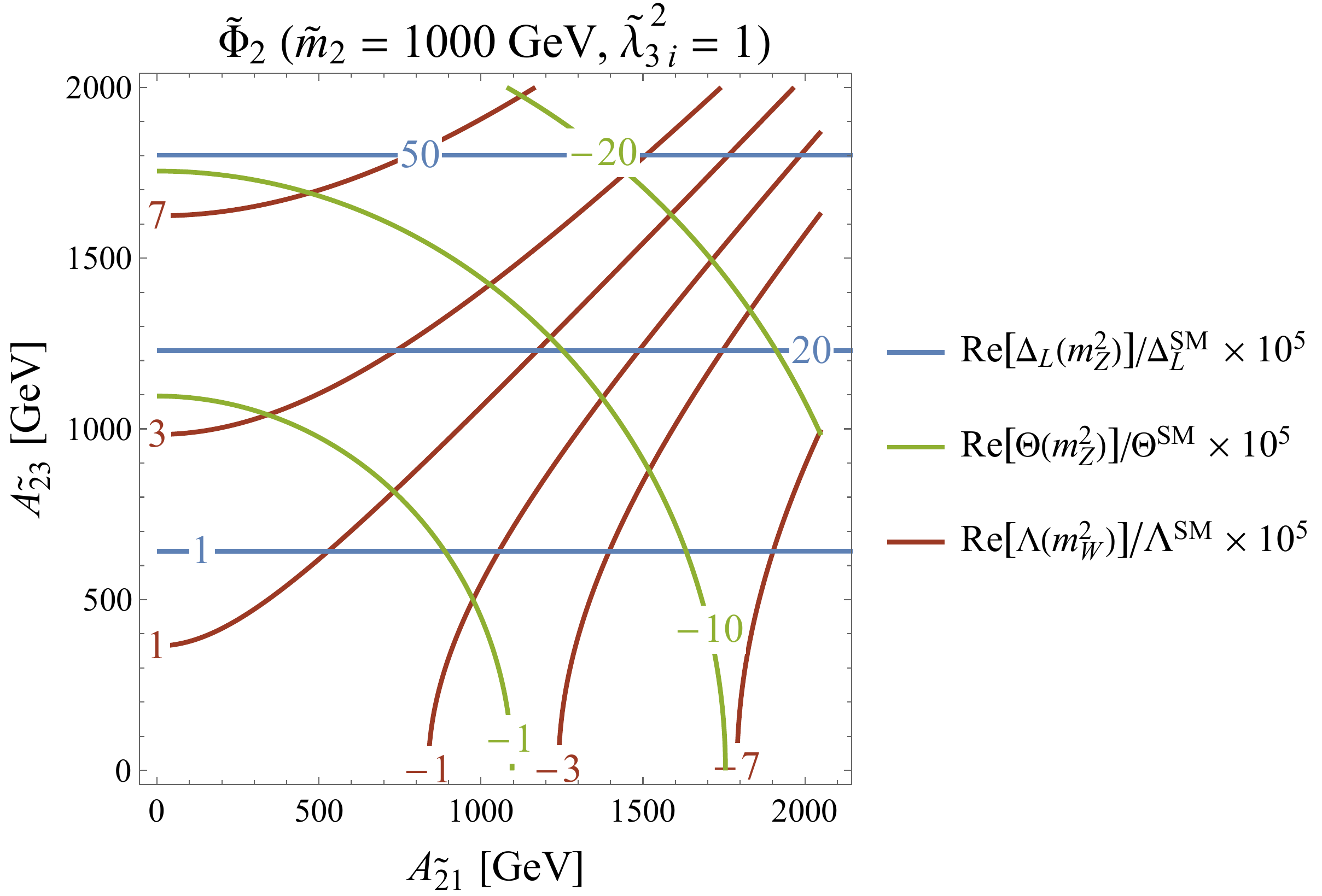}
	\caption{Modified $Z\ell\ell$, $Z\nu\nu$ and $W\ell\nu$ couplings in the  $A_{\tilde{2}1}$-$A_{\tilde{2}3}$ plane (in units of GeV) for $\tilde{m}_{2}=m_1=m_3=1\,$TeV and $|\tilde{\lambda}_{3\ell}^{2}|=1$.}
	\label{fig:doublet_mixing_2}
\end{figure}

\begin{boldmath}
	\subsection{Electroweak Gauge-Boson Couplings to Leptons: $Z\ell\ell$, $Z\nu\nu$ and $W\ell\nu$}
\end{boldmath}

We start our phenomenological analysis by considering the effects of scalar LQs in $Z\ell\ell$, $Z\nu\nu$ and $W\ell\nu$ effective couplings (at $q^2=0$) and the associated gauge boson decays (at $q^2=m_Z^2\,,m_W^2$), calculated in Sec.~\ref{sec:Zll}. While among $Z\to\ell^+\ell^-$ decays NP effects are strongly bounded by LEP~\cite{ALEPH:2005ab} measurements, the effective $W\ell\nu$ couplings are best constrained by low-energy observables, testing LFU of the charged current (see Ref.~\cite{Pich:2013lsa} for an overview).
\smallskip

We first focus on the LQ representations which generate an $m_{t}^2/m^{2}_{\rm LQ}$ effect in EW gauge-boson couplings to leptons, i.e. $\Phi_{1}$, $\Phi_{2}$ and $\Phi_{3}$. In the absence of LQ mixing, we can expect this effect to be dominant and couplings to third generation quarks are well motivated by the flavor anomalies. Note that we nonetheless included the \mbox{$q^2=\{m_Z^2,m_W^2\}$} terms which, due to $SU(2)_{L}$ invariance, can also arise from bottom loops for some of the representations shown in Fig.~\ref{fig:eff_couplings}. In order to keep the number of free parameters small, we did not include mixing among the LQs and assumed that only couplings to one lepton flavor $\ell=e,\mu,\tau$ at a time exist. This avoids limits from charged lepton flavor violating observables, which we consider later in this article. Furthermore, we normalized the LQ effect to the respective SM coupling and the LQ-quark-lepton coupling to one ( i.e. $\lambda_{3\ell}=1$) while all other couplings are zero. Note that the effect in Fig.~\ref{fig:eff_couplings}, given for couplings of unit strength, are consistent with $Z\to\ell^+\ell^-$ bounds for masses around $1.5\,$TeV or more. Furthermore, $Z\to\nu\bar \nu$ is constrained by the number of neutrino families
\begin{align}
N_{\nu}=\sum_{f,i}\bigg|\delta_{fi}+\frac{\sum_{\Phi}\Theta_{fi}^{\Phi}\big(q^2\big)+\tilde{\Theta}_{fi}}{\Theta_{\rm{SM}}(m_Z^2)}\bigg|^2\,,
\label{eq:number_neutrinos}
\end{align}
where the experimental value lies at~\cite{ALEPH:2005ab}
\begin{align}
N_\nu = 2.9840 \pm 0.0082 \,,
\label{eq:Nnu_measurement}
\end{align}
while the LQ effect is predicted to be constructive. Future colliders are expected to reach a 20 times better precision~\cite{Abada:2019lih}.
\smallskip

Let us now turn to the case of non-vanishing LQ couplings to the SM Higgs. We study as an example the scalar doublet $\tilde{\Phi}_{2}$ which couples only down-type quarks to leptons such that the $v^2/\tilde{m}_{2}^2$ effects from the mixing with $\Phi_{1}$ (generated by $A_{\tilde{2}1}$) and/or $\Phi_{3}$ (generated by $A_{\tilde{2}3}$) are expected to be dominant compared to the $m_Z^2/\tilde{m}_{2}^2$ effects. In Fig.~\ref{fig:doublet_mixing_2} we present the impact of LQs on on-shell $Z$ and $W$ couplings. Again, we set $\tilde{\lambda}^{2}_{3\ell}=1$ and we assume $\tilde{m}_{2}=m_{1}=m_{3}=1\,\text{TeV}$, which is compatible with current LHC limits~\cite{Sirunyan:2018ruf,Aaboud:2019jcc,Aaboud:2019bye}. Note that a non-zero $A_{\tilde{2}1}$ yields a destructive effect in $Z\ell\ell$ and $W\ell\nu$ couplings while the terms with $A_{\tilde{2}3}$ are constructive. 
\medskip

\begin{figure}
	\centering
	\includegraphics[width=0.59\textwidth]{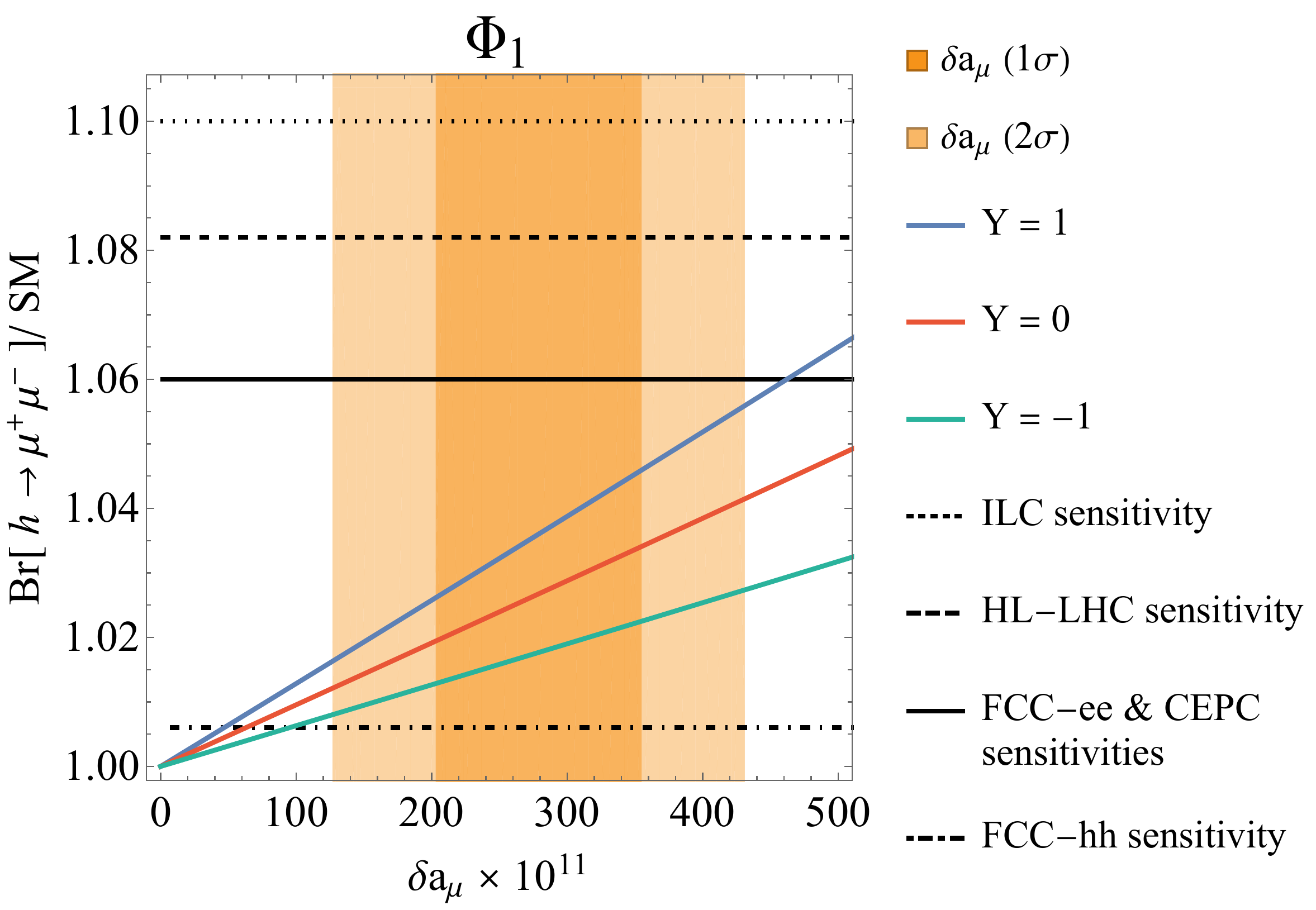}
	\includegraphics[width=0.395\textwidth]{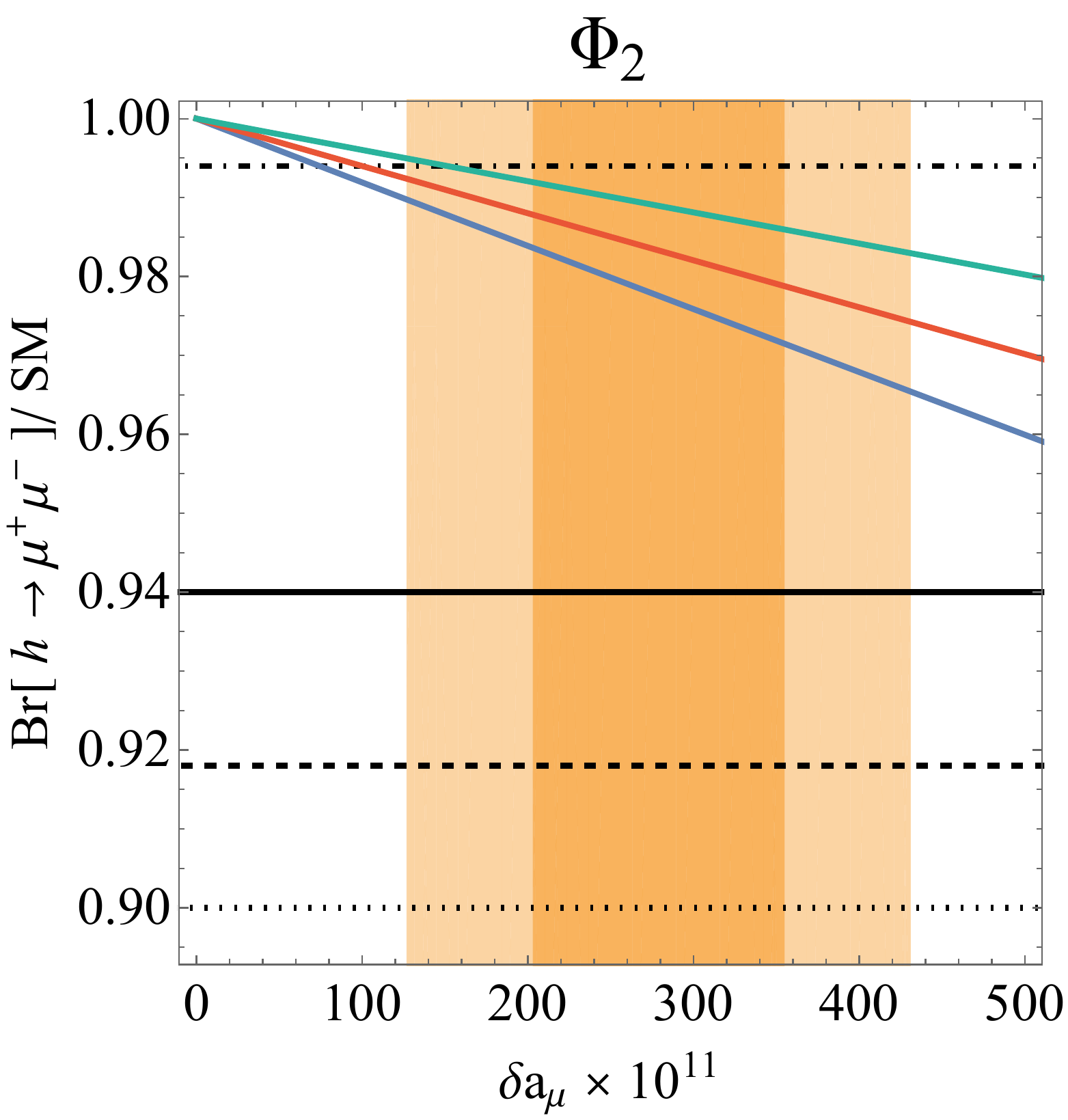}
	\caption{Correlations between ${\rm Br}[h\to\mu^+\mu^-]$, normalized to its SM value, and the NP contribution to the AMM of the muon ($\delta a_\mu$) for scenario $\Phi_1$ (left) and $\Phi_2$ (right) with \mbox{$m_{1,2}=1.5\,$TeV}. The predictions for different values of the LQ couplings to the Higgs are shown, where for $\Phi_1$ $Y=Y_1$ while for $\Phi_2$ $Y=Y_2+Y_{22}$. Even though the current ATLAS and CMS results are not yet constraining this model, sizeable effects are predicted, which can be tested at future colliders. Furthermore, $\Phi_1$ yields a constructive effect in $h\to\mu^+\mu^-$ while the one of $\Phi_2$ is destructive such that they can be clearly distinguished with increasing experimental precision.}
	\label{S1S2}
\end{figure}

\begin{boldmath}
	\subsection{Correlating the AMM of the Muon with $Z\to\ell^{+}\ell^{-}$ and $h\to\mu^{+}\mu^{-}$}
\end{boldmath}

In this sub-section, we focus on possible LQ explanations of the long-standing anomaly in the AMM of the muon. The discrepancy between its measurement~\cite{Bennett:2006fi} and the SM prediction~\cite{Aoyama:2020ynm}\footnote{This result is based on Refs.~\cite{Aoyama:2012wk,Aoyama:2019ryr,Czarnecki:2002nt,Gnendiger:2013pva,Davier:2017zfy,Keshavarzi:2018mgv,Colangelo:2018mtw,Hoferichter:2019gzf,Davier:2019can,Keshavarzi:2019abf,Kurz:2014wya,Melnikov:2003xd,Masjuan:2017tvw,Colangelo:2017fiz,Hoferichter:2018kwz,Gerardin:2019vio,Bijnens:2019ghy,Colangelo:2019uex,Blum:2019ugy,Colangelo:2014qya}. The recent lattice result of the Budapest-Marseille-Wuppertal collaboration (BMWc) for the hadronic vacuum polarization (HVP)~\cite{Borsanyi:2020mff} on the other hand is not included. This result would render the SM prediction of $a_\mu$ compatible with experiment. However, the BMWc results are in tension with the HVP determined from $e^+e^-\to$ hadrons data~\cite{Davier:2017zfy,Keshavarzi:2018mgv,Colangelo:2018mtw,Hoferichter:2019gzf,Davier:2019can,Keshavarzi:2019abf}. Furthermore, the HVP also enters the global EW fit~\cite{Passera:2008jk}, whose (indirect) determination is below the BMWc result~\cite{Haller:2018nnx}. Therefore, the BMWc determination of the HVP would increase the tension in EW fits~\cite{Crivellin:2020zul,Keshavarzi:2020bfy} and we opted for using the community consensus of Ref.~\cite{Aoyama:2020ynm}.} amounts to
\begin{align}
\delta a_\mu = (279 \pm 76) \times 10^{-11} \ ,
\label{eq:AMM}
\end{align}
corresponding to a $3.7\sigma$ tension. Note that this tension is quite large, i.e. of the order of the EW contribution of the SM. Since LQs are colored, the LHC bounds rule out masses significantly below 1 TeV such that an enhancement in $a_\mu$ is needed to compensate for the mass suppression. In fact, there are LQ representations that are able to generate $m_{t}/m_{\mu}$ enhanced contributions, see Eq.~\eqref{eq:CL_AMM}. These NP effects enter the AMM of the  muon as
\begin{align}
a_{\mu}=\frac{m_{\mu}}{4\pi^2}\text{Re} \big[C^{R}_{\mu\mu}\big]\,,
\end{align} 
with the Wilson coefficient defined in Eq.\eqref{eq:Heff_llgamma}.
\smallskip

First of all, we can expect a direct correlation with $h\to\mu^{+}\mu^{-}$~\cite{Crivellin:2020tsz} since both processes are chirality changing and therefore involve the same couplings of LQs to fermions\footnote{Similar results for $\tau\to\mu\gamma$ were obtained in Refs.~\cite{Dorsner:2015mja,Baek:2015mea, Cheung:2015yga}.}. We can express the NP effect in terms of $\Upsilon_{L}^{\Phi}$ and $\tilde{\Upsilon}_{L}$, defined in \eqref{eq:amplitude_hll}, as
\begin{align}
\dfrac{\text{Br}\big[h\to\mu^{+}\mu^{-}\big]\phantom{_{\text{SM}}}}{\text{Br}\big[h\to\mu^{+}\mu^{-}\big]_{\text{SM}}}=\Big|1+\sum_{\Phi}\Upsilon_{L,\mu\mu}^{\Phi}+\tilde{\Upsilon}_{L,\mu\mu}\Big|^2\,.
\end{align}
The resulting correlations are shown in Fig.~\ref{S1S2} for $\Phi_1$ and $\Phi_2$. Note that even though the current CMS and ATLAS measurements~\cite{Aad:2020xfq,CMS:2020eni} are not able to constrain these models yet, a FCC-hh~\cite{Benedikt:2018csr} can test them.
\smallskip

\begin{figure}
	\centering
	\includegraphics[width=0.59\textwidth]{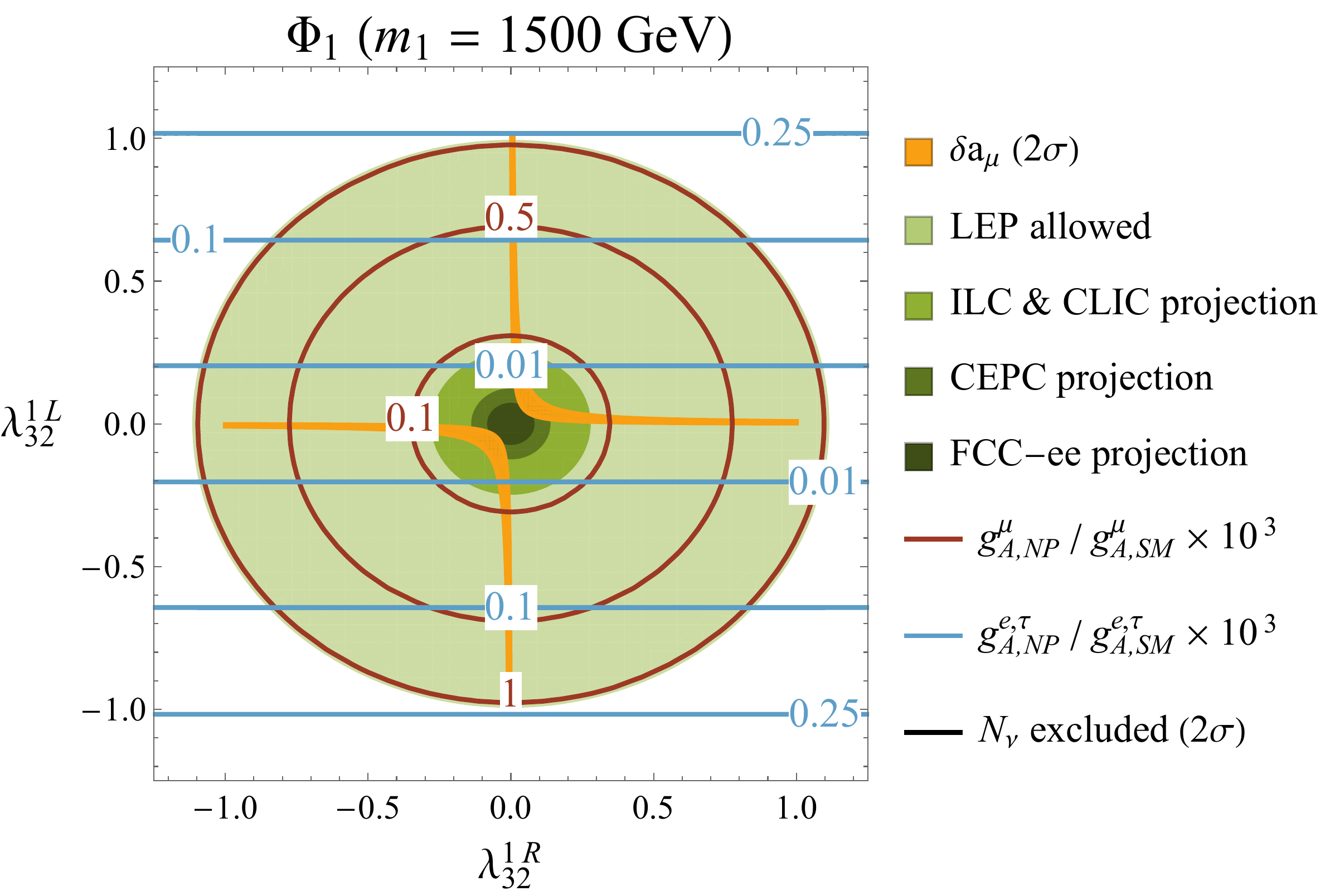}
	\includegraphics[width=0.395\textwidth]{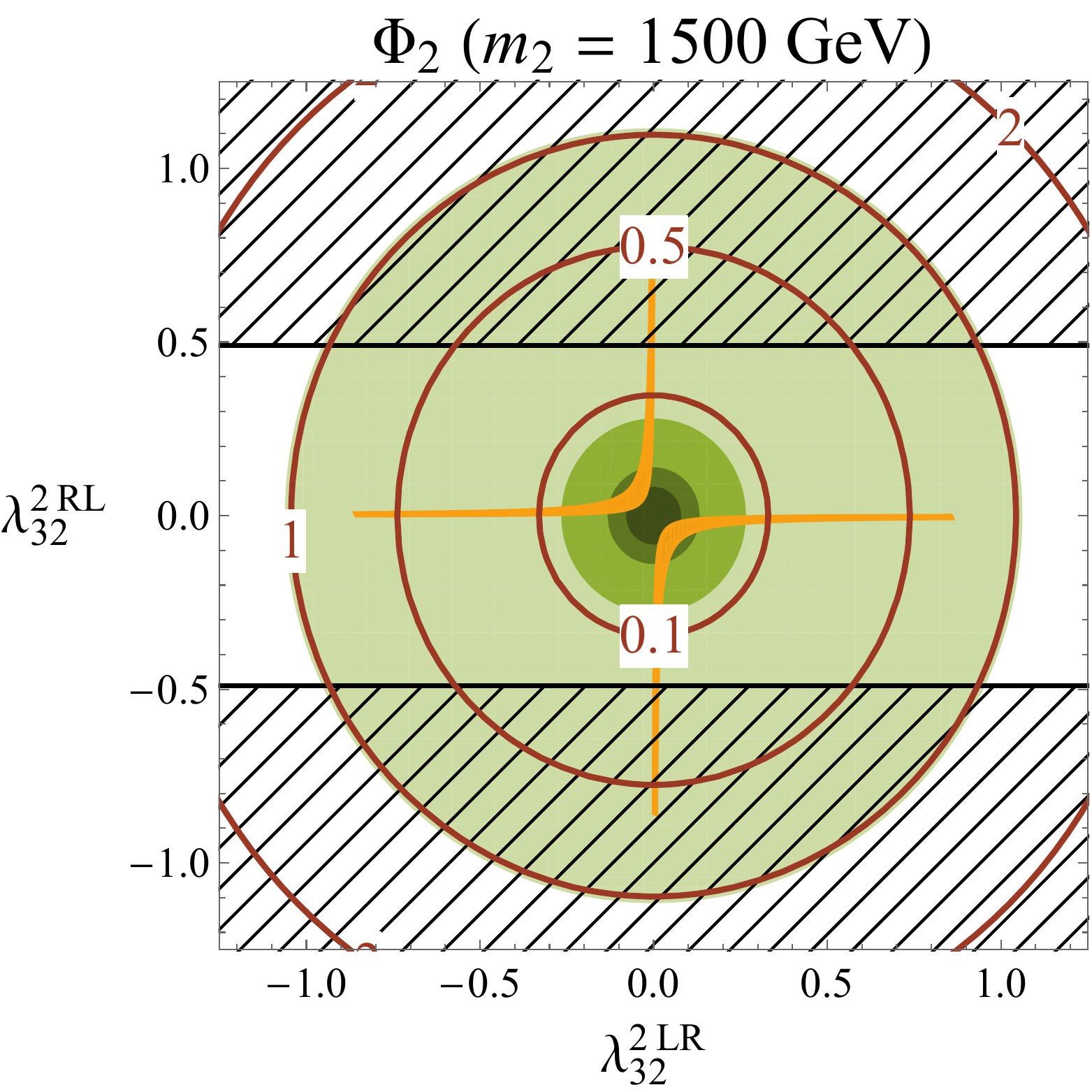}
	\caption{Allowed parameter space by LEP \cite{ALEPH:2005ab} (light green) for the couplings to left- and right-handed muons. In addition, we give the expected sensitivities of future collider experiments, see Tab.~\ref{tab:experimental_limits_Z}. The finite renormalization of $g_{2}$, induced by the effect in the Fermi constant, yields a LFU effect which is depicted by the blue lines in the plot on the left.}
	\label{fig:Zmumu_AMM}
\end{figure}

The LQ interactions with top quarks and muons also generate effects in $Z\mu\mu$ couplings. Therefore, let us as a next step consider the correlations of $a_\mu$ with $Z\to\ell^+\ell^-$ where we refine the analysis of Ref.~\cite{ColuccioLeskow:2016dox} by including the indirect effect, originating from the finite renormalization of the very precisely measured Fermi constant \cite{Zyla:2020zbs}
\begin{align}
G_{F}=1.166\,378\,7(6)\times 10^{-5}\,\text{GeV}^{-2} \,,
\end{align}
which can be expressed in terms of the SM parameters
\begin{align}
G_{F}=\frac{\sqrt{2}g_{2}^2}{8m_{W}^2}\,.
\end{align}
Since $m_W$ itself is measured in $W$ decays, $g_2$ can be determined once $G_{F}$ is measured via the muon lifetime. However, also NP contributions enter such that
\begin{align}
G_{F}\rightarrow G_{F}\big(1+\Lambda_{\mu\nu_{\mu}}^{W*}(0)\big)\big(1+\Lambda_{e\nu_{e}}^{W}(0)\big)\,,
\end{align}
resulting in a redefinition of $g_2$.

The $SU(2)_{L}$ singlet $\Phi_1$ with non-zero real couplings $\lambda_{32}^{1L}$ and $\lambda_{32}^{1R}$ affects $Z\mu\mu$ as well as $W\mu\nu_{\mu}$, while the effect on $Z\nu{\nu}$ is very small, see Fig.~\ref{fig:Zmumu_AMM}. The modified $W$ coupling by $\lambda_{32}^{1L}$ then yields a finite, LFU renormalization of $g_{2}$. This has been included in our analysis depicted in Fig.~\ref{fig:Zmumu_AMM}, leading to the allowed, green region deviating slightly from a circled shape.
\smallskip

The $SU(2)_{L}$ doublet $\Phi_2$ with non-zero couplings $\lambda_{32}^{2LR}$ and $\lambda_{32}^{2RL}$ only yields a negligible contribution to $W\mu\nu_{\mu}$. However, there is an $m_{t}$ effect in $Z\to\nu\bar\nu$, affecting $N_{\nu}$, which has been precisely measured, see Eq.~\eqref{eq:Nnu_measurement}. This then constrains $\lambda_{32}^{2RL}$ as we show in the plot in the right-hand side of Fig.~\ref{fig:Zmumu_AMM}. We additionally show in Fig.~\ref{fig:Zmumu_AMM} the expected sensitivities of future experiments for $Z\mu\mu$, which are summarized in Tab.~\ref{tab:experimental_limits_Z}.
\medskip

\begin{table}
\centering
\renewcommand{\arraystretch}{1.3}
	\resizebox{\textwidth}{!}{
	\begin{tabular}{|c|ccccc|}
	\hline
	$\ell$&$g_A^\ell/{g_A}_{\rm{SM}}$ LEP~\cite{ALEPH:2005ab} & FCC-ee~\cite{Abada:2019zxq} & ILC~\cite{Baer:2013cma} & CEPC~\cite{An:2018dwb} & CLIC~\cite{Aicheler:2012bya} \\
	\hline
$e$&$0.999681\pm 0.000698227 $&$ \pm 4.1 \times 10^{-6} $&$ \pm 4.1 \times 10^{-5} $&$ \pm 8.7 \times 10^{-6}$ & $\pm 4.4 \times 10^{-5}$ \\
$\mu$&$0.99986\pm 0.00107726 $&$ \pm 6.3 \times 10^{-6} $&$ \pm 6.3 \times 10^{-5} $&$ \pm 1.3 \times 10^{-5}$ & $\pm 6.7 \times 10^{-5}$\\
$\tau$&$1.00154 \pm 0.00127676 $&$ \pm 7.5 \times 10^{-6} $&$ \pm 7.5 \times 10^{-5} $&$ \pm 1.6 \times 10^{-5} $&$ \pm 8.0 \times 10^{-5}$\\
LFU & $0.99992 \pm 0.000518683$ & $\pm 3.1 \times 10^{-6}$ & $\pm 3.0 \times 10^{-5}$ & $\pm 6.5 \times 10^{-6}$ & $\pm 3.2 \times 10^{-5}$\\
\hline
	\end{tabular}
	}
	\caption{Experimental values for $Z\ell\ell$ couplings, extracted from LEP~\cite{ALEPH:2005ab} data and normalized to their SM values with $g_A^\ell=\Lambda^L_{\ell \ell}(m_Z^2) - \Lambda^R_{\ell \ell}(m_Z^2)$. We further show various expected sensitivities for future colliders (second to fifth row) under the assumption that the measurements of $g_A$ are improved by the same factor as $s_w^2$.}
	\label{tab:experimental_limits_Z}
\end{table}

\begin{boldmath}
	\subsection{Charged Lepton Flavor Violation}
\end{boldmath}

Let us now correlate different charged lepton flavor violating observables, i.e. $\ell\to\ell^\prime\gamma$, $Z\to\ell\ell^\prime$ and $\ell\to3\ell^\prime$. We do not study $\mu\to e$ conversion in nuclei, which could be dominant in case of couplings to first generation quarks, but rather again assume only couplings to third generation quarks.
\smallskip

The branching ratios for lepton flavor violating radiative lepton decays, as a function of the (effective) Wilson coefficients in Eq. \eqref{eq:Heff_llgamma}, are given by
\begin{align}
{\mathrm{Br}}\left[\ell_{i}\rightarrow\ell_{f}\gamma\right]=\frac{\alpha m_{\ell_{i}}^3}{256\pi^4} \tau_{\ell_i}\Big(\big|C_{\ell_{f}\ell_{i}}^{L}\big|^2+\big|C_{\ell_{f}\ell_{i}}^{R}\big|^2\Big)~,
\label{eq:Br_ell_ellgamma}
\end{align}
with $\tau_{\ell_i}$ as the life time of the decaying lepton. Similarly, the branching ratio for $Z\to\ell^{+}\ell^{\prime-}$ is given by
\begin{align}
{\mathrm{Br}}\big[Z\to \ell_{i}^{+}\ell_{f}^{-}\big]&=\frac{G_F m_Z^{3}}{3\sqrt{2}\pi\Gamma_{Z}^{\text{tot}}}\left(\big|\Lambda_{\ell_{f}\ell_{i}}^{L}(m_{Z}^2)\big|^{2}+\big|\Lambda_{\ell_{f}\ell_{i}}^{R}(m_{Z}^2)\big|^{2}\right)\,,
\end{align}
with $\Gamma_{Z}^{\text{tot}}=2.495\,\text{GeV}$~\cite{Abbiendi:2000hu} being the total $Z$ boson decay width and the $\Lambda^{L,R}_{\ell_{f}\ell_{i}}(q^2)$ are defined in \eqref{eq:Zll_amplitude}. 
\smallskip

For the three body decays we have
\begin{align}
\begin{split}
\mathrm{Br}\left[\tau^\mp\to \mu^\mp\mu^{+}\mu^{-}\right] &=\frac{m_{\tau}^3}{768\pi^3}\tau_{\tau}\bigg[ \frac{\alpha^2}{\pi^2} \big|C^{L}_{\mu\tau}\big|^2 \Big(\!\log\!\Big(\frac{m_{\tau}^2}{m_{\mu}^2}\Big)-\frac{11}{4}\Big)\\
&\quad+ \frac{m_{\tau}^2}{4}\bigg(\big|C_{\mu\mu\mu\tau}^{SLL}\big|^2+16\big|C_{\mu\tau\mu\mu}^{VLL}\big|^2+4\big|C_{\mu\tau\mu\mu}^{VLR}\big|^2+4\big|C_{\mu\mu\mu\tau}^{VLR}\big|^2\bigg)\\
&\quad-\frac{2\alpha}{\pi} m_{\tau}\, \text{Re} \Big[C_{\mu\tau}^{L*} \big(C_{\mu\tau\mu\mu}^{VRL}+2C_{\mu\tau\mu\mu}^{VRR}\big)\Big]+L\leftrightarrow R\,\bigg]\,,
\label{brmu3e}
\end{split}
\end{align}
with the Wilson coefficients defined in \eq{eq:4l_effective_hamiltonian}. The analogous expression for $\mu\to 3e$ can be obtained by obvious replacements. These rates have to be compared to the experimental limits given in Tab.~\ref{tab:experimental_limits} where we also quote the expected future sensitivities. We do not consider decays like $\tau^\mp\to \mu^\mp e^\pm e^\mp$ as the experimental constraints are slightly worse.
\smallskip

\begin{table}
	\centering
	\renewcommand{\arraystretch}{1.3}
	\resizebox{\textwidth}{!}{
	\begin{tabular}{|cr|cr|cr|}
	\hline
	${\rm Br}\left[Z\to\ell\ell^{\prime}\right]$ & Ref. & ${\rm Br}\left[\ell\to\ell^{\prime}\gamma\right]$ & Ref. & ${\rm Br}\left[\ell\to 3 \ell\right]$ & Ref.\\
	\hline
	$Z\to e^{\pm}\mu^{\mp}<7.5\times 10^{-7}$& \cite{Aad:2014bca} & $\mu \to e \gamma < 4.2 \times 10^{-13}$ & \cite{TheMEG:2016wtm}&$\mu \to 3e < 1.0 \times 10^{-12}$& \cite{Bellgardt:1987du}\\
	$Z\to e^{\pm}\tau^{\mp} <9.8\times 10^{-6}$ & \cite{Abreu:1996mj} & $\tau \to e \gamma <  3.3 \times 10^{-8}$ & \cite{ Aubert:2009ag}&$\tau \to \mu ee < 1.5 \times 10^{-8}$&\cite{Hayasaka:2010np}\\
	$Z\to \mu^{\pm}\tau^{\mp}<1.2\times 10^{-5}$ & \cite{Akers:1995gz} & $\tau \to \mu \gamma <  4.4 \times 10^{-8}$ & \cite{ Aubert:2009ag}&$\tau \to 3\mu <2.1 \times 10^{-8}$&\cite{Hayasaka:2010np}
	\\
	\hline
	$ Z \to \mu^{\pm} \tau^{\mp}<1.0 \times 10^{-8}$& \cite{Baer:2013cma} & $ \mu \to e \gamma < 6.0 \times 10^{-14}$ & \cite{Baldini:2018nnn}&$\mu \to 3e < 5.5 \times 10^{-15}$& \cite{Berger:2014vba}\\
	$ Z \to \mu^{\pm} \tau^{\mp}<1.0 \times 10^{-9}$& \cite{Abada:2019zxq} &$\tau \to \mu \gamma < 1.0 \times 10^{-9}$& \cite{Kou:2018nap}&$\tau \to 3 \mu < 1.0 \times10^{-9}$&\cite{Cerri:2018ypt}\\
	&&&&$\tau \to 3 \mu < 3.3 \times 10^{-10}$&\cite{Kou:2018nap}
	\\ \hline
	\end{tabular}
	}
	\caption{Current experimental limits (top panel) and projected future experimental sensitivities (bottom panel) on lepton flavor violating decays of charged leptons.}
	\label{tab:experimental_limits}
\end{table}

\begin{figure}
	\centering
	\includegraphics[width=0.8\textwidth]{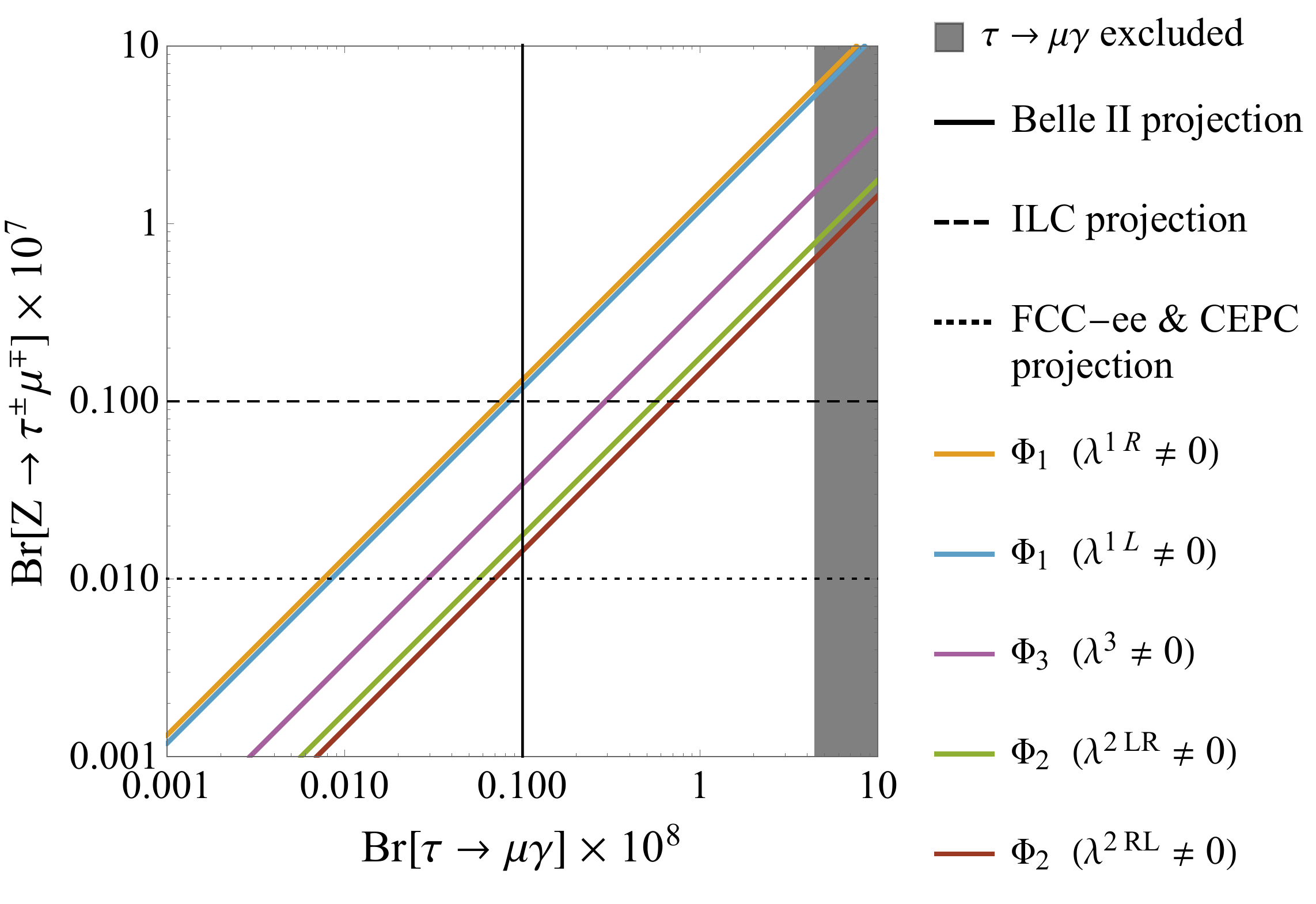}
	\caption{Correlations between $\tau\to\mu\gamma$ and $Z\to\tau\mu$ for the three LQ representations which generate an $m_{t}^2/m_{\rm LQ}^2$ effect in $Z\ell\ell$ couplings. We assume that $\Phi_1$ and $\Phi_2$ couple either to left or to right-handed leptons only such that chirally enhanced effects (which would result in dominant effects in $\tau\to\mu\gamma$) are absent.}
	\label{Ztaumutaumugamma}
\end{figure}

In our numerical analysis, we again assume that the LQs only couple to third generation quarks but now allow for the possibility that they couple to more than one lepton flavor at the same time. Let us start by examining the correlations between $\tau\to\mu\gamma$ and $Z\to\tau\mu$ in Fig.~\ref{Ztaumutaumugamma}. One can see that this correlation is very direct under the assumption that only one representation contributes and that for $\Phi_1$ and $\Phi_2$ only either the left- or the right-handed couplings to leptons are non-zero at the same time such that chirality enhanced effects in $\tau\to\mu\gamma$ are absent. Although currently $\tau\to\mu\gamma$ is more constraining, even in the absence of chirality enhanced contributions, in the future $Z\to\tau\mu$ can provide competitive or even superior bounds. The situation for $\tau\to e$ transitions is very similar and therefore not shown explicitly. 

\begin{figure}
	\centering
	\includegraphics[width=0.35\textwidth]{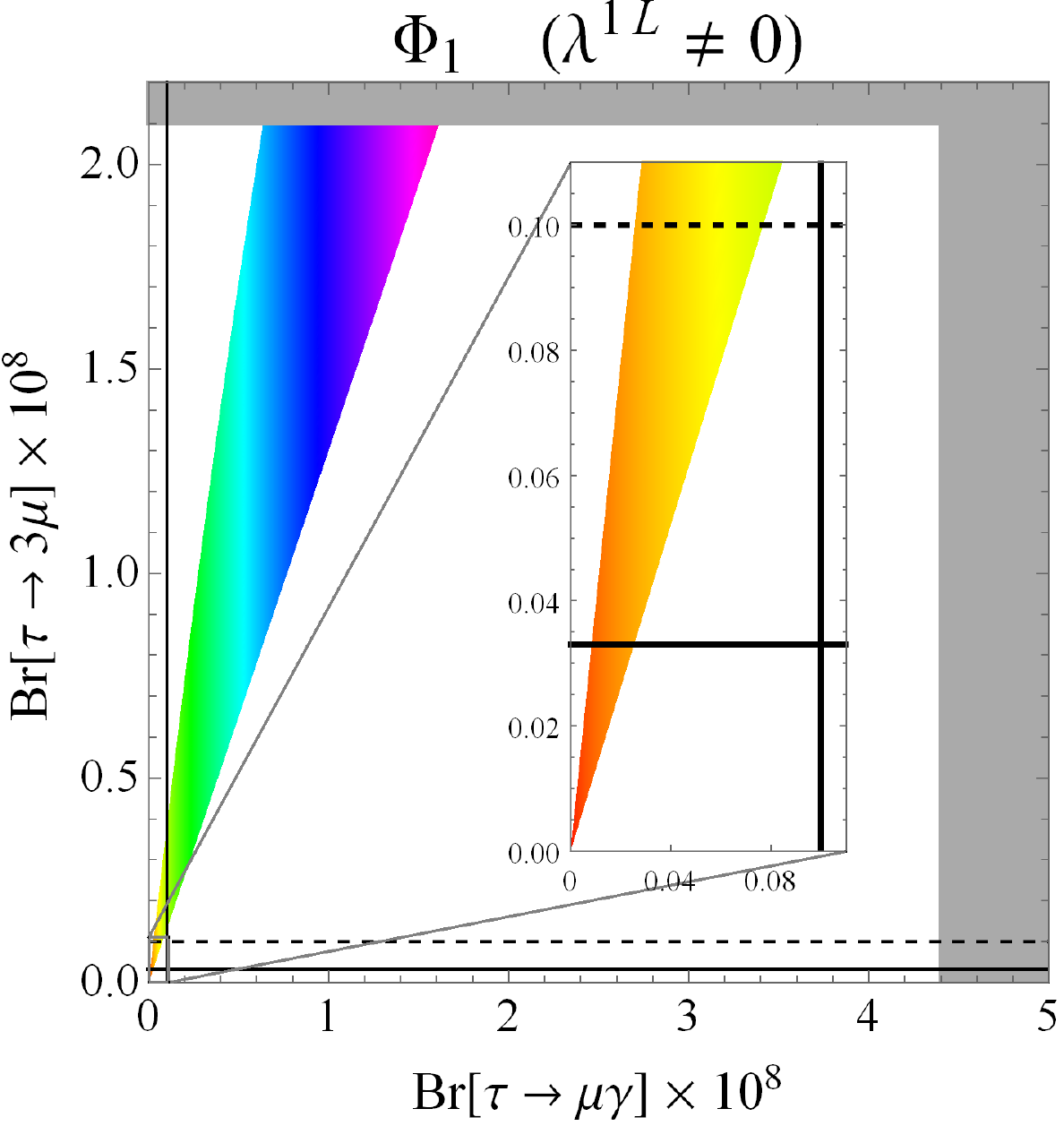}
	\hspace{1cm}
	\vspace{0.2cm}
	\includegraphics[width=0.35\textwidth]{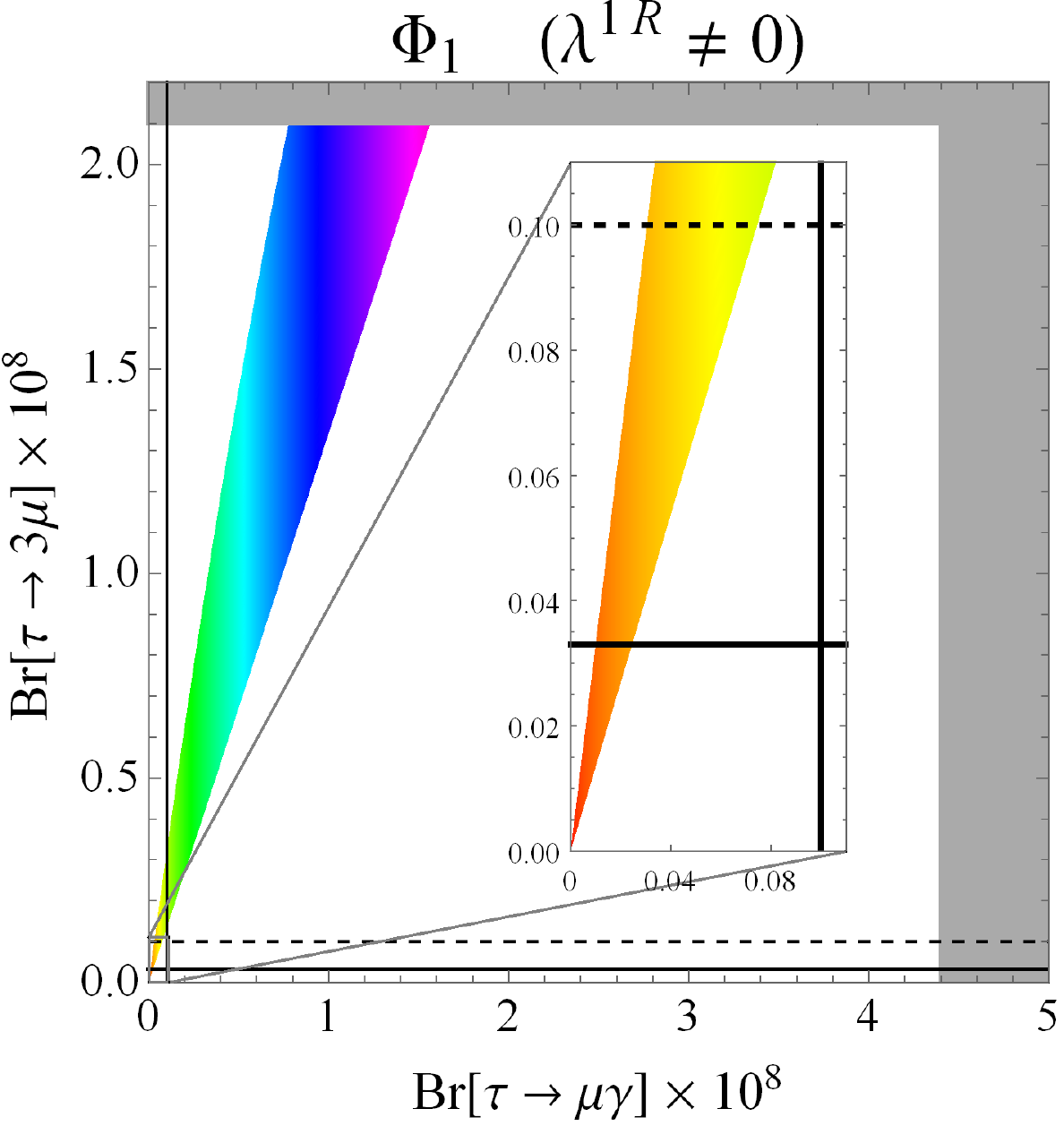}
	\vspace{0.2cm}
	\includegraphics[width=0.35\textwidth]{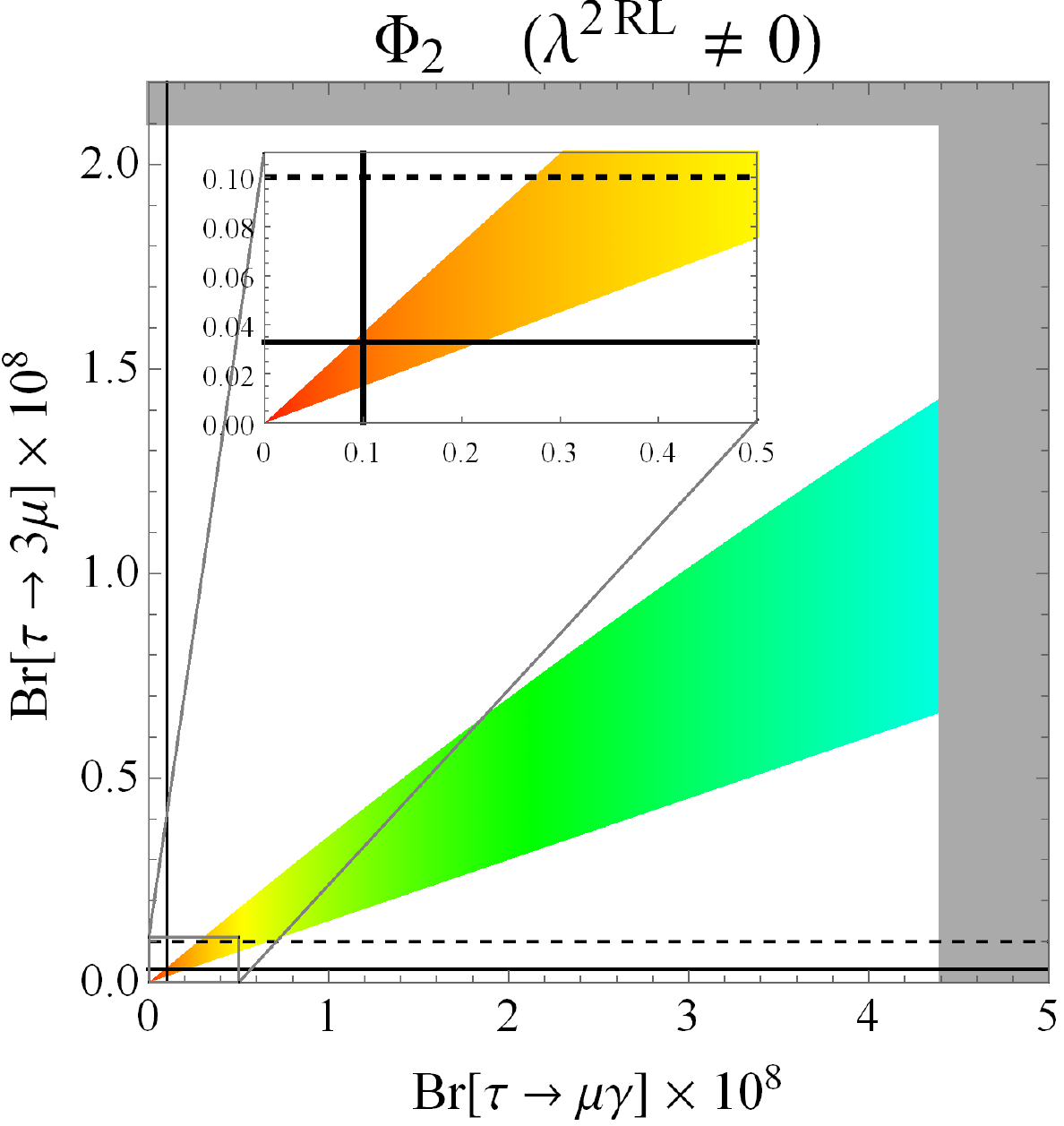}
	\hspace{1cm}
	\includegraphics[width=0.35\textwidth]{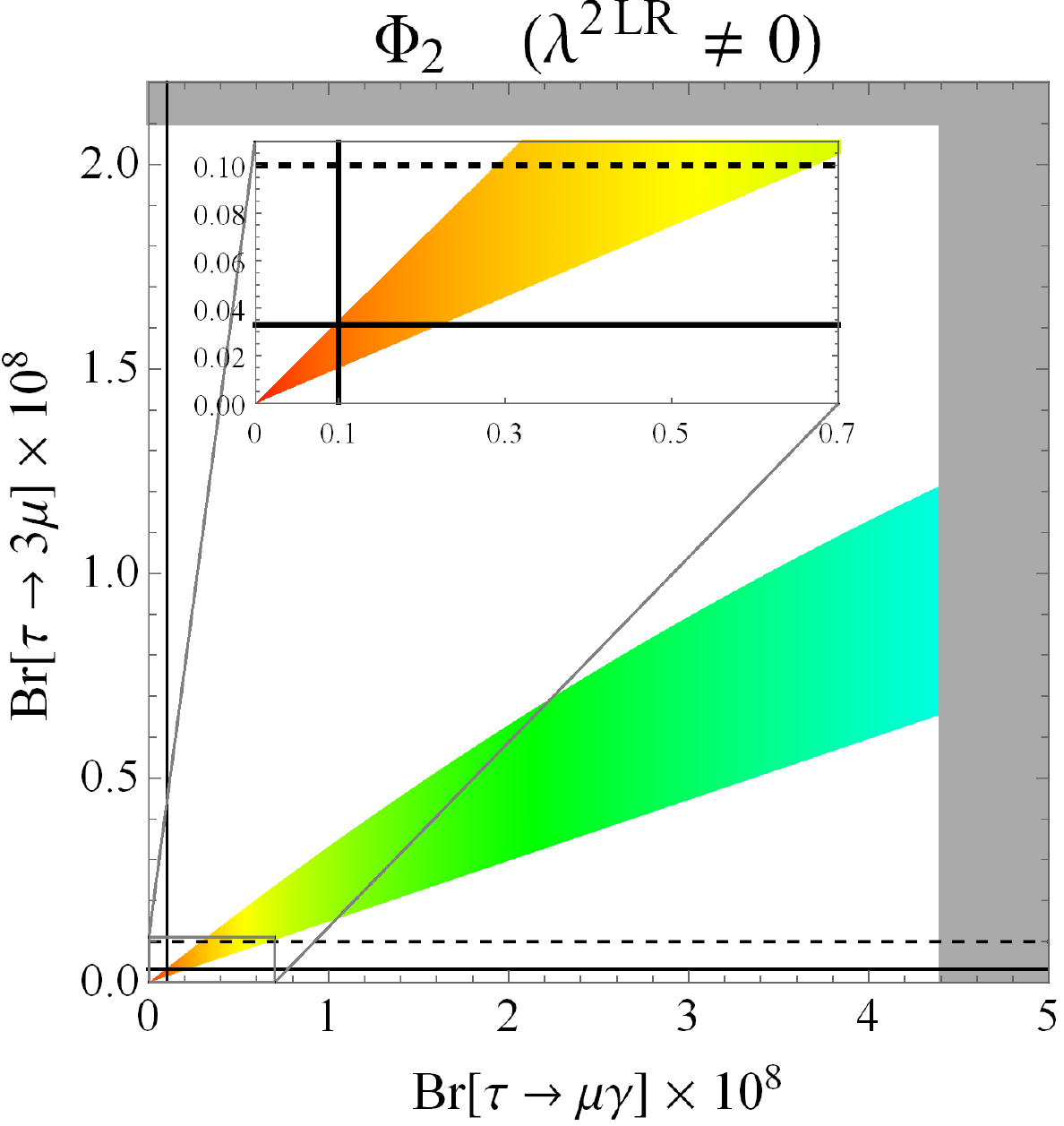}
	\includegraphics[width=0.35\textwidth]{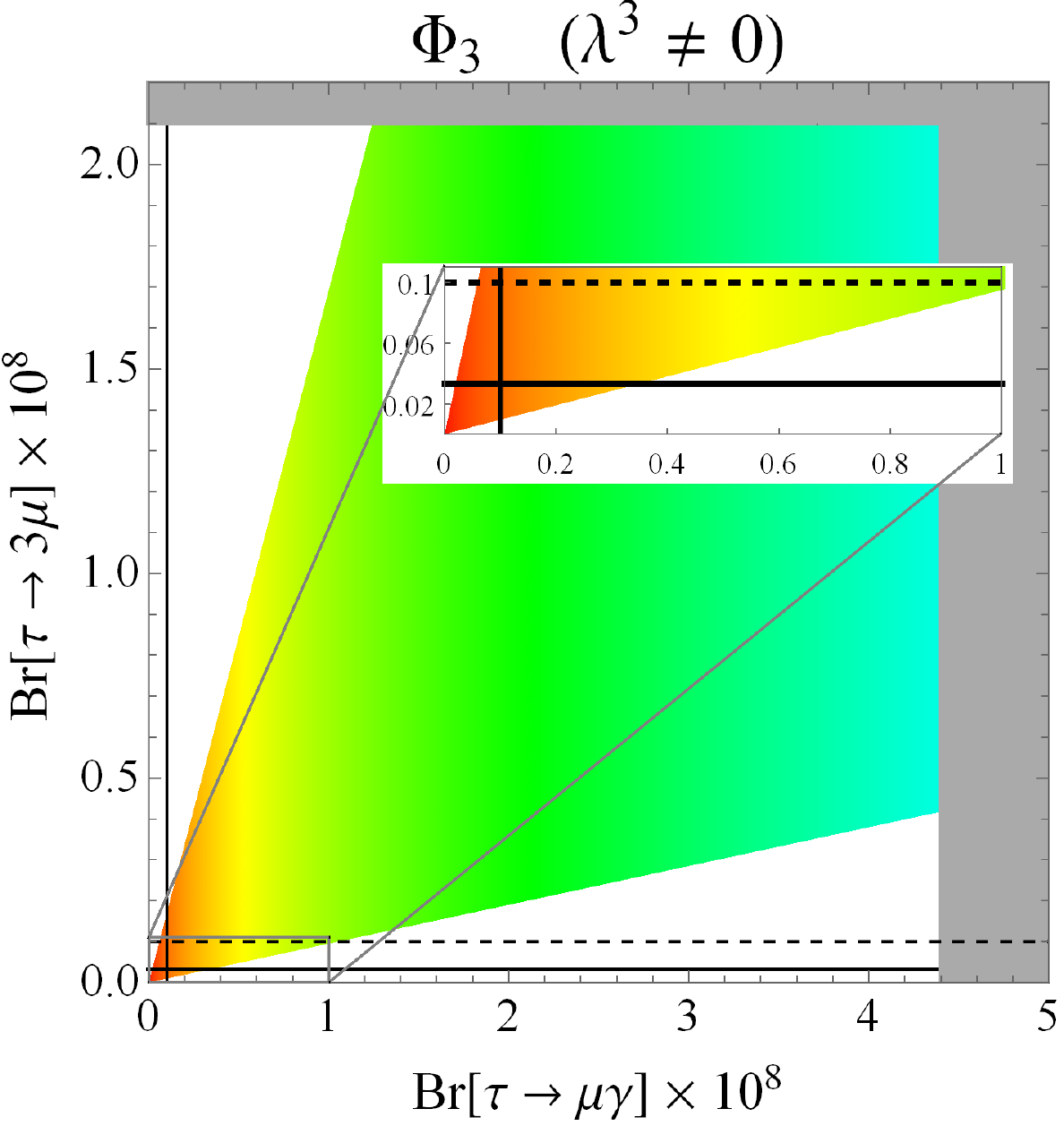}
	\hspace{1cm}
	\includegraphics[width=0.37\textwidth]{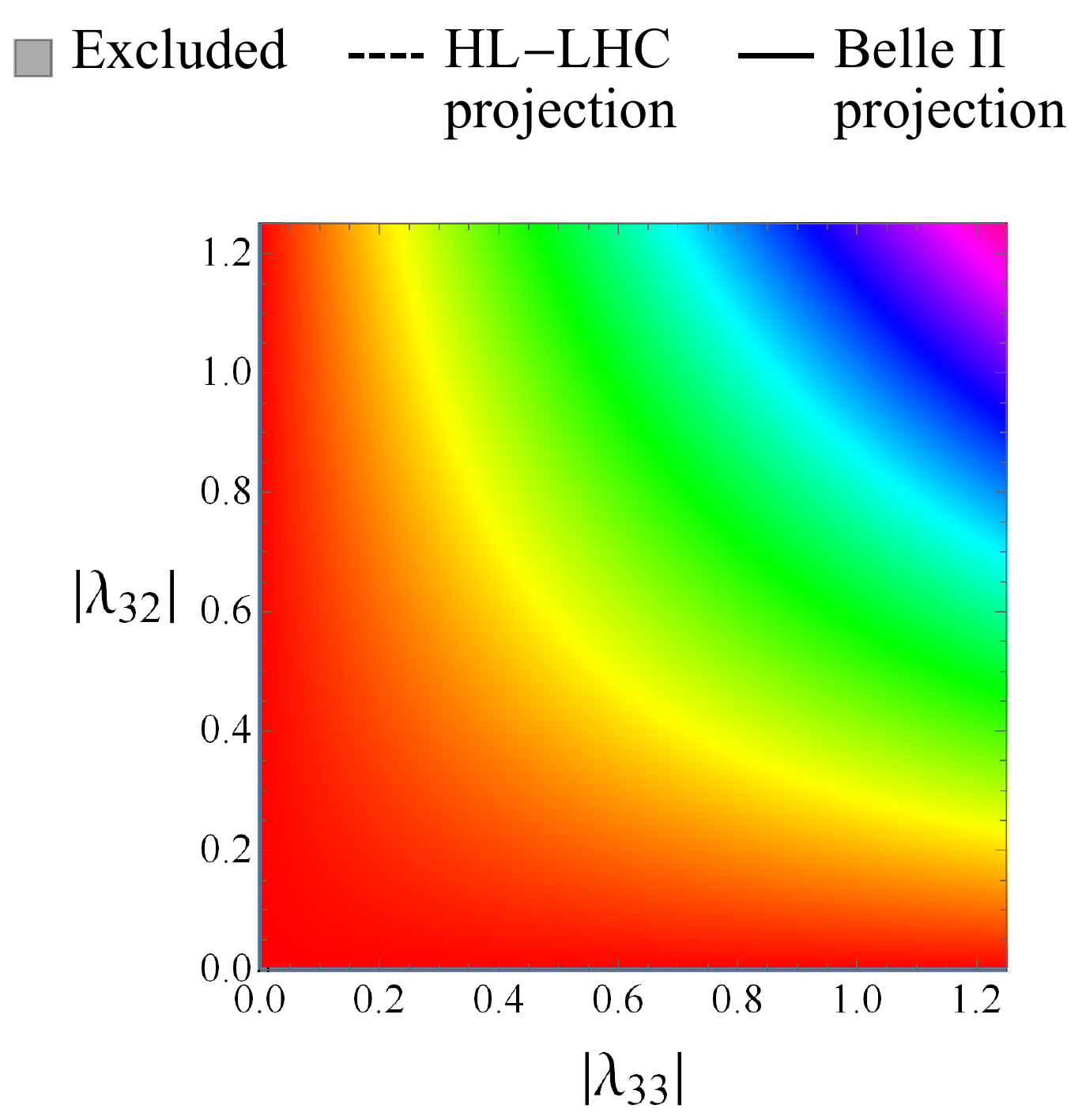}
	\caption{The correlations between $\tau\to\mu\gamma$ and $\tau\to 3\mu$ for a LQ mass of 1.5 TeV where we scanned $\lambda_{33}$ and $\lambda_{32}$ in the range $[-1.5,1.5]$. The gray regions are currently excluded by experiment. The dashed (solid) lines show the projected sensitivities for the HL-LHC (Belle II), see Tab.~\ref{tab:experimental_limits} for the numerical values.}
	\label{taumugammatau3mu}
\end{figure}

\begin{figure}
	\centering
	\includegraphics[width=0.35\textwidth]{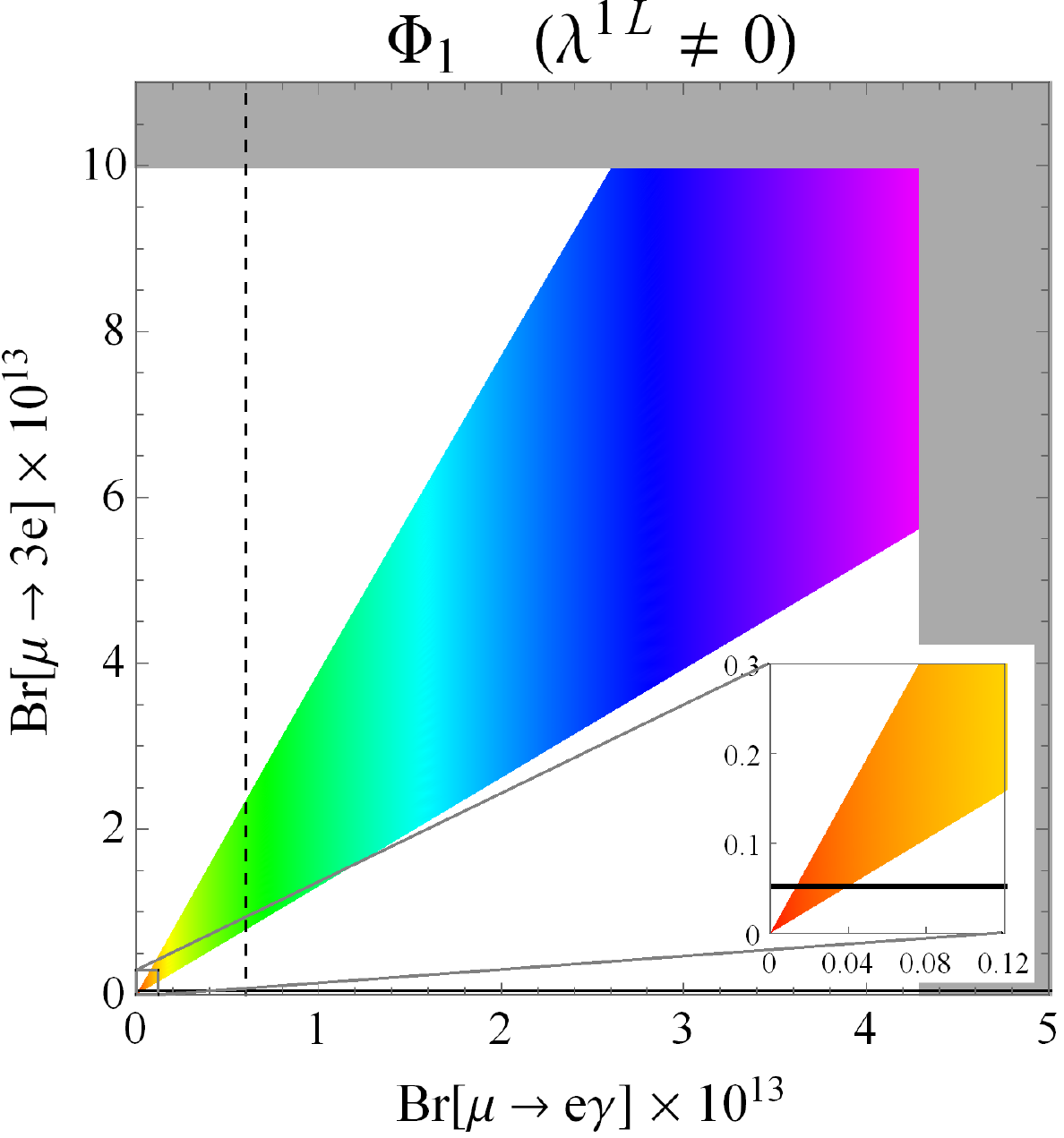}
	\hspace{1cm}
	\vspace{0.2cm}
	\includegraphics[width=0.35\textwidth]{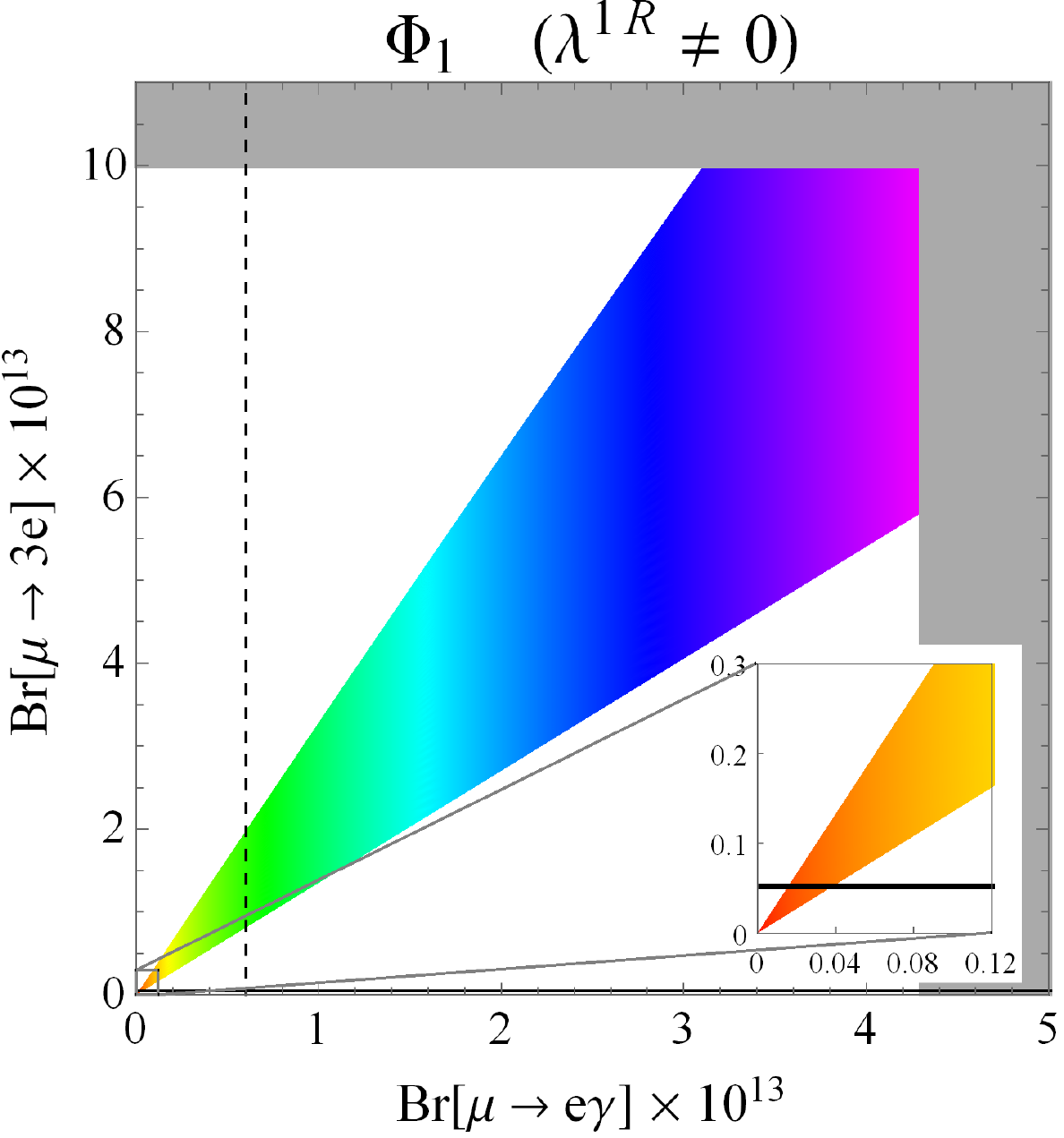}
	\vspace{0.2cm}
	\includegraphics[width=0.35\textwidth]{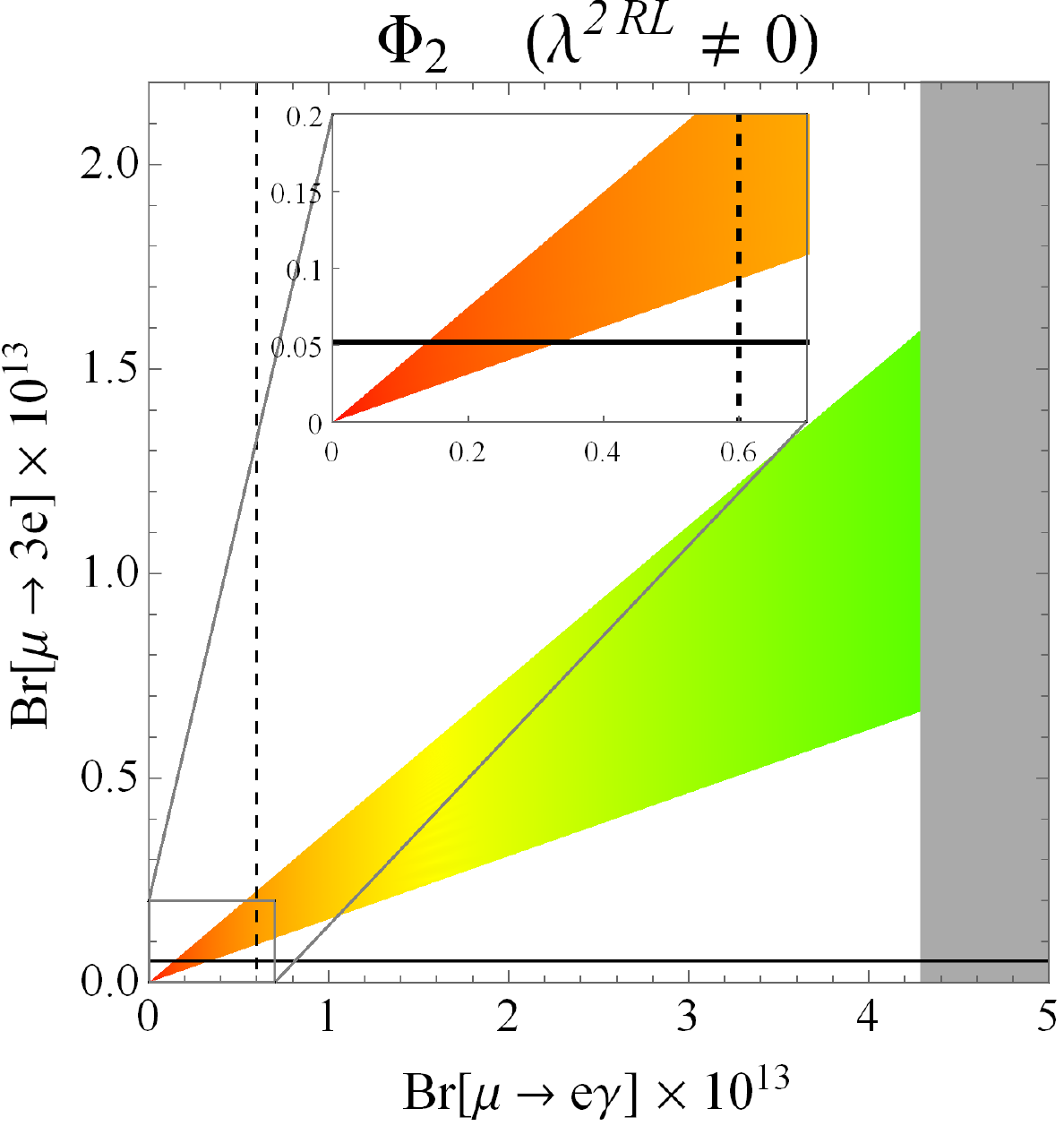}
	\hspace{1cm}
	\includegraphics[width=0.35\textwidth]{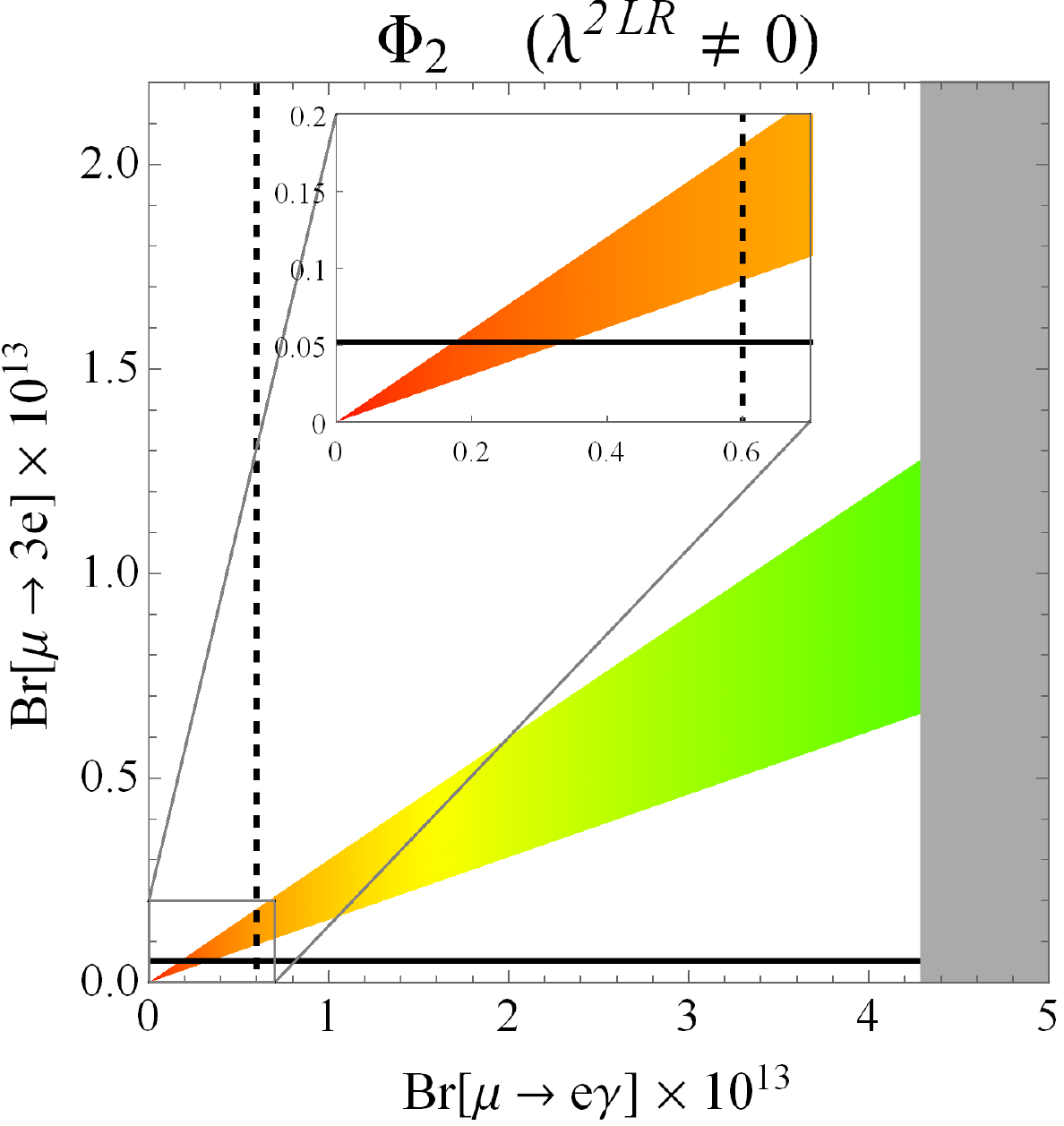}
	\includegraphics[width=0.35\textwidth]{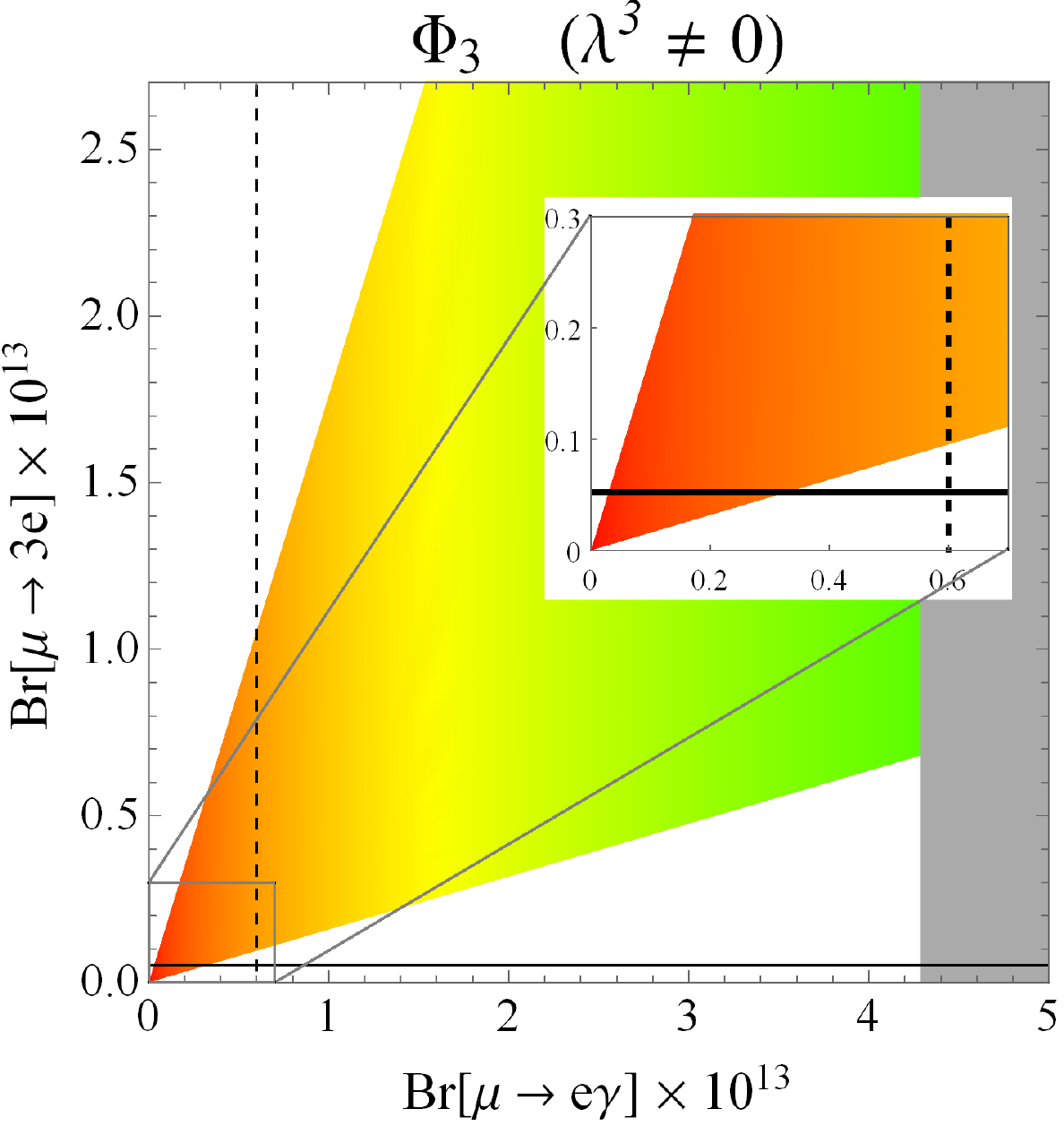}
	\hspace{1cm}
	\includegraphics[width=0.37\textwidth]{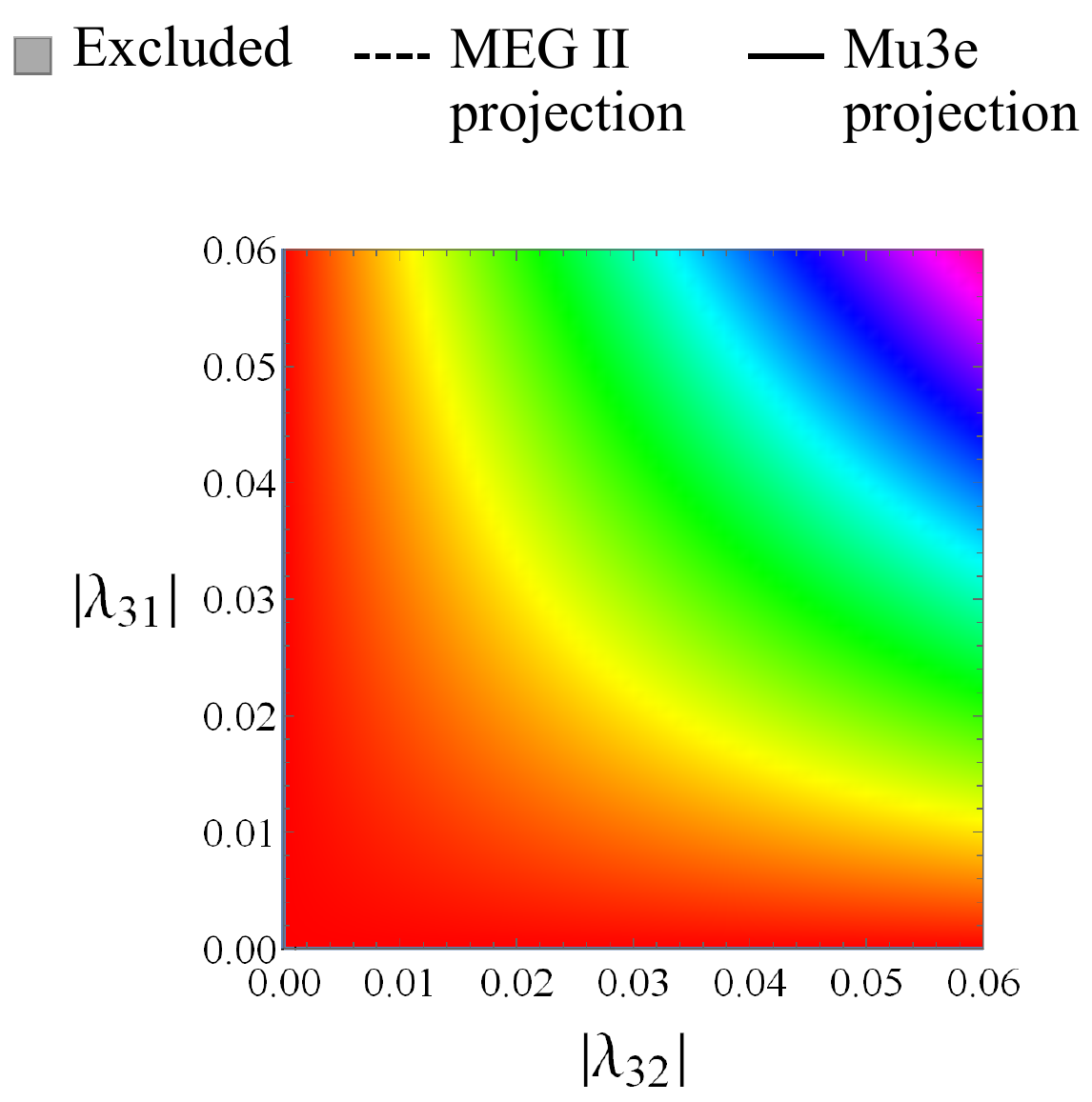}
	\caption{The analogue to the plots above for the $\mu\to e$ transition. The dashed lines depict the expected sensitivity from MEG II \cite{Baldini:2018nnn} and the solid line the one of Mu3e \cite{Berger:2014vba}. {Note that the color scaling shows the product of two couplings, as can be seen from the legend in the bottom-right.}}
	\label{muegammamu3e}
\end{figure}

In Fig.~\ref{taumugammatau3mu} we show the correlations between $\tau\to\mu\gamma$ and $\tau\to3\mu$. These correlations are not as clear as in the case of $Z\to\tau\mu$ due to the additional box contributions to $\tau\to3\mu$. Therefore, one obtains a cone instead of a straight line. Interestingly, for $\Phi_1$ the effect in $\tau\to\mu\gamma$ is smallest among the LQ representations due to the electric charge of the LQ. Hence, even though phase space suppressed, $\tau\to3\mu$ is more sensitive to this particular LQ than $\tau\to\mu\gamma$. Again, the situation in $\tau\to e$ transitions is very similar and therefore not shown explicitly. However, we show our analysis for $\mu\to e$ transitions in Fig.~\ref{muegammamu3e}. For the $\mu\to e\gamma$ scenario we do not show $Z\to\mu e$ since the low energy bounds are so stringent that the former cannot compete, even when taking into account future prospects. {The (lower) upper boundary of the cone is obtained for a hierarchic flavor structure, i.e. $\lambda_{33} \,  (\gg) \ll \lambda_{32}$ for $\tau \to \mu$ and $\lambda_{32}\, (\gg) \ll \lambda_{31}$ for $\mu \to e $ transitions, respectively, such that the box contributions are (sub)dominant. The opening angle of the cone is determined by the size of the box contributions to $\ell \to 3 \ell'$. For example, the LQ triplet yields the biggest box contribution, which can easily be seen from \eq{eq:WC_l->3l}.}
\medskip

\section{Conclusions}
\label{conclusions}

Leptoquarks are prime candidates to explain the flavor anomalies, i.e. the discrepancies between measurements and the SM predictions in $b\to s\ell^+\ell^-$, $b\to c\tau\nu$ and the AMM of the muon. With this motivation in mind, we calculated the one-loop amplitudes generated by scalar LQs for the purely leptonic transitions, involving:
\begin{itemize}
\begin{minipage}{\linewidth}
	\item $\ell\ell\gamma$
\vspace{-1mm}
\item $Z\ell\ell$ and $Z\nu\nu$
\vspace{-1mm}
\item $W\ell\nu$
\vspace{-1mm}
\item $h\ell\ell$
\vspace{-1mm}
\item $4\ell$
\vspace{-1mm}
\item $2\ell 2\nu$
\end{minipage}
\end{itemize}
Taking into account the most general set of interactions of the LQs with the SM Higgs doublet, we obtained relatively simple analytic expressions for the amplitudes by expanding the LQ mixing matrices in $v/m_{\rm{LQ}}$, corresponding to a mass insertion approximation. 
\smallskip

In our phenomenological analysis, we illustrated the results of our calculation by studying:
\begin{itemize}
	\item LQ effects in effective $Z\ell\ell$, $Z\nu\nu$ and $W\ell\nu$ couplings and the associated gauge boson decays. Here we found for the three representations which generate $m_t^2/m_{\rm LQ}^2$ enhanced effects ($\Phi_1$, $\Phi_2$ and $\Phi_2$) that $Z\to \ell^+\ell^-$ is smaller than within the SM while $Z\to\nu\nu$ is enhanced. For order one couplings, the effect is at the percent level for TeV scale LQs.
	\item Correlations between the AMM of the muon, $Z\to\ell^{+}\ell^{-}$, effective $W\mu\nu$ couplings and $h\to\mu^{+}\mu^{-}$. Here we found that, since an explanation of the $(g-2)_\mu$ anomaly requires a $m_t/m_\mu$ enhanced effect, also the contribution in $h\to\mu^{+}\mu^{-}$ is pronounced by the same factor. Furthermore, effects scaling like $m_t^2/m^2_{\rm LQ}$ in $Z\to\mu^{+}\mu^{-}$ are generated which are most relevant in case where the left-handed couplings are much larger than the right handed ones and vice versa.
	\item Correlations between $\tau\to\mu\gamma$, $Z\to\tau\mu$ and $\tau\to 3\mu$, as well as the analogues in $\mu\to e$ transitions. Here we observed that $\tau\to\mu\gamma$ and $Z\to\tau\mu$ can be directly correlated under the assumption the LQs couple only to left-handed or to right-handed leptons (but not to both of the same time). Furthermore, in this setup $\tau\to\mu\gamma$ and $\mu\to e\gamma$ do not receive chirally enhanced effects such that $\tau\to3\mu$ and $\mu\to 3e$ can give competitive bounds, which is in particular the case for $\Phi_1$.
\end{itemize}
These interesting correlations can be tested at future precision experiments and high-energy colliders. 
\medskip

{\it Acknowledgments} --- {We thank David Marzocca, Martin Hoferichter, Ulrich Nierste, Adrian Signer and Michael Spira for useful discussions.  The work of A.C. and D.M. is supported by a Professorship Grant (PP00P2\_176884) of the Swiss National Science Foundation and the one of C.G. and F.S. by the Swiss National Science Foundation grant 200020\_175449/1.}
\medskip

\appendix
\section{Appendix}

\subsection{Self-Energies}
\label{app:self_energies}
Focusing on the non-decoupling, momentum-independent parts of the self-energies, we have generically
\begin{subequations}
\label{app:self_energies_full_result}
\begin{align}
\Sigma_{fi}^{\ell LR}&=\frac{-m_{q_j}N_c}{16\pi^2}\Gamma_{q_{j}\ell_{f}}^{L,a*}\Gamma_{q_{j}\ell_{i}}^{R,a}\mathcal{I}_{0}\Big(\frac{\mu^2}{M_a^2},\frac{m_{q_j}^2}{M_a^2}\Big)\,,\\
\Sigma_{fi}^{\ell LL}&=\frac{-N_c}{32\pi^2}\Gamma_{q_{j}\ell_{f}}^{L,a*}\Gamma_{q_{j}\ell_{i}}^{L,a}\mathcal{I}_{1}\Big(\frac{\mu^2}{M_a^2},\frac{m_{q_j}^2}{M_a^2}\Big)\,,
\end{align}
\end{subequations}
with $\Sigma_{fi}^{\ell RR}$ and $\Sigma_{fi}^{\ell RL}$ obtained by interchanging chiralities and $\Sigma_{fi}^{\nu LL}$ by replacing $\ell$ with $\nu$. We set all quark masses within the loop equal to zero, except for the top mass. Additionally, one has to sum over all internal quarks $u_{j},d_{j},u_{j}^c$ and $d_{j}^c$, as well as over their flavors $j=\{1,2,3\}$. The loop functions take the simple form
\begin{subequations}
\label{app:self_energies_exact}
\begin{align}
\mathcal{I}_{0}(x,y)&=\frac{1}{\varepsilon}+1+\log(x)+y \log(y)\,,\\
\mathcal{I}_{1}(x,y)&=\frac{1}{\varepsilon}+\frac{1}{2}+\log(x)-y\,,
\end{align}
\end{subequations}
where the last terms in $\mathcal{I}_0$ and $\mathcal{I}_1$ are only relevant for the top quark and can be neglected in all other cases.
\smallskip

Now we expand the expressions in Eq.~\eqref{app:self_energies_full_result} in terms of $v/m_{\rm{LQ}}$ up to $\mathcal{O}(v^2/m_{\rm{LQ}}^2)$
\begin{subequations}
\allowdisplaybreaks
\begin{align}
\Sigma_{\ell,fi}^{LL}&\approx
\frac{-N_c}{32\pi^2}\sum_{j=1}^3\bigg[V_{jk}\lambda_{kf}^{1L*}V_{jl}^{*}\lambda_{li}^{1L}\bigg(\mathcal{I}_{1}\Big(\frac{\mu^2}{m_{1}^2},\frac{m_{u_j}^2}{m_{1}^2}\Big)-\frac{v^2Y_{1}}{m_{1}^2}+\frac{v^2|A_{\tilde{2}1}|^2}{\tilde{m}_{2}^4}\mathcal{I}_{4}\Big(\frac{m_{1}^2}{\tilde{m}_{2}^2}\Big)\bigg)\nonumber\\
&+V_{jk}\lambda_{kf}^{3*}V_{jl}^{*}\lambda_{li}^{3}\bigg(\!\mathcal{I}_{1}\Big(\frac{\mu^2}{m_{3}^2},\frac{m_{u_j}^2}{m_{1}^2}\Big)\!+2\mathcal{I}_{1}\Big(\frac{\mu^2}{m_{3}^2},0\Big)\!-\frac{v^2(3Y_{3}-2Y_{33})}{m_{3}^2}+\frac{v^2|A_{\tilde{2}3}|^2}{\tilde{m}_{2}^4}\mathcal{I}_{4}\Big(\frac{m_{3}^2}{\tilde{m}_{2}^2}\Big)\!\bigg)\nonumber\\
&+v^2\bigg(\frac{\lambda_{jf}^{3*}\lambda_{ji}^{1L}Y_{13}^{*}+\lambda_{jf}^{1L*}\lambda_{ji}^{3}Y_{13}}{m_{3}^2}\mathcal{H}_{1}\Big(\frac{m_{1}^2}{m_{3}^2}\Big)\nonumber\\
&+\frac{\lambda_{jf}^{3*}\lambda_{ji}^{1L}A_{\tilde{2}1}A_{\tilde{2}3}^{*}+\lambda_{jf}^{1L*}\lambda_{ji}^{3}A_{\tilde{2}1}^{*}A_{\tilde{2}3}}{\tilde{m}_{2}^4}\mathcal{I}_{5}\Big(\frac{m_{1}^2}{\tilde{m}_{2}^2},\frac{m_{3}^2}{\tilde{m}_{2}^2}\Big)\bigg)\nonumber\\
&+\tilde{\lambda}_{jf}^{2*}\tilde{\lambda}_{ji}^{2}\bigg(\mathcal{I}_{1}\Big(\frac{\mu^2}{\tilde{m}_{2}^2},0\Big)-\frac{v^2(Y_{\tilde{2}}+Y_{\tilde{2}\tilde{2}})}{\tilde{m}_{2}^2}+\frac{2v^2|A_{\tilde{2}3}|^2}{\tilde{m}_{2}^4}\mathcal{I}_{6}\Big(\frac{m_{3}^2}{\tilde{m}_{2}^2}\Big)\bigg)\nonumber\\
&+\lambda_{jf}^{2RL*}\lambda_{ji}^{2RL}\bigg(\mathcal{I}_{1}\Big(\frac{\mu^2}{m_{2}^2},\frac{m_{u_j}^2}{m_{2}^2}\Big)-\frac{v^2(Y_{2}+Y_{22})}{m_{2}^2}\bigg)
\bigg]\,,\\
\Sigma_{\ell,fi}^{RR}&\approx\sum_{j=1}^3\frac{-N_c}{32\pi^2}\bigg[\lambda_{jf}^{1R*}\lambda_{ji}^{1R}\bigg(\mathcal{I}_{1}\Big(\frac{\mu^2}{m_{1}^2},\frac{m_{u_j}^2}{m_{1}^2}\Big)-\frac{v^2Y_{1}}{m_{1}^2}+\frac{v^2|A_{\tilde{2}1}|^2}{\tilde{m}_{2}^4}\mathcal{I}_{4}\Big(\frac{m_{1}^2}{\tilde{m}_{2}^2}\Big)\bigg)\nonumber\\
&+V_{jk}^{*}\lambda_{kf}^{2LR*}V_{jl}\lambda_{li}^{2LR}\bigg(2\mathcal{I}_{1}\Big(\frac{\mu^2}{m_{2}^2},\frac{m_{u_j}^2}{m_{2}^2}\Big)-\frac{v^2(2Y_{2}+Y_{22})}{m_{2}^2}\bigg)\nonumber\\
&+\tilde{\lambda}_{jf}^{1*}\tilde{\lambda}_{ji}^{1}\bigg(\mathcal{I}_{1}\Big(\frac{\mu^2}{\tilde{m}_{1}^2},0\Big)-\frac{v^2Y_{\tilde{1}}}{\tilde{m}_{1}^2}\bigg)\bigg]\,,\\
\Sigma_{fi}^{\ell RL}&\approx\sum_{j=1}^3\frac{-m_{u_j} N_c}{16\pi^2}\bigg[\lambda_{jf}^{1R*}V_{jk}^{*}\lambda_{ki}^{1L}\bigg(\mathcal{I}_{0}\Big(\frac{\mu^2}{m_{1}^2},\frac{m_{u_j}^2}{m_{1}^2}\Big)-\frac{v^2Y_{1}}{m_{1}^2}+\frac{v^2|A_{\tilde{2}1}|^2}{\tilde{m}_{2}^4}\mathcal{I}_{4}\Big(\frac{m_{1}^2}{\tilde{m}_{2}^2}\Big)\bigg)\nonumber\\
&+v^2\lambda_{jf}^{1R*}V_{jk}^{*}\lambda_{ki}^{3}\bigg(\frac{Y_{13}}{m_{3}^2}\mathcal{H}_{1}\Big(\frac{m_{1}^2}{m_{3}^2}\Big)+\frac{A_{\tilde{2}1}^{*}A_{\tilde{2}3}}{\tilde{m}_{2}^4}\mathcal{I}_{5}\Big(\frac{m_{1}^2}{\tilde{m}_{2}^2},\frac{m_{3}^2}{\tilde{m}_{2}^2}\Big)\bigg)\nonumber\\
&+V_{jk}^{*}\lambda_{kf}^{2LR*}\lambda_{ji}^{2RL}\bigg(\mathcal{I}_{1}\Big(\frac{\mu^2}{m_{2}^2},\frac{m_{u_j}^2}{m_{2}^2}\Big)-\frac{v^2(Y_{2}+Y_{22})}{m_{2}^2}\bigg)\bigg]\nonumber\\
&+\sum_{j=1}^3\frac{m_{d_j}}{16\pi^2}\bigg[\frac{v^2\lambda_{jf}^{2LR*}\tilde{\lambda}_{ji}^{2}Y_{2\tilde 2}^{*}}{\tilde{m}_{2}^2} \mathcal{H}_{1}\Big(\frac{m_{2}^2}{\tilde{m}_{2}^2}\Big)-\frac{2v^2\tilde{\lambda}_{jf}^{1*}\lambda_{ji}^{3}Y_{3\tilde{1}}^{*}}{m_{3}^2}\mathcal{H}_{1}\Big(\frac{\tilde{m}_{1}^2}{m_{3}^2}\Big)\bigg]
\,,
\end{align}
\end{subequations}
with $\Sigma_{fi}^{\ell LR}=\Sigma_{if}^{\ell RL*}$ and
\begin{subequations}
\begin{align}
\mathcal{I}_{4}(x)&=\frac{1-x+x\log(x)}{x(x-1)^2}\\
\mathcal{I}_{5}(x,y)&=\frac{\log(x)}{(x-1)(x-y)}+\frac{\log(y)}{(y-1)(y-x)}\\
\mathcal{I}_{6}(x)&=\frac{x-1-\log(x)}{(x-1)^2} \,.
\end{align}
\end{subequations}
For neutrinos we have
\begin{align}
\begin{aligned}
\Sigma_{fi}^{\nu LL}&\approx\sum_{j=1}^3 \frac{-N_c}{32\pi^2}\bigg[\lambda_{jf}^{1L*}\lambda_{ji}^{1L}\bigg(\mathcal{I}_{1}\Big(\frac{\mu^2}{m_{1}^2},0\Big)-\frac{v^2Y_{1}}{m_{1}^2}+\frac{v^2|A_{\tilde{2}1}|^2}{\tilde{m}_{2}^4}\mathcal{I}_{4}\Big(\frac{m_{1}^2}{\tilde{m}_{2}^2}\Big)\bigg)\\
&+V_{jk}\lambda_{kf}^{3*}V_{jl}\lambda_{li}^{3}\bigg(\!\mathcal{I}_{1}\Big(\frac{\mu^2}{m_{3}^3},\frac{m_{u_j}^2}{m_{3}^2}\Big)+\mathcal{I}_{1}\Big(\frac{\mu^2}{m_{3}^3},0\Big)\!-\frac{v^2(3Y_{3}+2Y_{33})}{m_{3}^2}\!+\frac{5v^2|A_{\tilde{2}3}|^2}{\tilde{m}_{2}^4}\mathcal{I}_{4}\Big(\frac{m_{3}^2}{\tilde{m}_{2}^2}\Big)\!\bigg)\\
&-v^2\bigg(\frac{\lambda_{jf}^{1L*}\lambda_{ji}^{3}Y_{13}+\lambda_{jf}^{3*}\lambda_{ji}^{1L}Y_{13}^{*}}{m_{3}^2}\mathcal{H}_{1}\Big(\frac{m_{1}^2}{m_{3}^2}\Big)\\
&+\frac{\lambda_{jf}^{1L*}\lambda_{ji}^{3}A_{\tilde{2}1}^{*}A_{\tilde{2}3}+\lambda_{jf}^{3*}\lambda_{ji}^{1L}A_{\tilde{2}1}A_{\tilde{2}3}^{*}}{\tilde{m}_{2}^4}\mathcal{I}_{5}\Big(\frac{m_{1}^2}{\tilde{m}_{2}^2},\frac{m_{3}^2}{\tilde{m}_{2}^2}\Big)\bigg)\\
&+\tilde{\lambda}_{jf}^{2*}\tilde{\lambda}_{ji}^{2}\bigg(\mathcal{I}_{1}\Big(\frac{\mu^2}{\tilde{m}_{2}^2},0\Big)-\frac{v^2Y_{\tilde{2}}}{\tilde{m}_{2}^2}+\frac{v^2|A_{\tilde{2}1}|^2}{\tilde{m}_{2}^4}\mathcal{I}_{6}\Big(\frac{m_{1}^2}{\tilde{m}_{2}^2}\Big)+\frac{v^2|A_{\tilde{2}3}|^2}{\tilde{m}_{2}^4}\mathcal{I}_{6}\Big(\frac{m_{3}^2}{\tilde{m}_{2}^2}\Big)\bigg)\\
&+\lambda_{jf}^{2RL*}\lambda_{ji}^{2RL}\bigg(\mathcal{I}_{1}\Big(\frac{\mu^2}{m_{2}^2},\frac{m_{u_j}^2}{m_{2}^2}\Big)-\frac{v^2Y_{2}}{m_{2}^2}\bigg)\bigg]\,.
\end{aligned}
\end{align}
\medskip

\subsection{Loop Functions}
\label{app:loop_functions}
The loop-functions for $\ell\ell\gamma$ with on-shell photons  read
\begin{subequations}
\begin{align}
\mathcal{E}_{1}(x)&=7+4\log(x)\\
\mathcal{E}_{2}(x)&=11+4\log(x)\\
\mathcal{E}_{3}(x)&=1+4\log(x)\\
\mathcal{E}_{4}(x)&=5+4\log(x)\\
\mathcal{E}_{5}(x,y)&=\frac{4}{y}\log(x)+\frac{7y-11}{y(y-1)}+4\frac{2y-1}{y(y-1)^2}\log(y)\\
\mathcal{E}_{6}(x,y)&=\mathcal{E}_{1}(x)+4\frac{\log(y)}{y-1}\\
\mathcal{E}_{7}(x,y,z)&=-\frac{1}{z}\mathcal{E}_{1}(x)+\frac{4}{y-z}\bigg(\frac{\log(y)}{y-1}-\frac{x}{z}\frac{\log(z)}{(z-1)}\bigg)\\
\mathcal{E}_{8}(x,y)&=\frac{5}{y}+\frac{2 \log(x)}{y} +\frac{2\log(y)}{y(y-1)}\\
\mathcal{E}_{9}(x,y)&=-\frac{2}{y}+\frac{4\log(x)}{y}+\frac{4\log(y)}{y-1} \,,
\end{align}
\end{subequations}
see Eqs.~\eqref{eq:CL_AMM} and~\eqref{eq:CL_AMM_mixing}.
\smallskip

For off-shell photons, the results are given in Eqs.~\eqref{eq:llgamma_LQ} and~\eqref{eq:llgamma_LQ_mixing}, we have
\begin{subequations}
	\begin{align}
	\mathcal{F}_{1}(x)&=5+\log(x)\\
	\mathcal{F}_{2}(x)&=9+4\log(x)\\
	\mathcal{F}_{3}(x,y)&=\frac{4}{y}\log(x)+\frac{5y-9}{y(y-1)}+\frac{4(2y-1)\log(y)}{y(y-1)^2}\\
	\mathcal{F}_{4}(x)&=2+\log(x)\\
	\mathcal{F}_{5}(x)&=7+4\log(x)\\
	\mathcal{F}_{6}(x)&=11+4\log(x)\\
	\mathcal{F}_{7}(x)&=1+\log(x)\\
	\mathcal{F}_{8}(x)&=3+\log(x)\\
	\mathcal{F}_{9}(x,y)&=\frac{2}{y}\log(x)+\frac{2(2y-1)}{y(y-1)}-\frac{2\log(y)}{y(y-1)^2}\\
	\mathcal{F}_{10}(x,y)&=\frac{4}{y}\log(x)+\frac{5y-5+4\log(y)}{y(y-1)}\,,\\
	\mathcal{F}_{11}(y,z)&=-\frac{4}{yz}\log(x)-\frac{5}{yz}+\frac{4\log(y)}{y(y-1)(y-z)}-\frac{4\log(z)}{z(z-1)(y-z)}\,.
	\end{align}
\end{subequations}
The $\mathcal{H}$-functions are defined for the $Z$ and $W$ decays in Sections~\ref{sec:Zll} and~\ref{sec:Wlnu}
\begin{subequations}
	\label{eq:H_functions}
	\begin{align}
	\mathcal{H}_{0}(x)&=x\big(1+\log(x)\big)\\
	\mathcal{H}_{1}(x)&=\frac{\log(x)}{x-1}\\
	\mathcal{H}_{2}(x)&=\frac{3\big(x^2-1-2x\log(x)\big)}{(x-1)^3}\\
	\mathcal{H}_{3}(x)&=\frac{2-2x+(1+x)\log(x)}{(x-1)^3}\\
	\mathcal{H}_{4}(x,y)&=\frac{1}{(x-1)(y-1)}+\frac{x\log(x)}{(x-1)^2(x-y)}+\frac{y\log(y)}{(y-1)^2(y-x)}\\
	\mathcal{H}_{5}(x)&=\frac{2-2x+(x+1)\log(x)}{(x-1)^2}\,,\\
	\mathcal{H}_{6}(x,y)&=\frac{2-x-y}{(x-1)(y-1)(x-y)}+\frac{(2x^2-x-y)\log(x)}{(x-1)^2(x-y)^2}-\frac{(2y^2-x-y)\log(y)}{(y-1)^2(x-y)^2} \,.
	\end{align}
\end{subequations}
Finally, we have the loop functions used for Higgs decays in Section~\ref{sec:hll}
\begin{subequations}
	\begin{align}
	\mathcal{J}_{t}(x,y)&=2(x-4)\log(y)-8+\frac{13}{3}x-\frac{x^2y}{5}\\
	\mathcal{J}_{1}(x)&=\frac{x-1-x\log(x)}{x(x-1)^2}\\
	\mathcal{J}_{2}(x,y)&=\frac{\log(x)}{(x-1)(x-y)}-\frac{y}{x}\frac{\log(y)}{(y-1)(x-y)} \\
	\mathcal{J}_{3}(x)&=\frac{x-1-x\log(x)}{(x-1)^2}\,.
	\end{align}
\end{subequations}
\medskip

\begin{boldmath}
\subsection{Exact Result for $\ell\ell\gamma$}
\label{app:ellellgamma}
\end{boldmath}
If we expand the amplitudes obtained from the diagrams in Fig.~\ref{fig:AMM_LQ} up to first non-vanishing order in the external masses and momenta, we obtain~\cite{Crivellin:2017dsk,Crivellin:2018qmi}
\begin{align}
\begin{aligned}
C^{L}_{\ell_{f}\ell_{i}}&=\frac{-N_{c}}{4}\sum_{q}\bigg[\frac{m_{\ell_f}\Gamma^{L,a*}_{q_{j}\ell_{f}}\Gamma^{L,a}_{q_{j}\ell_{i}}+m_{\ell_i}\Gamma^{R,a*}_{q_{j}\ell_{f}}\Gamma^{R,a}_{q_{j}\ell_{i}}}{6 M_{a}^2}\big(1+3Q_{q}\big)\\ &\quad+\frac{m_{q_j}\Gamma_{q_{j}\ell_{f}}^{R,a*}\Gamma_{q_{j}\ell_{i}}^{L,a}}{M_{a}^2}\Big(1-2Q_{q}-2Q_{q}\log\!\Big(\frac{\mu_{\ell}^2}{M_{a}^2}\Big)\Big)\bigg]\,,
\label{eq:llgamma_CL}
\end{aligned}
\end{align}
where $q$ can be, depending on the LQ representations, either a (charge-conjugated) up- or down-type quark and $Q_{q}$ refers to its electric charge, i.e. $Q_{q}=\{\pm 1/3,\pm 2/3\}$. The quark flavour index $j$ runs from 1 to 3.
\smallskip

Note that we naively integrated out the internal LQs and quarks at the same scale. Therefore, in the case of light internal quarks, i.e. all except the top quark, the contribution contains both the hard matching part, the mixing within the effective theory and the soft contribution. For this reason, care is required if the internal quarks are lighter than the incoming lepton (e.g. the charm contribution to $\tau\to\mu\gamma$) since the RGE only contributes from the LQ scale down to the scale of the process and not to the scale of the internal quark. Therefore, we defined $\mu_{\ell}$ in \eq{eq:llgamma_CL} as follows
\begin{align}
\mu_{\ell}=
\begin{cases}
m_{q_{j}} & m_{q_{j}}>m_{\ell_i}\\
m_{\ell_i} & m_{q_{j}}\leq m_{\ell_i}
\end{cases}\,.
\end{align}
Next we give the exact results for off-shell photons, whereof the expanded expressions are given in Eqs.~\eqref{eq:llgamma_LQ} and~\eqref{eq:llgamma_LQ_mixing}. They read
\begin{subequations}
\label{eq:l3l_off_photon}
\begin{align}
\widehat{\Xi}_{fi}^{L}&=\sum_{q}\frac{N_{c}\Gamma_{q_{j}\ell_{f}}^{L,a*}\Gamma_{q_{j}\ell_{i}}^{L,a}}{576\pi^2 M_{a}^2}\mathcal{F}_{Q_{q}}\Big(\frac{m_{q_j}^2}{M_{a}^2}\Big)\,,
\\
\widehat{\Xi}_{fi}^{R}&=\sum_{q}\frac{N_{c}\Gamma_{q_{j}\ell_{f}}^{R,a*}\Gamma_{q_{j}\ell_{i}}^{R,a}}{576\pi^2 M_{a}^2}\mathcal{F}_{Q_{q}}\Big(\frac{m_{q_j}^2}{M_{a}^2}\Big)\,,
\end{align}
\end{subequations}
with $\widehat{\Xi}^{L(R)}_{fi}$ defined in \eqref{eq:offshell_photon_amp} and
\begin{align}
\mathcal{F}_{Q_{q}}(y)&=2+18Q_{q}+12Q_{q}\log(y)\,.
\end{align}
Again, $j$ runs from 1 to 3.
\medskip

\begin{boldmath}
\subsection{Exact Results for $Z\ell\ell$, $Z\nu\nu$, $W\ell\nu$ and $h\ell\ell$}
\end{boldmath}
In this section we give the exact expressions for the $Z$ and $W$ decays. The $\tilde{T}^{Q}$ and $\tilde{B}^{W_i}$ matrices, used in this section, are given in Appendix~\ref{app:matrices}. In this whole section, the $M_i$ stand for the diagonal bilinear mass terms in the charge eigenstates, given in \eq{eq:masses_diagonal}. It is implied by the corresponding coupling matrix $\Gamma^i$ with same index $i$ which of the eigenstates is concerned, e.g. $\Gamma_{u^c \ell}$ corresponds to $M^{-1/3}$.
\smallskip
For the $Z$ decays, we use the conventions defined in Eq.~\eqref{eq:def_Zll_and_Zvv}, this time with
\begin{subequations}
\begin{align}
\Lambda_{\ell_f\ell_i}^{L(R)}\!\big(q^2\big)&=\Lambda_{\text{SM}}^{L(R)}(q^2)\delta_{fi}+\sum_{Q}\Delta_{L(R),fi}^{Q}\big(q^2\big)+\tilde{\Delta}_{L(R),fi}^{Q}\,,\\
\Theta_{\nu_{f}\nu_{i}}\big(q^2\big)&=\Theta_{\text{SM}}(q^2)\delta_{fi} + \sum_{Q}\Theta_{fi}^{Q}\big(q^2\big)+\tilde{\Theta}_{fi}^{Q} \,,
\end{align}
\end{subequations}
where contrary to \eq{eq:effectice_Z_couplings} we show the results sorted by the charges of the LQs, since we do not distinguish between the cases with and without LQ mixing. Hence, the results cannot be grouped by representation. For $Q=-1/3$ we have for the $m_{t}$-enhanced contributions
\begin{subequations}
\begin{align}
\Delta^{-1/3}_{L,fi}\big(q^2\big)&=\frac{-N_{c}\Gamma_{t^{c}\ell_{f}}^{L,a*}\Gamma_{t^{c}\ell_{i}}^{L,b}}{32\pi^2}\bigg[\frac{m_{t}^2}{M_{a}^2}\Big[\delta_{ab}\Big(1+\log\!\Big(\frac{m_{t}^2}{M_{a}^2}\Big)\Big)+\tilde{T}^{-1/3}_{ab}\mathcal{H}_{1}\Big(\frac{M_{b}^2}{M_{a}^2}\Big)\Big]\nonumber\\
&\quad-\frac{q^2}{18M_{a}^2}\Big[2\tilde{T}_{ab}^{-1/3}\mathcal{H}_{2}\Big(\frac{M_{b}^2}{M_{a}^2}\Big)+\delta_{ab}\Big(11-10s_{w}^2+2(3-4s_{w}^2)\log\!\Big(\frac{m_{t}^2}{M_{a}^2}\Big)\Big)\Big]\bigg]\,,\\
\Delta^{-1/3}_{R,fi}\big(q^2\big)&=\frac{N_{c}\Gamma_{t^{c}\ell_{f}}^{R,a*}\Gamma_{t^{c}\ell_{i}}^{R,b}}{32\pi^2}\bigg[\frac{m_{t}^2}{M_{a}^2}\Big[\delta_{ab}\Big(1+\log\!\Big(\frac{m_{t}^2}{M_{a}^2}\Big)\Big)-\tilde{T}_{ab}^{-1/3}\mathcal{H}_{1}\Big(\frac{M_{b}^2}{M_{a}^2}\Big)\Big]\nonumber\\
&\quad+\frac{q^2}{18M_{a}^2}\Big[2\tilde{T}_{ab}^{-1/3}\mathcal{H}_{2}\Big(\frac{M_{b}^2}{M_{a}^2}\Big)-\delta_{ab}\Big(3+10s_{w}^2+8s_{w}^2\log\!\Big(\frac{m_{t}^2}{M_{a}^2}\Big)\Big)\Big]\bigg]\,,
\intertext{
and the case with light up-type quarks yields
}
\Delta^{-1/3}_{L,fi}\big(q^2\big)&=\sum_{j=1}^2\frac{N_{c}\Gamma_{u_{j}^{c}\ell_{f}}^{L,a*}\Gamma_{u_{j}^{c}\ell_{i}}^{L,b}}{864\pi^2}\frac{q^2}{M_{a}^2}\bigg[3\tilde{T}_{ab}^{-1/3}\mathcal{H}_{2}\Big(\frac{M_{b}^2}{M_{a}^2}\Big)\nonumber\\
&\quad-\delta_{ab}\Big(3-3i\pi(4s_{w}^2-3)-5s_{w}^2+3(4s_{w}^2-3)\log\!\Big(\frac{q^2}{M_{a}^2}\Big)\Big)\bigg]\,,\\
\Delta^{-1/3}_{R,fi}\big(q^2\big)&=\sum_{j=1}^{2}\frac{N_{c}\Gamma_{u_{j}^{c}\ell_{f}}^{R,a*}\Gamma_{u_{j}^{c}\ell_{i}}^{R,b}}{864\pi^2}\frac{q^2}{M_{a}^2}\bigg[3\tilde{T}_{ab}^{-1/3}\mathcal{H}_{2}\Big(\frac{M_{b}^2}{M_{a}^2}\Big)\nonumber\\
&\quad+\delta_{ab}\Big(5s_{w}^2+12i\pi s_{w}^2-12s_{w}^2\log\!\Big(\frac{q^2}{M_{a}^2}\Big)\Big)\bigg]\,.
\end{align}
\end{subequations}
The terms, induced by LQ mixing, read
\begin{subequations}
\begin{align}
\tilde{\Delta}^{-1/3}_{L,fi}&=\sum_{j=1}^{3}\frac{N_{c}\Gamma_{u_{j}^{c}\ell_{f}}^{L,a*}\Gamma_{u_{j}^{c}\ell_{i}}^{L,b}}{64\pi^2}\tilde{T}_{ab}^{-1/3}\bigg(3+2\log\!\Big(\frac{\mu^2}{M_{a}^2}\Big)-2\mathcal{H}_{1}\Big(\frac{M_{a}^2}{M_{b}^2}\Big)\bigg)\,,\\
\tilde{\Delta}^{-1/3}_{R,fi}&=\sum_{j=1}^{3}\frac{N_{c}\Gamma_{u_{j}^{c}\ell_{f}}^{R,a*}\Gamma_{u_{j}^{c}\ell_{i}}^{R,b}}{64\pi^2}\tilde{T}_{ab}^{-1/3}\bigg(3+2\log\!\Big(\frac{\mu^2}{M_{a}^2}\Big)-2\mathcal{H}_{1}\Big(\frac{M_{a}^2}{M_{b}^2}\Big)\bigg)\,.
\end{align}
\end{subequations}
For $Q=2/3$ we have (massless) down-type quark effects
\begin{subequations}
\begin{align}
\Delta^{2/3}_{L,fi}\big(q^2\big)&=\sum_{j=1}^{3}\frac{-N_{c}\Gamma_{d_{j}\ell_{f}}^{L,b*}\Gamma_{d_{j}\ell_{i}}^{L,a}}{864\pi^2}\frac{q^2}{M_{a}^2}\bigg[3\tilde{T}_{ab}^{2/3}\mathcal{H}_{2}\Big(\frac{M_{b}^2}{M_{a}^2}\Big)\nonumber\\
&\quad-\delta_{ab}\Big(4s_{w}^2+6i\pi s_{w}^2 -6s_{w}^2\log\!\Big(\frac{q^2}{M_{a}^2}\Big)\Big)\bigg]\,,\\
\Delta^{2/3}_{R,fi}\big(q^2\big)&=\sum_{j=1}^{3}\frac{-N_{c}\Gamma_{d_{j}\ell_{f}}^{R,b*}\Gamma_{d_{j}\ell_{i}}^{R,a}}{864\pi^2}\frac{q^2}{M_{a}^2}\bigg[3\tilde{T}^{2/3}_{ab}\mathcal{H}_{2}\Big(\frac{M_{b}^2}{M_{a}^2}\Big)\nonumber\\
&\quad+\delta_{ab}\Big(3-3i\pi(2s_{w}^2-3)-4s_{w}^2+3(2s_{w}^2-3)\log\!\Big(\frac{q^2}{M_{a}^2}\Big)\Big)\bigg]\,,
\end{align}
\end{subequations}
again with $a,b=\{1,2,3\}$ and the mixing terms read
\begin{subequations}
\begin{align}
\tilde{\Delta}^{2/3}_{L,fi}&=\sum_{j=1}^{3}\!\frac{-N_{c}\Gamma_{d_{j}\ell_{f}}^{L,b*}\Gamma_{d_{j}\ell_{i}}^{L,a}}{128\pi^2}\bigg(\!2(2\tilde{T}^{2/3}_{ab}-\delta_{ab})\log\!\Big(\frac{\mu^2}{M_{a}^2}\Big)+6\tilde{T}_{ab}^{2/3}-\delta_{ab}\!-4\tilde{T}_{ab}^{2/3}\mathcal{H}_{1}\Big(\frac{M_{a}^2}{M_{b}^2}\Big)\!\bigg)\,,\\
\tilde{\Delta}^{2/3}_{R,fi}&=\sum_{j=1}^{3}\!\frac{-N_{c}\Gamma_{d_{j}\ell_{f}}^{R,b*}\Gamma_{d_{j}\ell_{i}}^{R,a}}{128\pi^2}\bigg(\!2(2\tilde{T}^{2/3}_{ab}+\delta_{ab})\log\!\Big(\frac{\mu^2}{M_{a}^2}\Big)+6\tilde{T}_{ab}^{2/3}+\delta_{ab}\!-4\tilde{T}_{ab}^{2/3}\mathcal{H}_{1}\Big(\frac{M_{a}^2}{M_{b}^2}\Big)\!\bigg)\,.
\end{align}
\end{subequations}
For the LQs with electric charge $Q=-4/3$ we have
\begin{subequations}
\begin{align}
\Delta^{-4/3}_{L,fi}\big(q^2\big)&=\sum_{j=1}^{3}\frac{N_{c}\Gamma_{d_{j}^{c}\ell_f}^{L,a*}\Gamma_{d_{j}^{c}\ell_i}^{L,b}}{864\pi^2}\frac{q^2}{M_{a}^2}\bigg[3\tilde{T}^{-4/3}_{ab}\mathcal{H}_{2}\Big(\frac{M_{b}^2}{M_{a}}\Big)\nonumber\\
&\quad+\delta_{ab}\Big(3-3i\pi(2s_{w}^2-3)+2s_{w}^2+3(2s_w^2-3)\log\!\Big(\frac{q^2}{M_{a}^2}\Big)\Big)\bigg]\,,\\
\Delta^{-4/3}_{R,fi}\big(q^2\big)&=\sum_{j=1}^{3}\frac{N_{c}\Gamma_{d_{j}^{c}\ell_f}^{R,a*}\Gamma_{d_{j}^{c}\ell_i}^{R,b}}{864\pi^2}\frac{q^2}{M_{a}^2}\bigg[3\tilde{T}^{-4/3}_{ab}\mathcal{H}_{2}\Big(\frac{M_{b}^2}{M_{a}}\Big)\nonumber\\
&\quad+\delta_{ab}\Big(2s_{w}^2-6i\pi s_{w}^2 +6s_{w}^2\log\!\Big(\frac{q^2}{M_{a}^2}\Big)\Big)\bigg]\,,
\end{align}
\end{subequations} 
with $a,b=\{1,2\}$ and
\begin{subequations}
\begin{align}
\tilde{\Delta}_{L,fi}^{-4/3}&=\sum_{j=1}^{3}\frac{N_{c}\Gamma_{d_{j}^{c}\ell_f}^{L,a*}\Gamma_{d_{j}^{c}\ell_i}^{L,b}}{64\pi^2}\bigg(2(\tilde{T}_{ab}^{-4/3}+\delta_{ab})\log\!\Big(\frac{\mu^2}{M_{a}^2}\Big)\nonumber\\
&\quad+3\tilde{T}_{ab}^{-4/3}+\delta_{ab}-2\tilde{T}_{ab}^{-4/3}\mathcal{H}_{1}\Big(\frac{M_{a}^2}{M_{b}^2}\Big)\!\bigg)\,,\\
\tilde{\Delta}_{R,fi}^{-4/3}&=\sum_{j=1}^{3}\frac{N_{c}\Gamma_{d_{j}^{c}\ell_f}^{R,a*}\Gamma_{d_{j}^{c}\ell_i}^{R,b}}{64\pi^2}\tilde{T}_{ab}^{-4/3}\bigg(2\log\!\Big(\frac{\mu^2}{M_{a}^2}\Big)+3-2\mathcal{H}_{1}\Big(\frac{M_{a}^2}{M_{b}^2}\Big)\bigg)\,,
\end{align}
\end{subequations}
and for $Q=5/3$
\begin{subequations}
\begin{align}
\Delta^{5/3}_{L,fi}\big(q^2\big)&=\frac{-N_{c}\Gamma_{t\ell_{f}}^{L*}\Gamma_{t\ell_{i}}^{L}}{32\pi^2}\bigg[\mathcal{H}_{0}\Big(\frac{m_{t}^2}{M^2}\Big)-\frac{q^2}{9M^2}\Big(1+7s_{w}^2+4s_{w}^2\log\!\Big(\frac{m_{t}^2}{M^2}\Big)\Big)\bigg]\,,\\
\Delta^{5/3}_{L,fi}\big(q^2\big)&=\sum_{j=1}^{2}\frac{-N_{c}\Gamma_{u_{j}\ell_{f}}^{L*}\Gamma_{u_{j}\ell_{i}}^{L}}{1728\pi^2}\frac{q^2}{M^2}\bigg[3-2s_{w}^2+24i\pi s_{w}^2-24s_{w}^2\log\!\Big(\frac{q^2}{M^2}\Big)\bigg]\,,\\
\Delta^{5/3}_{R,fi}\big(q^2\big)&=\frac{N_{c}\Gamma_{t\ell_{f}}^{R*}\Gamma_{t\ell_{i}}^{R}}{32\pi^2}\bigg[\mathcal{H}_{0}\Big(\frac{m_{t}^2}{M^2}\Big)-\frac{q^2}{9M^2}\Big(6-7s_{w}^2+(3-4s_{w}^2)\log\!\Big(\frac{m_{t}^2}{M^2}\Big)\Big)\bigg]\,,\\
\Delta^{5/3}_{R,fi}\big(q^2\big)&=\sum_{j=1}^{2}\frac{N_{c}\Gamma_{u_{j}\ell_{f}}^{R*}\Gamma_{u_{j}\ell_{i}}^{R}}{1728\pi^2}\frac{q^2}{M^2}\bigg[3+2s_{w}^2-6i\pi(4s_{w}^2-3)+6(4s_{w}^2-3)\log\!\Big(\frac{q^2}{M^2}\Big)\bigg]\,,
\end{align}
\end{subequations}
with $M^2=m_{2}^2+v^2(Y_{22}+Y_{2})$.
\smallskip

For $Z\to\nu_{i}\bar{\nu}_{f}$ with left-handed neutrinos only, we have for $Q=-1/3$ with (charge-conjugated) down-type quarks
\begin{align}
\begin{aligned}
\Theta_{fi}^{-1/3}(q^2)&=\sum_{j=1}^{3}\frac{N_{c}\Gamma_{d_{j}^{c}\nu_{f}}^{L,a*}\Gamma_{d_{j}^{c}\nu_{i}}^{L,b}}{864\pi^2}\frac{q^2}{M_{a}^2}\bigg[3\tilde{T}_{ab}^{-1/3}\mathcal{H}_{2}\Big(\frac{M_{b}^2}{M_{a}^2}\Big)\\
&\quad+\delta_{ab}\Big(3-s_{w}^2+3i\pi(3-2s_{w}^2)-3(3-2s_{w}^2)\log\!\Big(\frac{q^2}{M_{a}^2}\Big)\Big)\bigg]
\\
&\quad-\sum_{j=1}^{3}\frac{N_{c}\Gamma_{d_{j}\nu_{f}}^{L,a*}\Gamma_{d_{j}\nu_{i}}^{L,b}}{864\pi^2}\frac{q^2}{M_{a}^2}\bigg[3\tilde{T}_{ab}^{-1/3}\mathcal{H}_{2}\Big(\frac{M_{b}^2}{M_{a}^2}\Big)\\
&\quad-\delta_{ab}\Big(s_{w}^2\!+6i\pi s_{w}^2 -6s_{w}^2\log\!\Big(\frac{q^2}{M_{a}^2}\Big)\Big)\bigg]\,.
\end{aligned}
\end{align}
The LQ indices run like $a,b=\{1,2,3\}$. Analogously to $Z\to\ell_{f}^{-}\ell_{i}^{+}$ we have the $\tilde{\Theta}$ terms, originating from the LQ mixing
\begin{align}
\tilde{\Theta}_{fi}^{-1/3}&=\sum_{j=1}^3\frac{N_{c}\Gamma_{d_{j}^{c}\nu_{f}}^{L,b*}\Gamma_{d_{j}^{c}\nu_{i}}^{L,a}}{64\pi^2}\tilde{T}_{ab}^{-1/3}\bigg(3+2\log\!\Big(\frac{\mu^2}{M_{a}^2}\Big)-2\mathcal{H}_{1}\Big(\frac{M_{a}^2}{M_{b}^2}\Big)\bigg)\nonumber\\
&\quad-\sum_{j=1}^3\frac{N_{c}\Gamma_{d_{j}\nu_{f}}^{L,a*}\Gamma_{d_{j}\nu_{i}}^{L,b}}{128\pi^2}\bigg(2(2\tilde{T}_{ab}^{-1/3}+\delta_{ab})\log\!\Big(\frac{\mu^2}{M_{a}^2}\Big)\\
&\quad+6\tilde{T}_{ab}^{-1/3}+\delta_{ab}-4\tilde{T}_{ab}^{-1/3}\mathcal{H}_{1}\Big(\frac{M_{a}^2}{M_{b}^2}\Big)\!\bigg)\,.\nonumber
\end{align}
In case of $Q=2/3$, we have the diagrams which include a heavy top quark
\begin{align}
\begin{aligned}
\Theta_{fi}^{2/3}(q^2)&=\frac{N_{c}\Gamma_{t\nu_{f}}^{L,b*}\Gamma_{t\nu_{i}}^{L,a}}{32\pi^2}\bigg[\frac{m_{t}^2}{M_{a}^2}\Big[\tilde{T}_{ab}^{2/3}\mathcal{H}_{1}\Big(\frac{M_{b}^2}{M_{a}^2}\Big)-\delta_{ab}\Big(\frac{1}{2}+\log\!\Big(\frac{m_{t}^2}{M_{a}^2}\Big)\Big)\Big]
\\
&\quad-\frac{q^2}{18M_{a}^2}\Big[2\tilde{T}_{ab}^{2/3}\mathcal{H}_{2}\Big(\frac{M_{b}^2}{M_{a}^2}\Big)-\delta_{ab}\Big(3+12s_{w}^2+8s_{w}^2\log\!\Big(\frac{m_{t}^2}{M_{a}^2}\Big)\Big)\Big]\bigg]
\\
&\quad-\frac{N_{c}\Gamma_{t^{c}\nu_{f}}^{L,a*}\Gamma_{t^{c}\nu_{i}}^{L,b}}{32\pi^2}\bigg[\frac{m_{t}^2}{M_{a}^2}\Big[\tilde{T}_{ab}^{2/3}\mathcal{H}_{1}\Big(\frac{M_{b}^2}{M_{a}^2}\Big)+\delta_{ab}\log\!\Big(\frac{m_{t}^2}{M_{a}^2}\Big)
\Big]
\\
&\quad-\frac{q^2}{18M_{a}^2}\Big[2\tilde{T}_{ab}^{2/3}\mathcal{H}_{2}\Big(\frac{M_{b}^2}{M_{a}^2}\Big)+
\delta_{ab}\Big(11-12s_{w}^2+(6-8s_{w}^2)\log\!\Big(\frac{m_{t}^2}{M_{a}^2}\Big)\Big)\Big]\bigg]\,,
\end{aligned}
\end{align}
and light up-type quarks
\begin{align}
\begin{aligned}
\Theta_{fi}^{2/3}(q^2)&=\sum_{j=1}^{2}\frac{-N_{c}\Gamma_{u_{j}\nu_{f}}^{L,b*}\Gamma_{u_{j}\nu_{i}}^{L,a}}{864\pi^2}\frac{q^2}{M_{a}^2}\bigg[3\tilde{T}_{ab}^{2/3}\mathcal{H}_{2}\Big(\frac{M_{b}^2}{M_{a}^2}\Big)\\
&\quad+\delta_{ab}\Big(2s_{w}^2+12i\pi s_{w}^2-12s_{w}^2\log\!\Big(\frac{q^2}{M_{a}^2}\Big)\Big)\bigg]
\\
&\quad+\sum_{j=1}^2\frac{N_{c}\Gamma_{u_{j}^{c}\nu_{f}}^{L,a*}\Gamma_{u_{j}^{c}\nu_{i}}^{L,b}}{864\pi^2}\frac{q^2}{M_{a}^2}\bigg[3\tilde{T}_{ab}^{2/3}\mathcal{H}_{2}\Big(\frac{M_{b}^2}{M_{a}^2}\Big)
\\
&\quad-\delta_{ab}\Big(3-2s_{w}^2+3i\pi(3-4s_{w}^2)+3(4s_{w}^2-3)\log\!\Big(\frac{q^2}{M_{a}^2}\Big)\Big)\bigg]\,.
\end{aligned}
\end{align}
The LQ indices take the values $a,b=\{1,2,3\}$. We finally have
\begin{align}
\begin{aligned}
\tilde{\Theta}_{fi}^{2/3}&=\sum_{j=1}^{3}\frac{-N_{c}\Gamma_{u_{j}\nu_{f}}^{L,b*}\Gamma_{u_{j}\nu_{i}}^{L,a}}{128\pi^2}\bigg(2(2\tilde{T}_{ab}^{2/3}+\delta_{ab})\log\!\Big(\frac{\mu^2}{M_{a}^2}\Big)+6\tilde{T}_{ab}^{2/3}+\delta_{ab}-4\tilde{T}_{ab}^{2/3}\mathcal{H}_{1}\Big(\frac{M_{a}^2}{M_{b}^2}\Big)\bigg)
\\
&\quad+\sum_{j=1}^{3}\frac{N_{c}\Gamma_{u_{j}^{c}\nu_{f}}^{L,a*}\Gamma_{u_{j}^{c}\nu_{i}}^{L,b}}{64\pi^2}\bigg(2(\tilde{T}_{ab}^{2/3}-\delta_{ab})\log\!\Big(\frac{\mu^2}{M_{a}^2}\Big)+3\tilde{T}_{ab}^{2/3}-\delta_{ab}-2\tilde{T}_{ab}^{2/3}\mathcal{H}_{1}\Big(\frac{M_{a}^2}{M_{b}^2}\Big)\bigg)\,.
\end{aligned}
\end{align}
\smallskip

For $W\to\ell_{f}^{-}\bar{\nu}_{i}$ decays, our definition of the amplitude is given in \eq{eq:ampl_wlnu} and contrary to \eq{eq:wlnu_eff}, we use
\begin{align}
\Lambda^W_{\ell_{f}\nu_{i}}\!\big(q^2\big)&= \Lambda^{W}_{\text{SM}}(q^2)\delta_{fi} + \Lambda^{q}_{fi}\big(q^2\big)+\tilde{\Lambda}^{q}_{fi} +\Lambda^{q^{c}}_{fi}\big(q^2\big)+\tilde{\Lambda}^{q^c}_{fi}\ ,
\end{align}
where we choose to group the results by the fact whether a quark ($q$) or a charge-conjugated quark ($q^c$) runs in the loop. Because of obvious reasons, a grouping by representation is again not possible. We have
\begin{align}
\begin{aligned}
\Lambda_{fi}^{q^c}\big(q^2\big)&=\frac{N_{c}}{64\pi^2}\bigg[\frac{m_{t}^2}{M_{b}^2}\Gamma_{t^{c}\nu_{f}}^{L,b*}\Gamma_{t^{c}\nu_{i}}^{L,b}+\frac{m_{t}^2}{M_{a}^2}\Big[\Gamma_{t^{c}\ell_{f}}^{L,a*}\Gamma_{t^{c}\ell_{i}}^{L,a}-2V_{3k}^{*}\Gamma_{t^{c}\ell_{f}}^{L,a*}\Gamma_{d_{k}^{c}\nu_{i}}^{L,a}\log\!\Big(\frac{m_{t}^2}{M_{a}^2}\Big)\\
&\quad-2\tilde{B}_{ab}^{W_2}\Gamma_{t^{c}\ell_{f}}^{L,a*}\Gamma_{t^{c}\nu_{i}}^{L,b}\mathcal{H}_{1}\Big(\frac{M_{b}^2}{M_{a}^2}\Big)\Big]+\frac{2q^2}{9M_{a}^2}\Big[6V_{3k}^{*}\Gamma_{t^{c}\ell_{f}}^{L,a*}\Gamma_{d_{k}^{c}\nu_{i}}^{L,a}\Big(1+\log\!\Big(\frac{m_{t}^2}{M_{a}^2}\Big)\Big)\\
&\quad+\tilde{B}_{ca}^{W_1}\Gamma_{b^{c}\ell_{f}}^{L,c*}\Gamma_{b^{c}\nu_{i}}^{L,a}\mathcal{H}_{2}\Big(\frac{M_{c}^2}{M_{a}^2}\Big)+\tilde{B}_{ab}^{W_2}\Gamma_{t^{c}\ell_{f}}^{L,a*}\Gamma_{t^{c}\nu_{i}}^{L,b}\mathcal{H}_{2}\Big(\frac{M_{b}^2}{M_{a}^2}\Big)\Big]\bigg]\,,
\end{aligned}
\end{align}
and its massless case
\begin{align}
\begin{aligned}
\Lambda_{fi}^{q^c}\big(q^2\big)&=\sum_{j=1}^{2}\frac{N_{c}}{576\pi^2}\frac{q^2}{M_{a}^2}\bigg[4V_{jk}^{*}\Gamma_{u_{j}^{c}\ell_{f}}^{L,a*}\Gamma_{d_{k}^{c}\nu_{i}}^{L,a}\Big(\log\!\Big(\frac{q^2}{M_{a}^2}\Big)-1+3i\pi\Big)\\
&\quad+2\tilde{B}_{ca}^{W_1}\Gamma_{d_{j}^{c}\ell_{f}}^{L,c*}\Gamma_{d_{j}^{c}\nu_{i}}^{L,a}\mathcal{H}_{2}\Big(\frac{M_{c}^2}{M_{a}^2}\Big)+2\tilde{B}_{ab}^{W_2}\Gamma_{u_{j}^{c}\ell_{f}}^{L,a*}\Gamma_{u_{j}^{c}\nu_{i}}^{L,b}\mathcal{H}_{2}\Big(\frac{M_{b}^2}{M_{a}^2}\Big)\bigg]\,.
\end{aligned}
\end{align}
The LQ indices $a$ and $b$ run from 1 to 3, while $c=\{1,2\}$. In case of quarks in the loop we have
\begin{align}
\begin{aligned}
\Lambda_{fi}^{q}\big(q^2\big)&=\frac{N_{c}}{64\pi^2}\bigg[\frac{m_{t}^2}{M_{b}^2}\Big(\Gamma_{t\nu_{f}}^{L,b*}\Gamma_{t\nu_{i}}^{L,b}-2\tilde{B}_{b}^{W_3}\Gamma_{t\ell_{f}}^{L*}\Gamma_{t\nu_{i}}^{L,b}\mathcal{H}_{1}\Big(\frac{M^2}{M_{b}^2}\Big)\Big)+\frac{m_{t}^2}{M^2}\Gamma_{t\ell_{f}}^{L*}\Gamma_{t\ell_{i}}^{L}\\
&+\sum_{j=1}^{3}\frac{2q^2}{9M_{b}^2}\Big(\tilde{B}_{ab}^{W_2}\Gamma_{d_{j}\ell_{f}}^{L,b*}\Gamma_{d_{j}\nu_i}^{L,a}\mathcal{H}_{2}\Big(\frac{M_{a}^2}{M_{b}^2}\Big)+\tilde{B}_{b}^{W_3}\Gamma_{u_{j}\ell_{f}}^{L*}\Gamma_{u_{j}\nu_{i}}^{L,b}\mathcal{H}_{2}\Big(\frac{M^2}{M_{b}^2}\Big)\Big)\bigg]\,.
\end{aligned}
\end{align}
Here, the index $b$ runs from 1 to 3. The $\tilde{\Lambda}$'s read
\begin{subequations}
\begin{align}
\tilde{\Lambda}^{q^c}_{fi}&=\frac{N_{c}}{64\pi^2}\bigg[\Big(2\tilde{B}_{ca}^{W_{1}}\Gamma_{d^{c}_{j}\ell_{f}}^{L,c*}\Gamma_{d^{c}_{j}\nu_{i}}^{L,a}+2\tilde{B}_{ab}^{W_2}\Gamma_{u_{j}^{c}\ell_{f}}^{L,a*}\Gamma_{u_{j}^{c}\nu_{i}}^{L,b}-2V_{kj}\Gamma_{u_{k}^{c}\ell_{f}}^{L,a*}\Gamma_{d_{j}^{c}\nu_{i}}^{L,a}-\Gamma_{d_{j}^{c}\nu_{f}}^{L,a*}\Gamma_{d_{j}^{c}\nu_{i}}^{L,a}\nonumber\\
&\quad-\Gamma_{u_{j}^{c}\ell_{f}}^{L,a*}\Gamma_{u_{j}^{c}\ell_{i}}^{L,a}\Big)\log\!\Big(\frac{\mu^2}{M_{a}^2}\Big)-\Gamma_{u_{j}^{c}\nu_{f}}^{L,b*}\Gamma_{u_{j}^{c}\nu_{i}}^{L,b}\log\!\Big(\frac{\mu^2}{M_{b}^2}\Big)-\Gamma_{d_{j}^{c}\ell_{f}}^{L,c*}\Gamma_{d_{j}^{c}\ell_{i}}^{L,c}\log\!\Big(\frac{\mu^2}{M_{c}^2}\Big)\\
&\quad+2\tilde{B}_{ca}^{W_{1}}\Gamma_{d_{j}^{c}\ell_{f}}^{L,c*}\Gamma_{d_{j}^{c}\nu_{i}}^{L,a}\Big(1-\mathcal{H}_{1}\Big(\frac{M_{a}^2}{M_{c}^2}\Big)\Big)+2\tilde{B}_{ab}^{W_2}\Gamma_{u_{j}^{c}\ell_{f}}^{L,a*}\Gamma_{u_{j}^{c}\nu_{i}}^{L,b}\Big(1-\mathcal{H}_{1}\Big(\frac{M_{a}^2}{M_{b}^2}\Big)\Big)\bigg]\,,\nonumber\\
\tilde{\Lambda}^{q}_{fi}&=\frac{-N_{c}}{64\pi^2}\bigg[\Big(2\tilde{B}_{ab}^{W_2}\Gamma_{d_{j}\ell_{f}}^{L,b*}\Gamma_{d_{j}\nu_i}^{L,a}+\Gamma_{d_{j}\nu_{f}}^{L,a*}\Gamma_{d_{j}\nu_{i}}^{L,a}\Big)\log\!\Big(\frac{\mu^2}{M_{a}^2}\Big)+\Gamma_{u_{j}\ell_{f}}^{L*}\Gamma_{u_{j}\ell_{i}}^{L}\log\!\Big(\frac{\mu^2}{M^2}\Big)\nonumber\\
&\quad+\Big(2\tilde{B}_{b}^{W_3}\Gamma_{u_{j}\ell_{f}}^{L*}\Gamma_{u_{j}\nu_{i}}^{L,b}+\Gamma_{d_{j}\ell_{f}}^{L,b*}\Gamma_{d_{j}\ell_{i}}^{L,b}+\Gamma_{u_{j}\nu_{f}}^{L,b*}\Gamma_{u_{j}\nu_{i}}^{L,b}\Big)\log\!\Big(\frac{\mu^2}{M_{b}^2}\Big)\\
&\quad+2\tilde{B}_{ab}^{W_2}\Gamma_{d_{j}\ell_{f}}^{L,b*}\Gamma_{d_{j}\nu_i}^{L,a}\Big(1-\mathcal{H}_{1}\Big(\frac{M_{a}^2}{M_{b}^2}\Big)\Big)+2\tilde{B}_{b}^{W_3}\Gamma_{u_{j}\ell_{f}}^{L*}\Gamma_{u_{j}\nu_{i}}^{L,b}\Big(1-\mathcal{H}_{1}\Big(\frac{M_{b}^2}{M^2}\Big)\Big)\bigg]\,,\nonumber
\end{align}
\end{subequations}
again with $a,b=\{1,2,3\}$ and $c=\{1,2\}$.
\smallskip

Next, we give the results for the Higgs decay into a pair of charged leptons, where our amplitude is defined in Eq.~\eqref{eq:amplitude_hll}. Again, we sort the results by the charge eigenstate contributions
\begin{align}
\Upsilon_{\ell_{f}\ell_{i}}^{L(R)}(q^2)&=\delta_{fi}+\sum_{Q}\Upsilon_{L(R),fi}^{Q}(q^2)+\tilde{\Upsilon}_{L(R),fi}^{Q}\,.
\end{align}
In case of a heavy top quark in the loop, the results read
\begin{subequations}
\begin{align}
\Upsilon_{L,fi}^{-1/3}(q^2)&=\frac{-N_{c}\Gamma_{t^{c}\ell_{f}}^{R,a*}\Gamma_{t^{c}\ell_{i}}^{L,b}}{64\pi^2 M_{a}^2}\frac{m_{t}}{m_{fi}} \bigg(m_{t}^2\delta_{ab}\mathcal{J}_{t}\Big(\frac{q^2}{m_{t}^2},\frac{m_{t}^2}{M_{a}^2}\Big)-4v\tilde{\Gamma}_{ab}^{-1/3}\mathcal{H}_{1}\Big(\frac{M_{b}^2}{M_{a}^2}\Big)\bigg)\,,\label{eq:hll_Y_func}\\
\Upsilon_{L,fi}^{5/3}(q^2)&=\frac{-N_{c}\Gamma_{t\ell_{f}}^{R*}\Gamma_{t\ell_{i}}^{L}}{64\pi^2 M^2}\frac{m_{t}}{m_{fi}} \bigg(m_{t}^2\mathcal{J}_{t}\Big(\frac{q^2}{m_{t}^2},\frac{m_{t}^2}{M^2}\Big)-8\frac{v^2Y_{2}}{\sqrt{2}}\bigg) \,,
\end{align}
\end{subequations}
with $M^2=m_{2}^2+v^2(Y_{22}+Y_{2})$.
\smallskip

In all the scenarios where $q^2\equiv m_{h}^2\gg m_{q_j}^2\gg m_{fi}^2$ the Higgs mass gives the dominant contribution
\begin{subequations}
\begin{align}
\Upsilon_{L,fi}^{Q}(q^2)&=\sum_{j}\frac{N_{c}\Gamma_{q_{j}\ell_{f}}^{R,a*}\Gamma_{q_{j}\ell_{i}}^{L,b}}{64\pi^2 M_{a}^2}\frac{m_{q_j}}{m_{fi}}\mathcal{J}^{Q}_{ab}(q^2)\,,
\label{eq:Y_L_higgs}\\
\Upsilon_{L,fi}^{5/3}(q^2)&=\sum_{j=1}^2\frac{N_{c}\Gamma_{u_{j}\ell_{f}}^{R*}\Gamma_{u_{j}\ell_{i}}^{L}}{64\pi^2 M^2}\frac{m_{u_j}}{m_{fi}}\mathcal{J}^{5/3}(q^2)\,,
\end{align}
\end{subequations}
with
\begin{subequations}
\begin{align}
\mathcal{J}^{Q}_{ab}(q^2)&=q^2\delta_{ab}\Big(2\log\!\Big(\frac{q^2}{M_{a}^2}\Big)-2i\pi-1\Big)-4v\tilde{\Gamma}_{ab}^{Q}\mathcal{H}_{1}\Big(\frac{M_{b}^2}{M_{a}^2}\Big)\,, \label{eq:hll_J_func}\\
\mathcal{J}^{5/3}(q^2)&=q^2\Big(2\log\!\Big(\frac{q^2}{M^2}\Big)-2i\pi-1\Big)-8\frac{Y_{2}}{\sqrt{2}}v^2\,.
\end{align}
\end{subequations}
In Eq.~\eqref{eq:Y_L_higgs} the range of $j$ depends on whether we have up- ($j=\{1,2\}$) or down-type ($j=\{1,2,3\}$) quarks in the loop, since we treat the top separately. Finally, we consider as a last scenario $m_{fi}\gg m_{q_j}\sim 0$ and we have
\begin{subequations}
\begin{align}
\Upsilon_{L,fi}^{Q}&=\frac{vN_{c}\tilde{\Gamma}_{ab}^{Q}}{32\pi^2}\bigg(\frac{m_{\ell_f}}{m_{fi}}\frac{\Gamma_{q_{j}\ell_{f}}^{L,a*}\Gamma_{q_{j}\ell_{i}}^{L,b}}{M_{a}^2}\mathcal{J}_{3}\Big(\frac{M_{b}^2}{M_{a}^2}\Big)+\frac{m_{\ell_i}}{m_{fi}}\frac{\Gamma_{q_{j}\ell_{f}}^{R,a*}\Gamma_{q_{j}\ell_{i}}^{R,b}}{M_{b}^2}\mathcal{J}_{3}\Big(\frac{M_{a}^2}{M_{b}^2}\Big)\bigg)\,,\\
\Upsilon_{L,fi}^{5/3}&=\frac{-\sqrt{2}v^{2}N_{c}Y_{2}}{64\pi^2 M^2}\bigg(\frac{m_{\ell_{f}}}{m_{fi}}\lambda_{jf}^{2RL*}\lambda_{ji}^{2RL}+\frac{m_{\ell_i}}{m_{fi}}\lambda_{jf}^{2LR*}\lambda_{ji}^{2LR}\bigg)\,.
\end{align}
\end{subequations}
\medskip

\begin{boldmath}
\subsection{Higgs, $Z$ and $W$ Boson Coupling Matrices}
\label{app:matrices}
\end{boldmath}
The Higgs-LQ interaction matrices $\tilde{\Gamma}$, used in Eqs.~\eqref{eq:hll_J_func} and~\eqref{eq:hll_Y_func}, and $\tilde{\Lambda}$ expanded up to $\mathcal{O}(v^2/m_{\rm{LQ}}^2)$ read
\begin{subequations}
\begin{align}
\tilde{\Gamma}^{-1/3}&\approx\frac{1}{\sqrt{2}}
\begin{pmatrix}
2v\big(Y_{1}+\frac{|A_{\tilde{2}1}|^2}{m_{1}^2-\tilde{m}_{2}^2}\big) & A_{\tilde{2}1}^{*} & v\Big(2Y_{13}-\frac{A_{\tilde{2}3}A_{\tilde{2}1}^{*}(m_{1}^2+m_{3}^2-2\tilde{m}_{2}^2)}{(m_{1}^2-\tilde{m}_{2}^2)(\tilde{m}_{2}^2-m_{3}^2)}\Big)\\
A_{\tilde{2}1} & \tilde{\Gamma}_{22}^{-1/3} & A_{\tilde{2}3}\\
v\Big(2Y_{13}^{*}-\frac{A_{\tilde{2}1}A_{\tilde{2}3}^{*}(m_{1}^2+m_{3}^2-2\tilde{m}_{2}^2)}{(m_{1}^2-\tilde{m}_{2}^2)(\tilde{m}_{2}^2-m_{3}^2)}\Big)&A_{\tilde{2}3}^{*} & 2v\Big(Y_{3}+\frac{|A_{\tilde{2}3}|^2}{m_{3}-\tilde{m}_{2}^2}\Big)
\end{pmatrix}\,,\nonumber\\
\text{with}&\quad \tilde{\Gamma}_{22}^{-1/3}=2v\Big(Y_{\tilde{2}}-\frac{|A_{\tilde{2}1}|^2}{m_{1}^2-\tilde{m}_{2}^2}-\frac{|A_{\tilde{2}3}|^2}{m_{3}^2-\tilde{m}_{2}^2}\Big)\,,\\
\tilde{\Gamma}^{2/3}&\approx \frac{1}{\sqrt{2}}
\begin{pmatrix}
2v Y_{2} & 2v Y_{\tilde{2}2} & 0\\
2vY_{\tilde{2}2}^{*} & 2v\Big(Y_{\tilde{2}}+Y_{\tilde{2}\tilde{2}}-\frac{2|A_{\tilde{2}3}|^2}{m_{3}^2-\tilde{m}_{2}^2}\Big) & -\sqrt{2}A_{\tilde{2}3}\\
0 & -\sqrt{2}A_{\tilde{2}3}^{*} & 2v\Big(Y_{3}+Y_{33}+\frac{2|A_{\tilde{2}3}|^2}{m_{3}^2-\tilde{m}_{2}^2}\Big)
\end{pmatrix}\,,\\
\tilde{\Gamma}^{-4/3}&\approx \Gamma^{-4/3}\,,
\end{align}
\end{subequations}
and
\begin{subequations}
\begin{align}
\tilde{\Lambda}^{-1/3}&\approx \frac{1}{2}
\begin{pmatrix}
Y_{1}& v\Big(\frac{Y_{31}A_{\tilde{2}3}^{*}}{\tilde{m}_{2}^2-m_{3}^2}+\frac{(Y_{\tilde{2}}-Y_{1})A_{\tilde{2}1}^{*}}{m_{1}^2-\tilde{m}_{2}^2}\Big) & Y_{13} \\
v\Big(\frac{Y_{31}^{*}A_{\tilde{2}3}}{\tilde{m}_{2}^2-m_{3}^2}+\frac{(Y_{\tilde{2}}-Y_{1})A_{\tilde{2}1}}{m_{1}^2-\tilde{m}_{2}^2}\Big) & Y_{\tilde{2}} & v\Big(\frac{Y_{31}A_{\tilde{2}1}}{\tilde{m}_{2}^2-m_{1}^2}+\frac{(Y_{\tilde{2}}-Y_{3})A_{\tilde{2}3}}{m_{3}^2-\tilde{m}_{2}^2}\Big)\\
Y_{13}^{*} & v\Big(\frac{Y_{31}^{*}A_{\tilde{2}1}^{*}}{\tilde{m}_{2}^{2}-m_{1}^2}+\frac{(Y_{\tilde{2}}-Y_{3})A_{\tilde{2}3}^{*}}{m_{3}^2-\tilde{m}_{2}^2}\Big) & Y_{3}
\end{pmatrix}\,,\\
\tilde{\Lambda}^{2/3}&\approx \frac{1}{2}
\begin{pmatrix}
Y_{2} & Y_{\tilde{2}2} & \frac{\sqrt{2}vY_{\tilde{2}2}A_{\tilde{2}3}}{\tilde{m}_{2}^2-m_{3}^2} \\
Y_{\tilde{2}2}^{*} & Y_{\tilde{2}}+Y_{\tilde{2}\tilde{2}} & \frac{\sqrt{2}v(Y_{3}-Y_{\tilde{2}}-Y_{\tilde{2}\tilde{2}})A_{\tilde{2}3}}{m_{3}^2-\tilde{m}_{2}^2}\\
\frac{\sqrt{2}vY_{\tilde{2}2}^{*}A_{\tilde{2}3}^{*}}{\tilde{m}_{2}^2-m_{3}^2} & \frac{\sqrt{2}v(Y_{3}-Y_{\tilde{2}}-Y_{\tilde{2}\tilde{2}})A_{\tilde{2}3}^{*}}{m_{3}^2-\tilde{m}_{2}^2} & Y_{3} +Y_{33}
\end{pmatrix}\,,\\
\tilde{\Lambda}^{-4/3}&\approx\Lambda^{-4/3}\,.
\end{align}
\end{subequations}
\smallskip

Next, we will give the expressions for the weak isospin matrices $T^{Q}$, expanded in terms of $v$. They read in case of no LQ mixing
\begin{align}
\begin{aligned}
T^{-1/3}=
\begin{pmatrix}
0&0&0\\ 0&-\frac{1}{2}&0 \\ 0&0&0
\end{pmatrix}\,,
&&
T^{2/3}=
\begin{pmatrix}
-\frac{1}{2}&0&0\\ 0&\frac{1}{2}&0\\ 0&0&1
\end{pmatrix}\,,&&
T^{-4/3}=
\begin{pmatrix}
0&0\\ 0&-1
\end{pmatrix}\,,
&&
T^{5/3}=\frac{1}{2}\,,
\end{aligned}
\end{align}
using the basis defined in Eq. \eqref{eq:LQ_basis}. A unitary redefinition of the LQ fields in order to diagonalize the mass matrices in Eq. \eqref{eq:LQ_mixing_matrices} also affects the $T^{Q}$ matrices
\begin{align}
\tilde{T}^{Q}=W^{Q}T^{Q}W^{Q\dagger}\,.
\end{align}
Note that the LQ field redefinition has no impact the electromagnetic interaction, since the coupling matrix is proportional to the unit matrix and the $W^{Q}$ then cancel due to unitarity. If we use the perturbative diagonalization ansatz, we obtain 
\begin{subequations}
\label{eq:T3_mixing}
\begin{align}
\tilde{T}^{-1/3}&\!\approx\!
\begin{pmatrix}
\frac{-v^2|A_{\tilde 21}|^2}{2(m_{1}^2-\tilde{m}_{2}^2)^2} & \frac{vA_{\tilde 21}^{*}}{2(\tilde{m}_2^2 -m_{1}^2)} & \frac{v^2 A_{\tilde 2 3}A_{\tilde 21}^{*}}{2(m_{1}^2-\tilde{m}_{2}^2)(\tilde{m}_{2}^2-m_{3}^2)}\\
\frac{vA_{\tilde 21}}{2(\tilde{m}_{2}^2-m_1^2)}& -\frac{1}{2}\!+\!\frac{v^2}{2}\Big(\!\frac{|A_{\tilde 21}|^2}{(m_{1}^2-\tilde{m}_{2}^2)^2}\!+\!\frac{|A_{\tilde 2 3}|^2}{(\tilde{m}_{2}^2-m_{3}^2)^2}\!\Big) & \frac{v A_{\tilde 2 3}}{2(\tilde{m}_{2}^2-m_3^2)}\\
\frac{v^2 A_{\tilde 21}A_{\tilde 2 3}^{*}} {2(m_{1}^2-\tilde{m}_{2}^2)(\tilde{m}_{2}^2-m_{3}^2)}& \frac{vA_{\tilde 2 3}^{*}}{2(\tilde{m}_{2}^2-m_3^2)}& \frac{-v^2|A_{\tilde 2 3}|^2}{2(\tilde{m}_{2}^2-m_{3}^2)^2}
\end{pmatrix}\,,\\
\tilde{T}^{2/3}&\!\approx\!
\begin{pmatrix}
-\frac{1}{2} & \frac{v^2 Y_{\tilde 22}}{m_2^2-\tilde{m}_{2}^2} & 0\\
\frac{v^2 Y_{\tilde 22}^{*}}{m_2^2-\tilde{m}_{2}^2} & \frac{1}{2}\!+\!\frac{v^2|A_{\tilde 2 3}|^2}{(\tilde{m}_{2}^2-m_{3}^2)^2} &\frac{vA_{\tilde 2 3}}{\sqrt{2}(m_3^2-\tilde{m}_{2}^2)}\\
0 & \frac{v A_{\tilde 2 3}^{*}}{\sqrt{2}(m_3^2-\tilde{m}_{2}^2)} & 1\!-\!\frac{v^2 |A_{\tilde 2 3}|^2}{(\tilde{m}_{2}^2-m_{3}^2)^2}
\end{pmatrix}\,,\\
\tilde{T}^{-4/3}&\!\approx\!
\begin{pmatrix}
0 & \frac{\sqrt{2}v^2 Y_{3 \tilde 1}^{*}}{m_3^2-\tilde{m}_{1}^2}\\
\frac{\sqrt{2}v^2 Y_{3 \tilde 1}}{m_3^2-\tilde{m}_{1}^2} & -1
\end{pmatrix}\,,
\end{align}
\end{subequations}
valid up to $\mathcal{O}(v^2/m_{\rm{LQ}}^2)$. $T^{5/3}$ is not affected, since the LQ with charge $Q=5/3$ does not mix.
\smallskip

Analogously to the isospin coupling ot the $Z$ boson, different LQ generations mix under $W$ interactions. Without LQ mixing, the interactions with the $W$ boson can be written in terms of the following matrices
\begin{align}
B^{W_1}=
\begin{pmatrix}
0 & 0 & 0\\
0 & 0 & \sqrt{2}
\end{pmatrix}\,,&&
B^{W_2}=
\begin{pmatrix}
0 & 0 & 0 \\
0 & 1 & 0 \\
0 & 0 & -\sqrt{2}
\end{pmatrix}\,,&&
B^{W_3}=
\begin{pmatrix}
1 & 0 & 0
\end{pmatrix}\,,
\end{align}
arranging the LQ in their charge eigenstates according to Eq. \eqref{eq:LQ_basis}. $B^{W_1}$ describes the interaction of LQs with electric charges $Q=-4/3$ and $Q=-1/3$, $B^{W_2}$ the ones with $Q=-1/3$ and $Q=2/3$, $B^{W_3}$ with $Q=5/3$ and $Q=2/3$. If we include LQ mixing, the matrices expanded up to $\mathcal{O}(v^2/m_{\rm{LQ}}^2)$, then read
\begin{subequations}
\label{eq:BW_mixing}
\begin{align}
\tilde{B}^{W_1}&\!\approx\!
\begin{pmatrix}
0 & 0 & \frac{2v^2 Y_{3 \tilde 1}^{*}}{\tilde{m}_{1}-m_{3}^2}\\
\frac{\sqrt{2}v^2}{m_{1}^2-m_{3}^2}\Big(\!\frac{A_{\tilde 21}A_{\tilde 2 3}^{*}}{m_1^2-\tilde{m}_{2}^2}\!+\!Y_{31}^{*}\!\Big) & \frac{\sqrt{2}vA_{\tilde 2 3}^{*}}{\tilde{m}_{2}^2-m_{3}^2} & \sqrt{2}\!-\!\frac{v^2|A_{\tilde 2 3}|^2}{\sqrt{2}(\tilde{m}_{2}^2-m_{3}^2)^2}
\end{pmatrix}\,,\\
\tilde{B}^{W_2}&\!\approx\!
\begin{pmatrix}
0 & \frac{vA_{\tilde 21}^{*}}{m_{1}^2-\tilde{m}_{2}^2} & \frac{\sqrt{2}v^2}{m_{1}^2-m_{3}^2}\Big(\!\frac{A_{\tilde 2 3}A_{\tilde 21}^{*}}{m_{3}^2-\tilde{m}_{2}^2}\!-\!Y_{31}\!\Big)\\
\frac{v^2 Y_{\tilde 22}^{*}}{m_{2}^2-\tilde{m}_{2}^2} & 1\!-\!\frac{v^2}{2}\Big(\!\frac{|A_{\tilde 21}|^2}{(m_{1}^2-\tilde{m}_{2}^2)^2}\!-\!\frac{|A_{\tilde 2 3}|^2}{(\tilde{m}_{2}^2-m_{3}^2)^2}\!\Big) & 0\\
0 & \frac{vA_{\tilde 2 3}^{*}}{\tilde{m}_{2}^2-m_{3}^2} & -\sqrt{2}\!-\!\frac{v^2|A_{\tilde 2 3}|^2}{\sqrt{2}(m_{3}^2-\tilde{m}_{2}^2)^2}
\end{pmatrix}\,,\\
\tilde{B}^{W_3}&\!\approx\!
\begin{pmatrix}
1 & \frac{-v^2 Y_{\tilde 22}^{*}}{m_{2}^2-\tilde{m}_{2}^2} & 0
\end{pmatrix}\,.
\end{align}
\end{subequations}
\newpage

\begin{boldmath}
\subsection{$4\ell$}
\end{boldmath}

Besides the penguin diagrams, mediated by the off-shell photon and $Z$ boson, we also have the box diagrams. The matching results on $\ell_{i}^{-}\to\ell_{f}^{-}\ell_{p}^{-}\ell_{r}^{+}$ read
\begin{subequations}
\begin{align}
C_{fipr}^{V\,LL}&=\frac{-N_c}{256\pi^2}\bigg[\Big(
\Gamma_{u^{c}_{j}\ell_{f}}^{L,a*}\Gamma_{u_{j}^{c}\ell_{i}}^{L,b}\Gamma_{u^{c}_{k}\ell_{p}}^{L,b*}\Gamma_{u_{k}^{c}\ell_{r}}^{L,a}
+\Gamma_{u^{c}_{j}\ell_{p}}^{L,a*}\Gamma_{u_{j}^{c}\ell_{i}}^{L,b}\Gamma_{u^{c}_{k}\ell_{f}}^{L,b*}\Gamma_{u_{k}^{c}\ell_{r}}^{L,a}
\Big)D_{2}\big(m_{k}^2,m_{j}^2,M_{a}^2,M_{b}^{2}\big)\nonumber\\
&\quad+\Big(\Gamma_{u_{j}\ell_{f}}^{L*}\Gamma_{u_{j}\ell_{i}}^{L}\Gamma_{u_{k}\ell_{p}}^{L*}\Gamma_{u_{k}\ell_{r}}^{L}
+\Gamma_{u_{j}\ell_{p}}^{L*}\Gamma_{u_{j}\ell_{i}}^{L}\Gamma_{u_{k}\ell_{f}}^{L*}\Gamma_{u_{k}\ell_{r}}^{L}\Big)D_{2}\big(m_{k}^2,m_{j}^2,M^2,M^{2}\big)\nonumber\\
&\quad+\Big(\Gamma_{d_{j}\ell_{f}}^{L,a*}\Gamma_{d_{j}\ell_{i}}^{L,b}\Gamma_{d_{k}\ell_{p}}^{L,b*}\Gamma_{d_{k}\ell_{r}}^{L,a}
+\Gamma_{d_{j}\ell_{p}}^{L,a*}\Gamma_{d_{j}\ell_{i}}^{L,b}\Gamma_{d_{k}\ell_{f}}^{L,b*}\Gamma_{d_{k}\ell_{r}}^{L,a}\Big)C_{0}\big(0,M_{a}^2,M_{b}^2\big)\\
&\quad+\Big(\Gamma_{d^{c}_{j}\ell_{f}}^{L,a*}\Gamma_{d_{j}^{c}\ell_{i}}^{L,b}\Gamma_{d_{k}^{c}\ell_{p}}^{L,b*}\Gamma_{d_{k}^{c}\ell_{r}}^{L,a}+\Gamma_{d^{c}_{j}\ell_{p}}^{L,a*}\Gamma_{d_{j}^{c}\ell_{i}}^{L,b}\Gamma_{d_{k}^{c}\ell_{f}}^{L,b*}\Gamma_{d_{k}^{c}\ell_{r}}^{L,a}\Big)C_{0}\big(0,M_{a}^2,M_{b}^2\big)\bigg]\,,\nonumber\\
C_{fipr}^{V\,LR}&=\frac{-N_c}{128\pi^2}\Big[
\Gamma_{u^{c}_{j}\ell_{f}}^{L,a*}\Gamma_{u_{j}^{c}\ell_{i}}^{L,b}\Gamma_{u^{c}_{k}\ell_{p}}^{R,b*}\Gamma_{u_{k}^{c}\ell_{r}}^{R,a}D_{2}\big(m_{k}^2,m_{j}^2,M_{a}^2,M_{b}^{2}\big)\nonumber\\
&\quad-2\Gamma_{t^{c}\ell_{f}}^{L,a*}\Gamma_{t^{c}\ell_{i}}^{L,a}\Gamma_{t^{c}\ell_{p}}^{R,b*}\Gamma_{t^{c}\ell_{r}}^{R,b}m_{t}^2D_{0}\big(m_{t}^2,m_{t}^2,M_{a}^2,M_{b}^2\big)\nonumber\\
&\quad+\Gamma_{u_{j}\ell_{f}}^{L*}\Gamma_{u_{j}\ell_{i}}^{L}\Gamma_{u_{k}\ell_{p}}^{R*}\Gamma_{u_{k}\ell_{r}}^{R}D_{2}\big(m_{k}^2,m_{j}^2,M^2,M^{2}\big)\\
&\quad-2\Gamma_{t\ell_{f}}^{L}\Gamma_{t\ell_{i}}^{L}\Gamma_{t\ell_{p}}^{R}\Gamma_{t\ell_{r}}^{R}m_{t}^2D_{0}\big(m_{t}^2,m_{t}^2,M^2,M^2\big)\nonumber\\
&\quad+\Gamma_{d_{j}\ell_{f}}^{L,a*}\Gamma_{d_{j}\ell_{i}}^{L,b}\Gamma_{d_{k}\ell_{p}}^{R,b*}\Gamma_{d_{k}\ell_{r}}^{R,a}C_{0}\big(0,M_{a}^2,M_{b}^2\big)+\Gamma_{d^{c}_{j}\ell_{f}}^{L,a*}\Gamma_{d_{j}^{c}\ell_{i}}^{L,b}\Gamma_{d_{k}^{c}\ell_{p}}^{R,b*}\Gamma_{d_{k}^{c}\ell_{r}}^{R,a}C_{0}\big(0,M_{a}^2,M_{b}^2\big)\Big]\,,\nonumber\\
C_{fipr}^{S\,LL}&=\frac{-N_c \,m_{t}^2}{64\pi^2}\Big[2\Gamma_{t^{c}\ell_{f}}^{R,a*}\Gamma_{t^{c}\ell_{i}}^{L,b}\Gamma_{t^{c}\ell_{p}}^{R,b*}\Gamma_{t^{c}\ell_{r}}^{L,a}D_{0}(m_{t}^2,m_{t}^2,M_{a}^2,M_{b}^2)\nonumber\\
&\quad-\Gamma_{t^{c}\ell_{p}}^{R,a*}\Gamma_{t^{c}\ell_{i}}^{L,b}\Gamma_{t^{c}\ell_{f}}^{R,b*}\Gamma_{t^{c}\ell_{r}}^{L,a}D_{0}(m_{t}^2,m_{t}^2,M_{a}^2,M_{b}^2)\\
&\quad+\Gamma_{t\ell_{f}}^{R*}\Gamma_{t\ell_{i}}^{L}\Gamma_{t\ell_{p}}^{R}\Gamma_{t\ell_{r}}^{L}D_{0}(m_{t}^2,m_{t}^2,M^2,M^2)\Big] \,,\nonumber
\end{align}
\end{subequations}
where $C_{fipr}^{V\,RR}$, $C_{fipr}^{V\,RL}$ and $C_{fipr}^{S\,RR}$ are obtained by simply exchanging $\Gamma^{L}\leftrightarrow\Gamma^{R}$.
\medskip

\begin{boldmath}
\subsection{$2\ell 2\nu$}
\end{boldmath}

Here we show the box contributions, induced by the $Q=-1/3$ and $Q=2/3$ LQs. We obtain for $\ell_{i}^{-}\to\ell_{f}^{-}\nu_{p}\bar{\nu}_{r}$
\begin{subequations}
\begin{align}
D_{\ell_{f}\ell_{i}}^{L,pr}&=\frac{-N_c}{64\pi^2}\bigg[\Gamma_{u_{j}^{c}\ell_{f}}^{L,b*}\Gamma_{u_{j}^{c}\ell_{i}}^{L,a}\Big(\Gamma_{d_{k}^{c}\nu_{p}}^{L,a*}\Gamma_{d_{k}^{c}\nu_{r}}^{L,b}-\Gamma_{d_{k}\nu_{p}}^{L,b*}\Gamma_{d_{k}\nu_{r}}^{L,a}\Big)C_{0}(m_{u_j}^2,M_{a}^2,M_{b}^2)\nonumber\\
&\quad+\Gamma_{d_{k}\ell_{f}}^{L,b*}\Gamma_{d_{k}\ell_{i}}^{L,a}\Big(\Gamma_{u_{j}\nu_{p}}^{L,a*}\Gamma_{u_{j}\nu_{r}}^{L,b}-\Gamma_{u_{j}^{c}\nu_{p}}^{L,b*}\Gamma_{u_{j}^{c}\nu_{r}}^{L,a}\Big)C_{0}(m_{u_j}^2,M_{a}^2,M_{b}^2)\bigg]\,,\\
D_{\ell_{f}\ell_{i}}^{R,pr}&=\frac{-N_c}{64\pi^2}\bigg[\Gamma_{u_{j}^{c}\ell_{f}}^{R,b*}\Gamma_{u_{j}^{c}\ell_{i}}^{R,a}\Big(\Gamma_{d_{k}^{c}\nu_{p}}^{L,a*}\Gamma_{d_{k}^{c}\nu_{r}}^{L,b}-\Gamma_{d_{k}\nu_{p}}^{L,b*}\Gamma_{d_{k}\nu_{r}}^{L,a}\Big)C_{0}(m_{u_j}^2,M_{a}^2,M_{b}^2)\nonumber\\
&\quad+\Gamma_{d_{k}\ell_{f}}^{R,b*}\Gamma_{d_{k}\ell_{i}}^{R,a}\Big(\Gamma_{u_{j}\nu_{p}}^{L,a*}\Gamma_{u_{j}\nu_{r}}^{L,b}-\Gamma_{u_{j}^{c}\nu_{p}}^{L,b*}\Gamma_{u_{j}^{c}\nu_{r}}^{L,a}\Big)C_{0}(m_{u_j}^2,M_{a}^2,M_{b}^2)\bigg] \,.
\end{align}
\end{subequations}

\newpage

\bibliographystyle{JHEP}
\bibliography{BIB}

\end{document}